\documentclass[12pt]{article}
\usepackage{amsmath} 
\usepackage{amssymb}
\usepackage{bm}
\usepackage{bbm}
\usepackage{cite}
\usepackage{color} 
\definecolor{light-blue}{rgb}{0.8,0.85,1}
\definecolor{dark-blue}{rgb}{0,0,0.7}

\usepackage{graphicx}
\usepackage{slashed} 
\usepackage{tabularx}

\usepackage[colorlinks]{hyperref} 
\hypersetup{
citecolor=blue,
linkcolor=blue,
urlcolor=blue
}

\usepackage{epstopdf}
\usepackage{ifthen}
\newcommand{\draftmode}{off}
\ifthenelse{\equal{\draftmode}{on}}{
   
\setlength{\parindent}{0pt}                      
\setlength{\parskip}{5ex}                          

\pagestyle{headings}    
\pagestyle{myheadings}  
\markright{\hfill \today \hfill}

\usepackage[nofiglist,notablist,nomarkers]{endfloat} 
} 

\newcommand{\note}[1]{#1}

\newcommand{\mnote}[1]{}

\newcommand{\blue}{\color{blue}}

\newcommand{\Sec}[1]{Sec.~\ref{#1}}

\newcommand{\fig}[1]{Fig.~\ref{#1}}

\newcommand{\tab}[1]{Tab.~\ref{#1}}

\newcommand{\eq}[1]{Eq.~(\ref{#1})}

\begin{document}
\title{
\vspace{-1.0truecm}
\Large\bf
Cosmic ray propagation and dark matter in light of
the latest AMS-02 data}
\author{
Hong-Bo Jin$^{a,b}$, 
Yue-Liang Wu$^{a,c,d}$, \ 
and  Yu-Feng Zhou$^{a,c}$
\footnote{Emails: hbjin@bao.ac.cn,  ylwu@itp.ac.cn, yfzhou@itp.ac.cn}
\\ \\
\it $^{a)}$ State Key Laboratory of Theoretical Physics,
\it $^{b)}$ National   \\ 
\it Astronomical  Observatories, 
\it Chinese Academy of Sciences, \\
\it $^{c)}$ Kavli Institute for Theoretical Physics China,
\it Institute   of  \\
\it Theoretical Physics Chinese Academy of Sciences,\\
\it $^{d)}$University of Chinese Academy of Sciences,
\\ 
\it Beijing, 100190, P.R. China
}

\date{}
\maketitle

\vspace{-1.0truecm}
\begin{abstract}
The AMS-02 experiment is measuring 
the high energy cosmic rays with 
unprecedented accuracy.
We explore the possibility of determining the cosmic-ray propagation models
using the AMS-02 data $alone$.
A global Bayesian analysis of
the constraints on the cosmic-ray propagation models from 
the \mnote{(1)}\note{preliminary} AMS-02 data on 
the Boron to Carbon nuclei flux ratio and proton flux is performed,
with the assumption that 
the primary nucleon source is a broken power law in rigidity.
The ratio of
the diffusion coefficient $D_{0}$ to the diffusive halo height $Z_{h}$ 
is  determined with high accuracy
$D_{0}/Z_{h}\simeq 2.00\pm0.07~\text{cm}^{2}\text{s}^{-1}\text{kpc}^{-1}$,
and 
the value of the halo width is found to be
$Z_{h}\simeq 3.3$ kpc with uncertainty less than $50\%$.
As a consequence,
the typical uncertainties in the positron fraction predicted from 
dark matter (DM) annihilation is reduced to  a factor of two,
and that in the antiproton flux is about an order of magnitude.
Both of them are significantly smaller than that from the analyses prior to AMS-02.
Taking into account the uncertainties and correlations in 
the propagation parameters,
we derive conservative upper limits on 
the cross sections for DM annihilating into 
various standard model final states from the current PAMELA antiproton data.
We also investigate 
the  reconstruction capability  of the future high precision AMS-02 antiproton data on 
the DM properties.
The results show that 
for DM particles lighter than $\sim 100$ GeV and 
with typical thermal annihilation cross section,
the cross section can be well reconstructed  with 
uncertainties about  a factor of two
for the AMS-02 three-year data taking.

\end{abstract}

\newpage
\section{Introduction}

Although compelling evidence from 
astronomical observations has indicated that 
dark matter (DM) contributes to $26.8\%$ of 
the total energy density  of the Universe
\cite{Ade:2013zuv},
the particle nature of DM remains largely unknown.
If  DM particles in the galactic halo can annihilate or decay into 
the standard model (SM) final states, 
they may contribute to 
primary sources of cosmic-ray particles,
which can be probed by precision DM indirect detection experiments.

Recently, 
the Alpha Magnetic Spectrometer (AMS-02) collaboration has updated 
its measurement of the cosmic-ray positron fraction, i.e.,
the  ratio between cosmic-ray positron flux and the total flux of electrons and positrons
in the energy range of 0.5--500 GeV
\cite{Accardo:2014lma}. 
The high precision data indicate that 
the positron fraction increases with energy  in the energy range 8--270 GeV,
consistent with the   previouse measurements by 
PAMELA~\cite{
Adriani:2008zr,
Adriani:2010ib
}
and Fermi-LAT~\cite{
FermiLAT:2011ab
}
but with much higher accuracy.
For the first time, 
it was shown that the positron fraction ceases to increase at the energy  $\sim 270$ GeV.
The rise and the existence of a maximum in
the positron fraction  is unexpected from 
the conventional astrophysics in which 
the majority of positrons are believed  to be from 
the collisions of primary cosmic-ray nuclei with interstellar gas.
Besides astrophysical explanations, 
an exciting   possibility is that
the observed positron fraction excess  is due to
DM annihilation or decay in the galactic halo.

In DM interpretations, 
through analysing  the cosmic-ray positron anomaly,
the properties of DM particle such as 
its mass  and annihilation cross section or decay life-time can be inferred, 
and different DM models can be distinguished or even excluded
 ( for recent global analyses on AMS-02 data, see e.g. Refs 
 \cite{
Kopp:2013eka,
DeSimone:2013fia,
Yuan:2013eja,
Cholis:2013psa,
Feng:2013zca, 
Jin:2013nta,Liu:2013vha,
Bergstrom:2013jra, 
Ibarra:2013zia, 
Geng:2013nda, 
DiMauro:2014iia, 
Lin:2014vja,Ibe:2014qya,Cao:2014cda,
Mertsch:2014poa
 } ).
However, 
the conclusions  are in general sensitive to 
the choice of cosmic-ray propagation model,
cosmic-ray background
as well as the
profile of DM halo density distribution.
The main source of the uncertainty is related to 
that  in the propagation models. 
Analyses based on the data prior to AMS-02 
have shown that 
the uncertainties of this type 
can reach 
$\mathcal{O}(10)$ in  the prediction for  positron flux 
\cite{Delahaye:2007fr} 
and $\mathcal{O}(100)$ for anti-proton flux
for DM annihilation
\cite{Donato:2003xg}. 
Note that the backgrounds of 
primary and secondary cosmic-ray particles 
which are of crucial importance in identifying  the DM signals
also depend on the propagation models.

In the diffusion models of cosmic-ray propagation,
the major propagation parameters involve 
the diffusion halo height $Z_{h}$, 
the spatial diffusion coefficient $D_{0}$,
the convection velocity $V_{c}$ related to the galactic wind, 
the  Alfv$\grave{\mbox{e}}$n speed $V_{a}$ related to the reaccelaration, 
and the primary source terms,
etc..
The propagation models and parameters can be 
constrained by a set of astrophysical observables.
The ratio between the fluxes of cosmic-ray secondary and primary nuclei 
such as 
that  of  Boron to Carbon nuclei (B/C)
and the ratio of the radioactive isotopes such as 
that of Beryllium nuclei  ${}^{10}\text{Be}/{}^{9}\text{Be}$ are 
commonly used to determine these parameters
without knowing the primary sources
( for recent global fits, see e.g. 
\cite{Putze:2010zn,Trotta:2010mx,Liu:2011re,Auchettl:2011wi}
).
The primary source terms can be determined separately 
by the fluxes of primary cosmic-ray nuclei 
such as that of cosmic-ray protons.

Recently, 
the AMS-02 collaboration has reported the measurement of 
the B/C ratio  in the kinetic energy interval from 0.5 to 670 GeV/nucleon
with an unprecedented accuracy
\cite{Oliva:icrc2013}.
The AMS-02 experiment also released the data of 
proton flux as a function of rigidity from 1 GV to 1.8 TV
\cite{Haino:icrc2013},
which is consistent with the previous measurement 
made by PAMELA in the low rigidity range from 20 to 100 GV
\cite{Adriani:2011cu}. 
In the high rigidity region above $\sim 100$ GV, 
the proton spectrum measured by  AMS-02 is consistent with a single power law spectrum.
Under the assumption that the primary source is a broken power law in rigidity,
the two type of data can be used together to 
determine the cosmic-ray propagation parameters.

In light of the recent significant experimental progresses,
it is of interest to 
revisit  the constraints on the cosmic-ray propagation models
and explore the potential of the AMS-02 experiment on the capability of DM discovery.
In this work, 
we first determine the main propagation parameters 
through a global Bayesian analysis to the 
\note{preliminary}\mnote{(1)} AMS-02 data.
We follow the strategy of determining {\it both} the propagation
parameters and the primary sources in the same framework, 
using the data of B/C ratio and the proton flux. 
We show that the combination of B/C ratio and proton flux can
lift the degeneracy in $Z_{h}$ and $D_{0}$,
and
both the  parameters can be well determined 
by the AMS-02 data  alone. 
We find that
the ratio of the diffusion coefficient $D_{0}$ to the diffusive halo height $Z_{h}$ 
is determined with high accuracy
$D_{0}/Z_{h}\simeq 2.00\pm0.07~\text{cm}^{2}\text{s}^{-1}\text{kpc}^{-1}$,
and 
the best-fit value of the halo width is  
$Z_{h}\simeq 3.3$ kpc with uncertainty within $50\%$.
From the allowed regions of parameter space, 
we estimate the uncertainties in 
the positron fraction and 
antiproton fluxes predicted by DM annihilation.
We show that  
the uncertainties in the predicted 
positron fraction is within a factor of two and
that in the antiproton flux is within an order of magnitude, 
which are significantly smaller than that  from the previous analyses prior to AMS-02 
( see e.g. \cite{Putze:2010zn,Trotta:2010mx}).
We construct reference propagation models corresponding to the minimal, median and 
maximal antiproton fluxes from DM annihilation into $b$-quarks.
Combined with the PAMELA antiproton data,
we derive conservative upper limits on  the cross sections of DM annihilating into 
typical SM final states. 
We further project the sensitivity of the forthcoming AMS-02 data  on the antiproton flux.
The results show that for DM particle lighter than $\sim 100$ GeV
with a typical thermal annihilation cross section,
the cross section can be reconstructed  with uncertainties within a factor of two
for the AMS-02 three-year data taking.

This paper is organized as  follows.
In \Sec{sec:frame},
we outline the formulas describing the propagation of cosmic-ray particles.
In \Sec{sec:Bayes}, 
we briefly overview the method of Bayesian inference used in our analysis.
In \Sec{sec:fit}, 
we present results on constraining  the propagation models from the AMS-02 data of
cosmic-ray B/C ratio and  proton flux.
In \Sec{sec:positron},
we discuss the uncertainties in the prediction for positron fraction
from DM annihilation into typical leptonic final states.
In \Sec{sec:minmax},
we select typical propagation models corresponding to the minimal, median and maximal
antiproton fluxes from DM annihilation into $b\bar b$.
In \Sec{sec:pbarlimits},
taking into account the uncertainties in the propagation parameters,
we derive upper limits on the DM annihilation cross sections for typical annihilation channels
from PAMELA antiproton data. 
The reconstruction capability for the future AMS-02 data on the DM mass and annihilation cross 
sections is discussed.
The conclusions are given in \Sec{sec:conclusion}.

\section{Propagation of cosmic-ray charged particles}\label{sec:frame}

It has been recognized that the propagation of cosmic rays in the Galaxy
can be  effectively described as a process of diffusion
\cite{Berezinskii:1990}.  
In this section, 
we briefly overview the main features of the cosmic-ray diffusion within the Galaxy.
Detailed reviews of the transportation of processes can be found in Ref.
\cite{Maurin:2002ua}. 
The Galactic halo within which  the diffusion processes occur is parametrized by 
a cylinder with radius $R_{h} = 20$ kpc and half-height $Z_{h}=1-20$ kpc.   
The diffusion  equation for the cosmic-ray charged particles reads
(see e.g. \cite{Ginzburg:1990sk})
\begin{align}\label{eq:propagation}
  \frac{\partial \psi}{\partial t} =&
  \nabla (D_{xx}\nabla \psi -\boldsymbol{V}_{c} \psi)
  +\frac{\partial}{\partial p}p^{2} D_{pp}\frac{\partial}{\partial p} \frac{1}{p^{2}}\psi
  -\frac{\partial}{\partial p} \left[ \dot{p} \psi -\frac{p}{3}(\nabla\cdot \boldsymbol{V}_{c})\psi \right]
  \nonumber \\
  & -\frac{1}{\tau_{f}}\psi
  -\frac{1}{\tau_{r}}\psi
  +q(\boldsymbol{r},p)  ,
\end{align}
where $\psi(\boldsymbol{r},p,t)$ is  
the number density per unit of total particle momentum, 
which is related to the phase space density $f(\boldsymbol{r},\boldsymbol{p}, t)$ as
$\psi(\boldsymbol{r},p,t)=4\pi p^{2}f(\boldsymbol{r},\boldsymbol{p},t) $. 
For steady-state diffusion, it is assumed that  $\partial  \psi/\partial t=0$.
The number densities of cosmic-ray particles are vanishing at the boundary of the halo,
i.e.,
$\psi(R_{h},z,p)=\psi(R, \pm Z_{h},p)=0$.
The spatial diffusion coefficient $D_{xx}$ is energy dependent and 
can be parametrized as
\begin{align}\label{eq:29}
D_{xx}=\beta D_{0} \left( \frac{\rho}{\rho_{0}} \right)^{\delta}  ,
\end{align}
where $\rho=p/(Ze)$ is the rigidity of the cosmic-ray particle with electric charge $Ze$.
The the power spectral index $\delta$ can have  different values 
$\delta=\delta_{1(2)}$  
when $\rho$ is below (above) a reference rigidity $\rho_{0}$.  
The coefficient  $D_{0}$ is a normalization constant, 
and $\beta=v/c$ is the velocity of the cosmic-ray particle
with $c$ the speed of light.
The convection term in the diffusion equation is related to 
the drift of cosmic-ray particles from 
the Galactic disc due to the Galactic wind.  
The direction of the wind is  assumed to be 
along the direction  perpendicular to the galactic disc plane and
have opposite  sign above and below the disc.
The  diffusion in momentum space is described by 
the reacceleration parameter $D_{pp}$ 
which is related to the  velocity of disturbances in the hydrodynamical plasma, 
the so called Alfv$\grave{\mbox{e}}$n speed $V_{a}$ as follows~\cite{Ginzburg:1990sk}
\begin{align}\label{eq:Dpp}
D_{pp}=
\frac{4V_{a}^{2} p^{2}}
{3D_{xx}\delta
\left(4-\delta^{2}\right)
\left(4-\delta\right)w},
\end{align}
where $w$ characterise the level of turbulence. 
We take $w=1$ as only $V_{a}^{2}/w$ is  relevant in the calculation.
In \eq{eq:propagation},
the momentum loss rate is denoted  by $\dot{p}$ which 
could  be due to 
ionization in the interstellar medium neutral matter,
Coulomb scattering off thermal  electrons in ionized plasma,
bremsstrahlung,
synchrotron radiation,
and 
inverse Compton scattering, etc..
The parameter $\tau_{f}(\tau_{r})$ is 
the time scale for fragmentation (radioactive decay) of 
the cosmic-ray nuclei as they interact with interstellar hydrogen and helium.

High energy electrons/positrons loss  energy due to 
the processes like inverse Compton scattering and synchrotron radiation.
The typical propagation length is around a few kpc for electron energy 
around 100 GeV.
In the calculation of energy loss rate, 
the interstellar magnetic field in  cylinder coordinates $(R,z)$ is assumed 
to have the form
\begin{align}
 B(R,z)=B_{0} 
 \exp\left(-\frac{R-r_{\odot}}{R_{B}}\right)
 \exp\left(-\frac{|z|}{z_{B}} \right),
 \end{align}
where
$B_{0}=5\times 10^{-10}$~Tesla,
$R_{B}=10$~kpc,
$z_{B}=2$~kpc~\cite{Strong:1998fr},
and 
$r_{\odot}\approx 8.5$ kpc is the distance from the Sun to the galactic center.
The spectrum of a primary source term for a cosmic-ray nucleus  $A$  is 
assumed to have a broken power low behaviour 
\begin{align}\label{eq:primarySource}
\frac{dq_{A}(p)}{dp} 
\propto
\left( 
\frac{\rho}{\rho_{As}}
\right)^{\gamma_{A}} ~,
\end{align}
with $\gamma_{A}=\gamma_{A1}(\gamma_{A2})$ for 
the nucleus rigidity $\rho$ below (above) a reference rigidity $\rho_{As}$.
For cosmic-ray electrons, 
sometimes two breaks $\rho_{es1}$, $\rho_{es2}$ are introduced with
three power law indices $\gamma_{e1}$, $\gamma_{e2}$ and $\gamma_{e3}$.
The radial distribution of the source term can be determined by 
independent observables.  
Based on the distribution of SNR,
the spatial distribution of the primary sources is assumed to have  the following form
\cite{Case:1996}
\begin{align}\label{eq:source-distribution}
q_{A}(R,z)
=
q_{0} \left( \frac{R}{r_{\odot}} \right)^{\eta}
\exp
\left[
-\xi \frac{R-r_{\odot}}{r_{\odot}}
-\frac{|z|}{0.2~\text{kpc}}
 \right]~,
\end{align}
where $\eta=1.25$ and $\xi=3.56$ are adapted to reproduce 
the Fermi-LAT gamma-ray data of the 2nd Galactic quadrant 
\cite{
Strong:1998pw, 
Trotta:2010mx,
Tibaldo:2009spa
},
and $q_{0}$ is a normalization parameter.
In the 2D diffusion model, 
one can use the realistic non-uniform interstellar gas distribution of 
$\text{H}_{\text{I,II}}$ and $\text{H}_{2}$ 
determined from 21cm and CO surveys.

Secondary  cosmic-ray particles are created in 
collisions of primary cosmic-ray particles with interstellar  gas.
The secondary antiprotons are created dominantly from inelastic $pp$- and $p$He-collisions.
The corresponding source term  reads
\begin{align}
q(p)=
\beta c n_{i} \sum_{i=\text{H,He}}
\int dp'   \frac{\sigma_{i}(p,p')}{dp'} n_{p}(p')
\end{align}
where 
$n_{i}$ is the number density of interstellar hydrogen (helium),
$n_{p}$ is the number density of primary cosmic-ray proton per total momentum, 
and $d\sigma_{i}(p,p')/dp'$ is the differential cross section
for  $p+\text{H(He)}\to \bar p + X$.

The primary source term of cosmic-ray particles from 
the annihilation of Majorana DM particles has the  following form
\begin{align}\label{eq:ann-source}
q(\boldsymbol{r},p)=\frac{\rho(\boldsymbol{r})^2}{2 m_{\chi}^2}\langle \sigma v \rangle 
\sum_X \eta_X \frac{dN^{(X)}}{dp} ,
\end{align}
where $\langle \sigma v \rangle$ is 
the velocity-averaged DM annihilation cross section multiplied by DM relative velocity
(referred to as cross section) 
which is the quantity appears in 
the Boltzmann equation for
calculating the evolution of DM number density.
$\rho(\boldsymbol{r})$ is the DM energy density distribution function,
and
$dN^{(X)}/dp$ is the injection energy spectrum  of  antiprotons
from DM annihilating into SM final states through 
all possible  intermediate states $X$ with  
$\eta_X$ the corresponding branching fractions. 
The injection spectra $dN^{(X)}/dp$ from DM annihilation are calculated using 
the numerical package PYTHIA~v8.175
\cite{
Sjostrand:2007gs 
},
in which the long-lived particles such as neutron and $K_{L}$ are allowed to decay
and the final state interaction are taken into account.
Since PYTHIA~v8.15 the polarization and correlation of final states in  $\tau$-decays 
has been  taken into account
\cite{Ilten:2012zb}.

The fluxes of cosmic-ray particles from DM annihilation depend also on 
the choice of DM halo profile.  
N-body simulations suggest a universal form of the DM profile 
\begin{align}
\rho(r)=\rho_\odot
\left( \frac{r}{r_\odot} \right)^{-\gamma}
\left(
\frac{1+(r_\odot/r_s)^\alpha}{1+(r/r_\odot)^\alpha}
\right)^{(\beta-\gamma)/\alpha} ,
\end{align}
where 
$\rho_\odot \approx 0.43 \text{ GeV}\text{ cm}^{-3}$ is
the local DM energy density 
\cite{Salucci:2010qr
}.
The values of the parameters $\alpha$, $\beta$, $\gamma$ and $r_{s}$
for the  Navarfro-Frenk-White (NFW) profile~\cite{Navarro:1996gj},
the isothermal profile \cite{Bergstrom:1997fj} 
and the Moore profile 
\cite{Moore:1999nt, 
Diemand:2004wh} 
are summarized in \tab{tab:DMprofile}.
\begin{table}[htb]
\begin{center}
\begin{tabular}{ccccc}
\hline\hline
             	&$\alpha$& $\beta$& $\gamma$& $r_{s}$(kpc)\\
\hline
NFW		& 1.0 	& 3.0	& 1.0 	& 20\\ 
Isothermal& 2.0	&2.0		& 0 		& 3.5\\
Moore	 & 1.5	&3.0		&1.5		&28.0\\
\hline\hline
\end{tabular}
\caption{
Values of parameters $\alpha$, $\beta$, $\gamma$ and $r_{s}$ for three
DM halo models,
NFW 
\cite{Navarro:1996gj}, 
Isothermal 
\cite{Bergstrom:1997fj}, 
and 
Moore 
\cite{Moore:1999nt, 
Diemand:2004wh}. 
}
\label{tab:DMprofile}
\end{center}
\end{table}
An other widely adopted  DM profile is the  Einasto profile 
\cite{Einasto:2009zd 
}
\begin{align}
\rho(r)=\rho_\odot \exp
\left[
-\left( \frac{2}{\alpha_E}\right)
\left(\frac{r^{\alpha_E}-r_\odot^{\alpha_E}}{r_s^{\alpha_E}} \right)
\right] ,
\end{align}
with $\alpha_E\approx 0.17$ and $r_s\approx 20$ kpc.

The   interstellar flux of the cosmic-ray particle is related to its density function as 
\begin{align}\label{eq:38}
\Phi= \frac{v}{4\pi} \psi(\boldsymbol{r},p)~.
\end{align}
For high energy nuclei $v\approx c$. 
At the top of the atmosphere (TOA) of the Earth, 
the fluxes of cosmic-rays  are affected  by solar winds 
and the helioshperic magnetic field. 
This effect is taken into account using 
the force-field approximation
\cite{Gleeson:1968zza}.
In this approach,
$\Phi^{\text{TOA}}$ the cosmic-ray nuclei flux at the top of the atmosphere of the Earth 
which is measured by the experiments is related to the interstellar flux as follows
\begin{align}
\label{eq:45}
\Phi^{\text{TOA}}(E_{\text{TOA}})=\left(\frac{2m  E_{\text{TOA}}+E_{\text{TOA}}^{2}}{2m  E_{\text{kin}}+E_{\text{kin}}^{2}}\right)\Phi(E_{\text{kin}}) ,
\end{align}
where $E_{\text{TOA}}=E_{\text{kin}}-\phi_{F}$ is the kinetic energy of the cosmic-ray nuclei
at the top of the atmosphere of the Earth.

\mnote{(4)}
Analytical solutions to the propagation equation can be obtained in 
a simplified two-zone diffusion model in which
the thin galactic disk is approximated  by a delta-function $\delta(z)$
(for reviews, see e.g. \cite{Maurin:2002ua}).
For an illustration, 
let us consider a simple case where 
the reacceleration and energy loss terms are negligible, 
and 
$V_{c}$ is a constant along the $z$-direction.
The steady state propagation equation in this case can be written as
\begin{align}
0=D_{xx}  \nabla^{2}\psi -V_{c} \nabla \psi
- 2h\delta(z)\frac{1}{\tau_{f}}\psi
-\frac{1}{\tau_{r}}\psi 
+2h\delta(z) q(R,z,p) .
\end{align}
where $h\approx 0.1$ kpc is the half-height of the galactic disk used as 
a normalization factor. 
Using the Bessel expansion of the number density
\begin{align}\label{eq:Bessel}
\psi(R,z,p)=\sum_{i=1}^{\infty} \psi_{i}(z,p) J_{0} \left(\zeta_{i}\frac{R}{R_{h}}\right) ,
\end{align}
where $J_{0}(x)$ is 
the zero-th order Bessel function of the first kind
and 
$\zeta_{i}$ is the $i$-th zero of the Bessel function,
the equation for the coefficients $\psi_{i}(z,p) $ can be written as
\begin{align}
0=D_{xx}\left( \frac{\partial^{2}}{\partial z^{2}}-\frac{\zeta_{i}^{2}}{R_{h}^{2}}\right) \psi_{i}
-V_{c} \frac{\partial}{\partial z} \psi_{i }
-2h\delta(z) \frac{1}{\tau_{f}} \psi_{i}
-\frac{1}{\tau_{r}} \psi_{i}
+2h\delta(z) q_{i},
\end{align}
where $q_{i}$ are the coefficients of the Bessel expansion of the source term $q(R,z,p)$
similar to $\psi_{i}$ in \eq{eq:Bessel}.
The solution of the above equation at $z=0$ is given by 
\cite{Maurin:2002ua}
\begin{align}\label{eq:two-zone-solution}
\psi_{i}(0)=\frac{2h q_{i}}{V_{c}+2h/\tau_{f}+D_{xx} S_{i}\coth(S_{i} Z_{h}/2)} ,
\end{align} 
where
\begin{align}
S_{i}^{2}=\frac{V_{c}^{2}}{D_{xx}^{2}} +\frac{4}{D_{xx} \tau_{r}} +\frac{4\zeta^{2}_{i}}{R_{h}^{2}} .
\end{align}
In the limit $S_{i} Z_{h} \ll 1$ 
which is valid at sufficiently high energy,
one can use the power expansion 
$\coth(x)\approx 1/x+x/3+\mathcal{O}(x^{3})$ and obtain
\begin{align}\label{eq:DS}
D_{xx} S_{i} \coth(S_{i} Z_{h}/2)
\approx
\left(\frac{D_{xx}}{Z_{h}} \right)
\left(
	2
	+ \frac{V_{c}^{2} Z_{h}^{2}}{6 D_{xx}^{2}}
	+ \frac{2Z_{h}^{2}}{3D_{xx}\tau_{r}}
	+ \frac{2Z_{h}^{2}}{3 R_{h}^{2}}\zeta_{i}^{2} 
\right) .
\end{align}
Since $D_{xx}\propto D_{0}$, 
the above expression shows the well-known behaviour  that
the parameters  $D_{0}$ and $Z_{h}$ are 
almost degenerate.
This degeneracy is however slightly lifted by the two subleading contributions.
One is related to the decay of the radioactive species, 
and
the other one is related to the fixed halo radius $R_{h}$ 
which is common to all the cosmic-ray species.
\mnote{(2)}
The values of $D_{0}$ and $Z_{h}$ can be determined by 
fitting simultaneously to  
the B/C flux ratio and  
the ratio of the isotopes of Beryllium nuclei 
${}^{10}\text{Be}/{}^{9}\text{Be}$,
as ${}^{10}\text{Be}$ is radioactive and 
its propagation is directly sensitive to $D_{0}$.
An advantage of using  such flux ratios is that
the propagation parameters can be determined without 
the knowledge of  the primary sources.
On the other hand, 
as shown in \eq{eq:DS},
for a fixed value of $D_{0}/Z_{h}$, 
an increase of $Z_{h}$ will result in 
a slight decrease of the flux $\psi_{i}$ even for stable cosmic-ray species.  
Therefore, 
the stable primary cosmic-ray fluxes such as the proton flux 
can also be used 
together with the $B/C$ flux ratio
to  determine the values of   $Z_{h}$,
provided that 
the primary sources are specified and 
the data are precision enough.

The energy spectrum of  the proton flux is known to 
follow a single power law  
$\psi(0)\propto \rho^{-\gamma_{\psi}}$ in 
the energy range $\mathcal{O}(20-10^{7})$ GeV with 
$\gamma_{\psi} \approx 2.7$.
Since $D_{xx} \propto \rho^{\delta}$,
according to the solution of \eq{eq:two-zone-solution}, 
if the rigidity  dependence of the source term  is also 
a single power law  $q_{i} \propto \rho^{-\gamma}$,
then at high energies the approximate relation 
$\gamma_{\psi} \approx \gamma+\delta$ follows, 
which means  that 
for the proton spectrum
the two parameters $\gamma$ and $\delta$ are nearly degenerate.
However, at lower energies $E_{\text{kin}}\lesssim 20$~ GeV, 
the single power-law approximation  of the proton energy spectrum breaks down.
The energy redistribution processes 
such as
reacceleration, convection and solar modulation, etc. 
contribute to the changes in the spectral shape of the proton flux.
Thus $\gamma$ and $\delta$ can be determined individually 
by the proton flux together with 
other propagation parameters. 
Furthermore, 
the primary proton source term can also be  a broken power law in rigidity
as widely adopted in the diffusive re-acceleration models
\cite{
Moskalenko:2001ya,
Ptuskin:2005ax,
Trotta:2010mx
},
which is  also suggested independently by 
the $\gamma$-ray observation of the nearby molecular clouds
\cite{Neronov:2011wi}.
As it will be shown in detail in \Sec{sec:fit}, 
the combination of proton flux plus B/C ratio can
break the degeneracies between the parameters,
and 
allows for a determination of the propagation parameters 
$D_{0}$, $Z_{h}$, $V_{a}$, $\gamma_{p1,p2}$ and $\delta$ etc. with 
reasonable precisions.

In our numerical calculations,
we shall solve the diffusion equation of \eq{eq:propagation} 
using the publicly available  code  GALPROP v54
\cite{astro-ph/9807150,astro-ph/0106567,astro-ph/0101068,astro-ph/0210480,astro-ph/0510335}
which utilizes  realistic astronomical information on 
the distribution of interstellar gas and other data as input, 
and considers various kinds of data including 
primary and secondary nuclei, electrons and positrons, $\gamma$-rays, synchrotron  radiation, etc. 
in a self-consistent way. 
Other approaches based on simplified assumptions on 
the Galactic gas distribution 
which  allow  for fast  analytic solutions can be found in Refs.
\cite{astro-ph/0103150,
astro-ph/0212111,
astro-ph/0306207,
1001.0551,
Cirelli:2010xx
}.
The propagation parameters shall be determined from a global fit 
using Bayesian inference with Markov Chain Monte-Carlo method.

\section{Bayesian inference}\label{sec:Bayes}
The Bayesian inference is based on 
calculating the  posterior probability distribution function (PDF) of  
the unknown parameter set 
$\boldsymbol{\theta}=\{\theta_{1},\dots,\theta_{m}\}$ in a given model,
which actually updates our state of belief from the prior PDF of $\boldsymbol{\theta}$ 
after taking into account the information provided by  the experimental data set $D$.
The posterior PDF is related to the prior PDF by  the Bayes's therom
\begin{align}\label{eq:Bayes}
p(\boldsymbol{\theta}|D)=\frac{\mathcal{L}(D|\boldsymbol\theta)\pi(\boldsymbol\theta)}{p(D)} ,
\end{align}
where $\mathcal{L}(D|\boldsymbol\theta)$ is the likelihood function,
and 
$\pi(\boldsymbol\theta)$ is the prior PDF which 
encompasses our state of knowledge on the values of the parameters before 
the observation of  the data.
The quantity $p(D)$ is the Bayesian evidence 
which is obtained by integrating the product  of the likelihood and the prior over
the whole volume of the parameter space
\begin{align}
p(D)=\int_{V} \mathcal{L}(D|\theta)\pi(\boldsymbol\theta) d\theta .
\end{align}
The evidence is an important  quantity for Bayesian model comparison. 
It is straight forward to obtain the marginal PDFs of interested parameters 
$\{\theta_{1},\dots,\theta_{n}\} (n<m) $ by 
integrating out other nuisance parameters $\{\theta_{n+1},\dots,\theta_{m}\}$
\begin{align}
p(\theta_{1},\dots,\theta_{n})_{\text{marg}}=\int p(\boldsymbol\theta|D) \prod_{i=n+1}^{m}d\theta_{i} .
\end{align}
The marginal PDF  is often used in visual presentation. 	
If there is no preferred value of $\theta_{i}$ in the allowed range ($\theta_{i,\text{min}}$, $\theta_{i,\text{max}}$),
the priors can be taken as a flat distribution
\begin{align}\label{eq:priors}
\pi(\theta_{i}) \propto
\left\{
\begin{tabular}{ll}
1, &  \text{for } $\theta_{i,\text{min}}<\theta_{i}<\theta_{i,\text{max}}$
\\
0, & \text{otherwise}
\end{tabular}
\right. 
.
\end{align}
The likelihood function is often assumed to be Gaussian
\begin{align}
\mathcal{L}(D|\boldsymbol\theta)=
\prod_{i}
\frac{1}{\sqrt{2\pi \sigma_{i}^{2}}}
\exp\left[
	-\frac{(f_{\text{th},i}(\boldsymbol\theta)-f_{\text{exp},i})^{2}}{2\sigma_{i}^{2}}  
\right]  ,  
\end{align}
where $f_{\text{th},i}(\boldsymbol\theta)$ are 
the predicted $i$-th observable from the model which 
depends on the parameter set $\boldsymbol\theta$, 
and
$f_{\text{exp},i}$ are the ones measured by the experiment with uncertainty $\sigma_{i}$.
For experiments with only a few events observed, 
the form of the likelihood function can be taken as Poisson. 
When the form of the likelihood function is specified, 
the posterior PDF can be determined by sampling the distribution 
according to the prior PDF and the likelihood function	
using  Markov Chain Monte Carlo (MCMC) methods.
A commonly adopted algorithm is  Metropolis-Hastings MCMC
which is implemented in the numerical package {\tt CosmoMC}
\cite{Lewis:2002ah}. 
Other advanced sampling methods such as the  {\tt MultiNest} algorithm
are also commonly adopted
\cite{Feroz:2007kg, 
Feroz:2008xx}. 

The statistic mean value of a parameter $\theta$ can be obtained from
the posterior PDF $P(\boldsymbol\theta|D)$ in a straight forward manner.
Using the MCMC sequence 
$\{\theta^{(1)}_{i}, \theta^{(2)}_{i},\dots,\theta^{(N)}_{i}\}$ 
of the parameter $\theta_{i}$
with $N$ the length of the Markov chain, 
the mean (expectation) value $\langle \theta_{i} \rangle $ is given by 
\begin{align}
\langle \theta_{i} \rangle
=
\int \theta_{i} P(\theta_{i}|D) d\theta_{i}
=\frac{1}{N}\sum_{k=1}^{N} \theta^{(k)}_{i}.
\end{align}
The $1\sigma$ standard  deviation of the parameter $\theta_{i}$ is given by
$\sigma^{2}=\sum^{N}_{k=1} (\theta^{(k)}_{i}-\langle \theta_{i} \rangle)^{2}/(N-1)$.

\section{Constraining propagation models using AMS-02 data}\label{sec:fit}
The propagation models can be constrained by cosmic-ray data.
Since 
the statistics  of the  AMS-02 data on charged cosmic-ray particles are now
much higher than  that of other experiments 
and will continue to increase,
it is of interest to  consider 
constraining  the propagation models using the AMS-02 data alone.
One  advantage of this strategy is that 
the complicities involving the combination of 
the systematics of different type of experiments can be avoided.
Furthermore, 
all the current AMS-02 data are taken in the same period of solar activity, 
which makes it easier to estimate the effect of solar modulation consistently. 

The AMS-02 data of which we shall include in the analysis are
the spectra of the cosmic-ray nuclei ratio B/C (18 data points)~\cite{Oliva:icrc2013}
and the proton flux (100 data points)~\cite{Haino:icrc2013},
namely, 
the whole data set is 
\begin{align}\label{eq:dataSetAMS}
D=\{ D^{\text{AMS}}_{B/C}, D^{\text{AMS}}_{p} \} .
\end{align}
\mnote{(1)}
\note{
Note that 
the current data released by the AMS-02 collaboration is still preliminary, 
which can be different from the final published results. 
 }

Since we are   focusing on determining the propagation parameters,
the AMS-02 data of positrons and electrons 
\cite{Aguilar:2014mma} 
are not considered for the moment,
as it is known that 
they are unlikely to be fully  consistent with the conventional backgrounds,
which calls for exotic  contributions either from nearby astrophysical sources or  from DM interactions.

We adopt the conventional diffusive reaccelaration (DR) models in which $V_{c} \simeq 0$.
It has been shown that in the GALPROP approach
a nonvanishing  $V_{c}$ results in the predicted  peak of 
B/C spectrum to be too wide in comparison with the data
\cite{Strong:1998pw,Moskalenko:2001ya}.
We consider the case where 
$R=20$ kpc and $\delta_{1}=\delta_{2}\equiv \delta$,
thus there are 4 free parameters related to the cosmic-ray propagation: 
$Z_{h}$, $D_{0}$, $\delta$ and  $V_{a}$.
Two additional parameters 
$\gamma_{p1}$ and $\gamma_{p2}$  are introduced 
for the power-law indices of the primary source terms.
The break in rigidity of the primary source is fixed at 
$\rho_{ps}=10^{4}~\text{MV}$.
In the GALPROP code, 
the primary nuclei source term is normalized in such a way that
the proton flux $N_{p}$  at 
a reference kinetic energy $E_{\text{kin}}=$100 GeV is reproduced.
We find 
$N_{p}=4.83\pm0.02~\text{cm}^{-2}\text{sr}^{-1}\text{s}^{-1}\text{MeV}^{-1}$ 
from interpolating the AMS-02 proton flux data at $100$ GeV.
The solar modulation amplitude $\phi$ 
which affects the low energy spectra of the cosmic ray particles
correlates strongly with the power-law index $\gamma_{p1}$.
Fitting both $\phi$ and $\gamma_{p1}$ simultaneously  will significantly slow down
the convergence of the MCMC sampling. 
Thus, in this work we  fix the value of $\phi$ at $\phi=550$ MV. 
As a cross check, after the global fit, 
we performed a number of fits with other choices of  $\phi$.
The result shows that the lowest $\chi^{2}$ corresponds to 
$\phi \approx 542$ MV, 
which is  close to the  value we adopted. 
Thus in total there are 6 free parameters
\begin{align}\label{eq:paramSet}
\boldsymbol\theta=\{Z_{h}, D_{0},\delta, V_{a}, \gamma_{p1}, \gamma_{p2} \} .
\end{align}
The priors of all the parameters are chosen to be uniform distributions  according to \eq{eq:priors} 
with the prior intervals shown in \tab{tab:param}.

In the GALPROP code, 
the diffusion equation is solved numerically on a 
spatial grid with widths $\Delta R=1$ kpc and $\Delta Z=0.2$ kpc.
The momentum grid is on a logarithmic scale with a scale factor 1.4.
For  sampling the posterior distributions and calculating the marginal distributions,
we use the numerical package {\tt CosmoMC}
\cite{Lewis:2002ah} 
which implements the Metropolis-Hastings algorithm in the MCMC scan of the whole parameter space. 
We have built 18 parallel MCMC chains with $\sim$1500 samples in each chain 
after  burn-in.
These chains satisfy the convergence condition that the ratio of the
inter-chain variance and intra-chain variance is less than 0.2
\cite{Roberts:2014}.
In total $2.6\times 10^{4}$ samples  were obtained from the MCMC scan.
The results of the best-fit values, statistical mean values, standard deviations
and allowed intervals at $95\%$ confidence level (CL)  for these parameters
are summaried in \tab{tab:param}.
\begin{table}[htb]
\begin{center}
\begin{tabular}{lllll | l}
  \hline\hline
Quantity 	&Prior & Best-fit &Posterior mean and 	&Posterior 95\% &Ref.\cite{Trotta:2010mx}		\\
		&range&value	&Standard deviation	& range		&							\\
\hline
$Z_h (\text{kpc})$
	&[1, 11]	&3.2  &3.3$\pm$0.6			&[2.1, 4.6] &5.4$\pm$1.4		\\
$D_0/Z_{h}$
	&[1, 3] &2.02 	&2.00$\pm$0.07		&[1.82, 2.18] & (1.54$\pm$0.48)	\\
$\delta $
	&[0.1, 0.6] &0.29 	&0.29$\pm$0.01		&[0.27, 0.32] &0.31$\pm$0.02	\\
$V_{a}(\text{km}\cdot\text{s}^{-1})$
	&[20, 70]  &44.7 &44.6$\pm$1.2		&[41.3, 47.5]	&38.4$\pm$2.1\\
$\gamma_{p1}$
	&[1.5, 2.1] &1.79  &1.78$\pm$0.01		&[1.75, 1.81]	&1.92$\pm$0.04\\
$\gamma_{p2}$
&[2.2,2.6] &2.46 	&2.45$\pm$0.01		&[2.43,2.47] &2.38$\pm$0.04	\\
  \hline\hline
\end{tabular}
\end{center}
\caption{
Constraints on the propagation models from the global Bayesian analyses to the AMS-02 data 
of B/C ratio and  proton flux.
The prior interval, best-fit value, statistic mean, standard deviation and 
the allowed range at $95\%$ CL are listed for each propagation parameter.
The parameter $D_{0}/Z_{h}$ is in units of $10^{28}\text{cm}^{2}\cdot\text{s}^{-1}\text{kpc}^{-1}$.
For a comparison, 
we also list the mean values and standard deviations of 
these parameters from 
a previous analysis in \cite{Trotta:2010mx}. 
The value of $D_{0}/Z_{h}$ in the parentheses 
is obtained from \cite{Trotta:2010mx} using a naive combination 
of $D_{0}$ and $Z_{h}$ without considering the correlation.
}
\label{tab:param}
\end{table}
{\blue

}
For a comparison, 
we also list the allowed ranges determined from 
a previous analysis in Ref.~\cite{Trotta:2010mx} 
which is based on the data prior to AMS-02 such as
the B/C ratio from HEAO-3
\cite{Engelmann:1990zz}, 
ATIC-2
\cite{Panov:2007fe} 
and CREAM-1
\cite{Ahn:2008my}, 
the data of ${}^{10}\text{B}/{}^{9}\text{Be}$ from
ACE\cite{Yanasak:2001ACE},
and the data of 
Carbon and Oxygen nuclei fluxes from ACE\cite{George:2009ACE}.
For an estimate of the goodness-of-fit, 
we evaluate the $\chi^{2}$ function 
which is defined as $\chi^{2}=-2\ln \mathcal{L}$.
Using the best-fit parameters, we find that in total $\chi^{2}=49.0$ in which 
the contribution from B/C is 6.1 and that from proton flux is 42.9.
Thus $\chi^{2}/\text{dof}=49.0/112$ which indicates a good 
agreement with the data.

As it can be seen from the table,
although the fitting strategy is quite different,
the parameters determined by the AMS-02 data
are similar to that in Ref.~\cite{Trotta:2010mx},
but  the uncertainties in the parameters are significantly smaller.
For instance, the ratio $D_{0}/Z_{h}$ is found to be
\begin{align}
\frac{D_{0}}{Z_{h}}=(2.00\pm 0.07)~\text{cm}^{2}\text{s}^{-1}\text{kpc}^{-1}.
\end{align}
The uncertainty  is within $5\%$,
which is mostly constrained by the  B/C data. 
Note that 
a relatively small halo height is favoured by 
the  AMS-02 data 
\begin{align}
Z_{h}=3.3\pm0.6~\text{kpc}.
\end{align}
Compared with 
$Z_{h}=5.4\pm 1.4~\text{kpc}$ obtained in Ref.~\cite{Trotta:2010mx},
the value of $Z_{h}$ from this work is $\sim 40 \%$ lower with 
the  uncertainty smaller by a factor of two.
A previous MCMC fit based on the two-zone diffusion model gives
$Z_{h}=8^{+8}_{-7}$~kpc
\cite{Putze:2010zn}.

\begin{figure}\begin{center}
\includegraphics[width=0.45\textwidth]{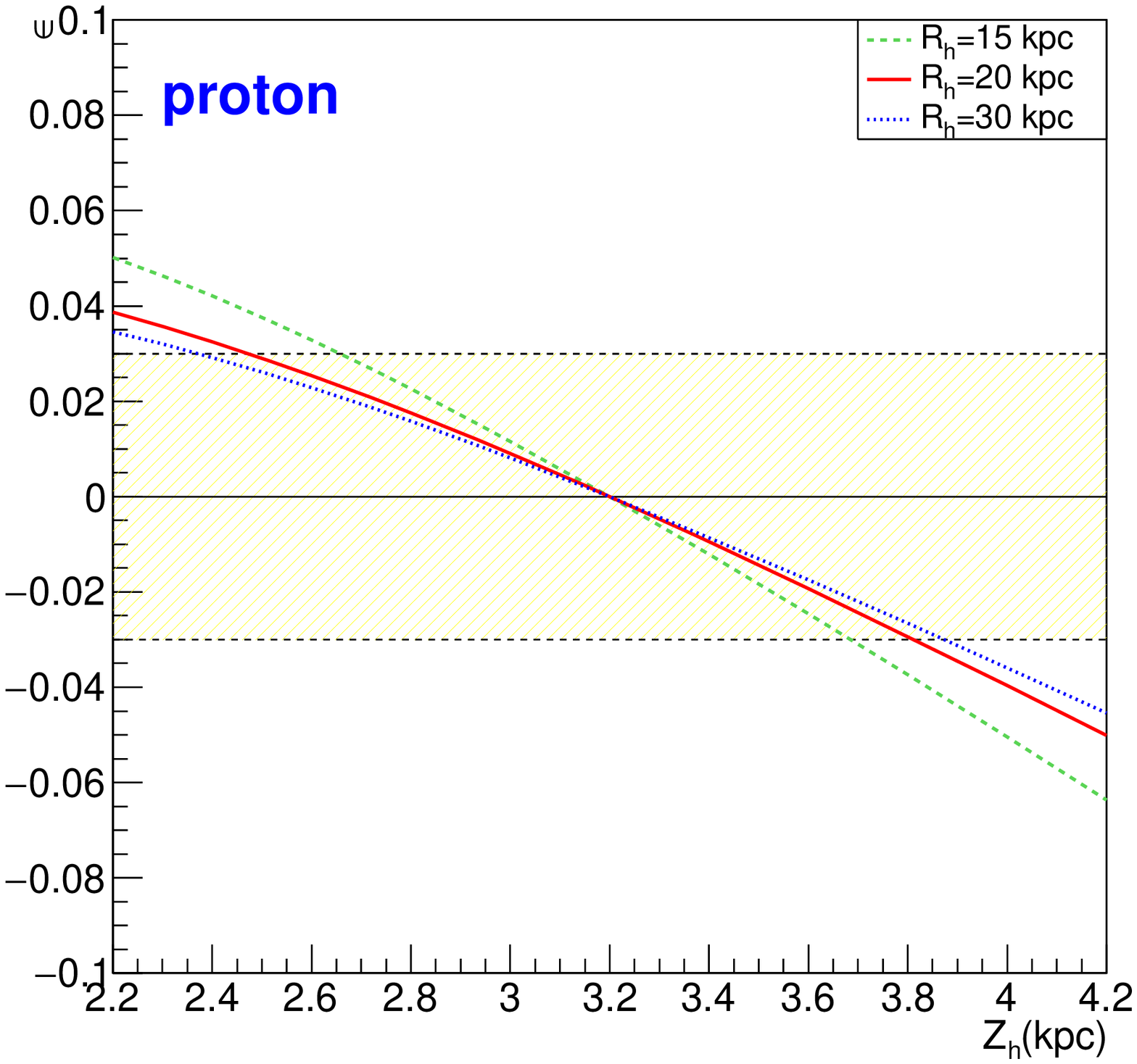}
\includegraphics[width=0.45\textwidth]{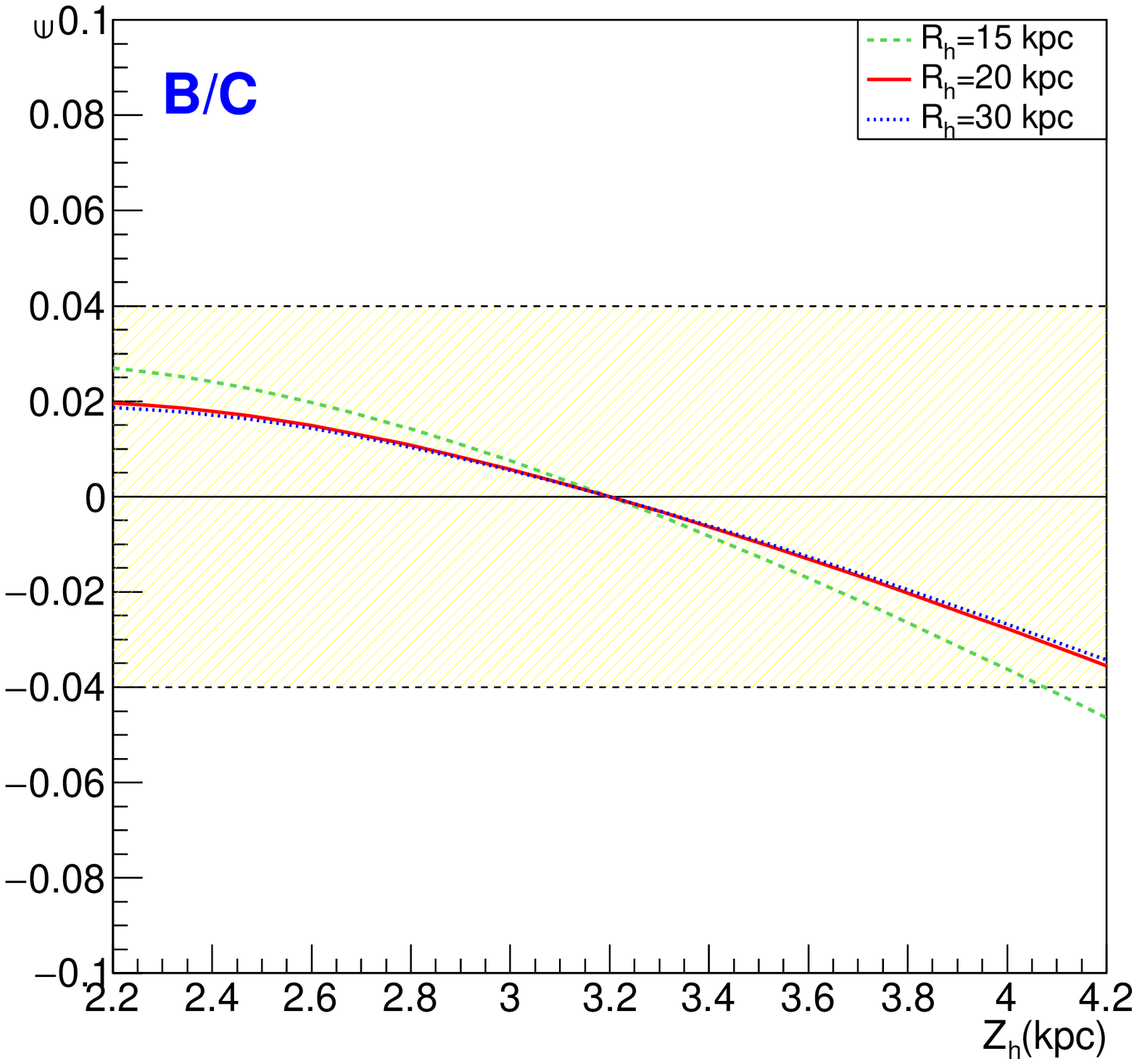}
\\
\includegraphics[width=0.45\textwidth]{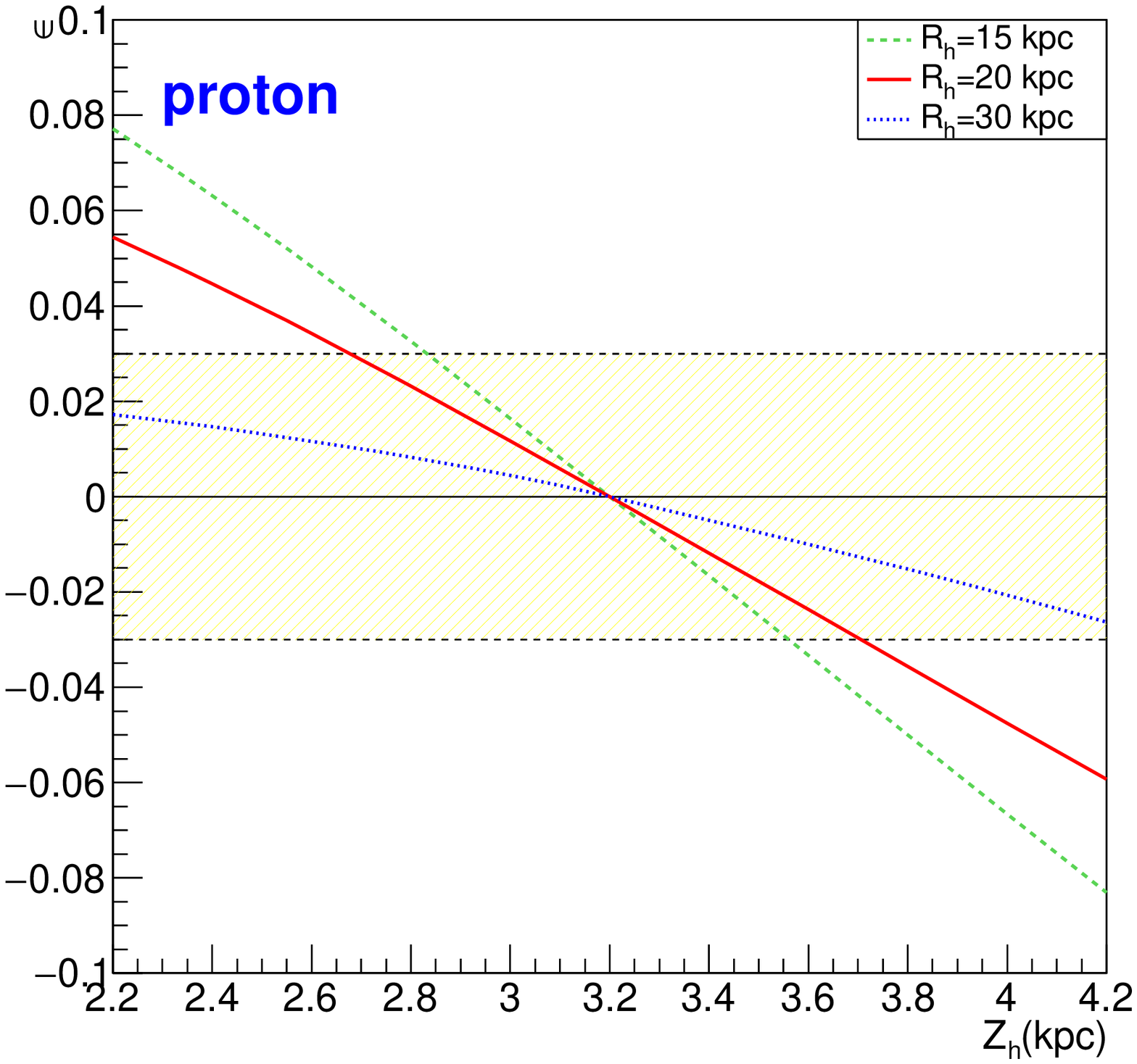}
\includegraphics[width=0.45\textwidth]{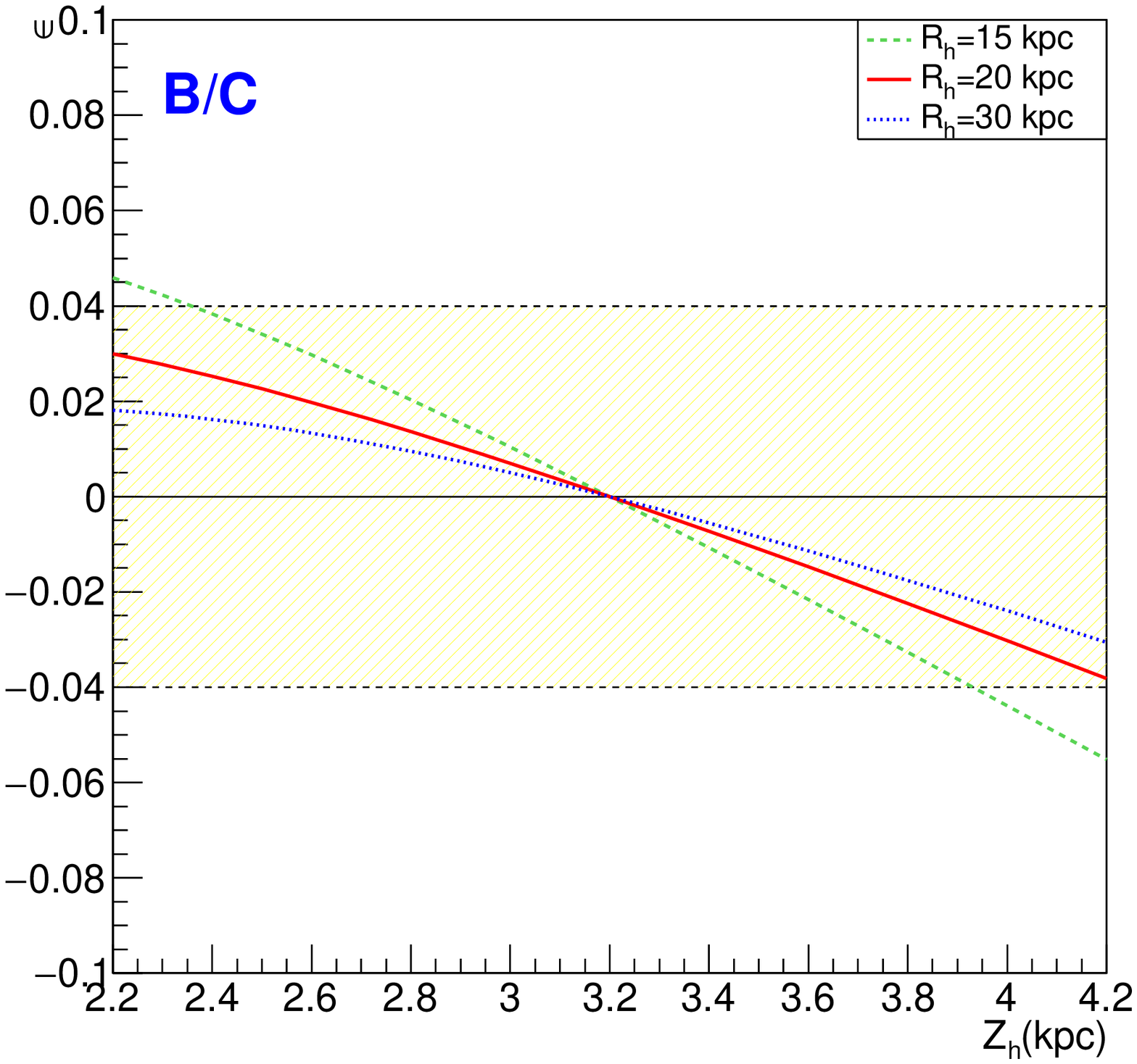}
\caption{
Upper panels)
Relative deviation $\epsilon_{1}$ of proton flux (upper left) and B/C ratio (upper right)
as a function of the halo half-height $Z_{h}$ 
at the kinetic energy $E_{\text{kin}}=24.2$~GeV/n, 
for three choices  of $R_{h}=15$, 20 and 30 kpc, respectively.
The ratio $D_{0}/Z_{h}$ and other parameters are fixed at 
their  best-fit values given in \tab{tab:param}.
The spatial distribution of the primary source is taken from 
\eq{eq:source-distribution}.
The calculation is done using the GALPROP code.
The horizontal bands represent
the uncertainties of the AMS-02 data at around 20 GeV~\cite{Haino:icrc2013}.
Lower panels)
The same as in the upper panels, 
but with an uniform distribution of the primary source term.
}
\label{fig:zhdependenc}
\end{center}\end{figure}

While  the ratio  $D_{0}/Z_{h}$ is sensitive to  the B/C ratio,
the absolute value of $Z_{h}$ is more sensitive to the  proton flux.
For an illustration  of the $Z_{h}$ dependence,
it is useful to define a relative deviation of an observable $\psi(D_{0},Z_{h})$ from 
a reference value $\psi(\hat{D}_{0}, \hat{Z}_{h})$ as follows
\begin{align}
\epsilon_{1}(D_{0},Z_{h})
\equiv 
\frac{\psi(D_{0}, Z_{h})-\psi(\hat{D}_{0}, \hat{Z}_{h})}{\psi(\hat{D}_{0}, \hat{Z}_{h})},
\end{align}
where $\psi$ can be the proton flux or the B/C flux ratio.
We choose $\hat{D_{0}}$ and $\hat{Z_{h}}$ to be 
the best-fit value of the diffusion coefficient and the halo half-height listed in \tab{tab:param}.
Using the GALPROP code,
we show in \fig{fig:zhdependenc} 
how the value of $\epsilon_{1}(D_{0},Z_{h})$ changes with $Z_{h}$ 
for the proton flux and B/C flux ratio,
under the constraint that $D_{0}/Z_{h}=\hat{D}_{0}/\hat{Z_{0}}$,
at a reference kinetic energy $E_{\text{kin}}=24.2$~GeV/n  with 
all the other parameters fixed at 
their best-fit values listed in \tab{tab:param}.
The option {\verb proton_norm_flux=0 } is used to 
prevent the GALPROP code from  
automatically  normalizing the proton flux to $N_{p}$.  
If there exists  an exact  $D_{0}/Z_{h}$ degeneration, 
it is expected that $\epsilon_{1}(D_{0},Z_{h})$ is vanishing for 
all the value of $Z_{h}$.
However, 
as shown in the upper panels of  \fig{fig:zhdependenc},
the proton flux decreases by $\sim 9\%$ in 
the $Z_{h}$ interval $2.2-4.2$~kpc for $R_{h}=20$  kpc.
The decrease of the proton flux with 
an increasing $Z_{h}$ is consistent with \eq{eq:DS}.
The uncertainties in the data of the proton flux are dominated by 
the systematic uncertainties in 
the acceptance ($\sim 2.8\%$), 
trigger efficiency ($\sim 1.0\%$) and 
proton track efficiency ($\sim 1.0\%$)
\cite{Haino:icrc2013}.
The total systematic uncertainty added up together is $\sim 3.1\%$,
which is also shown in \fig{fig:zhdependenc} for a comparison.
One can see that such a precision measurement on the proton flux
can place an useful constraint on $Z_{h}$.
On the other hand, 
the $Z_{h}$ dependence of the B/C ratio is relatively small.
The $\beta$-decay  of $^{10}\text{Be}\to$ $^{10}\text{B}$ may introduce 
another  $Z_{h}$ dependence  in the B/C ratio,
as discussed in Ref.
\cite{Putze:2010zn
}.
The  uncertainty in the data of  the B/C ratio is 
$\sim 4\%$ at $E_{\text{kin}}\sim 20 \text{ GeV/n}$~\cite{Oliva:icrc2013},
which is less stringent  in constraining the value of $Z_{h}$.

In this work, 
we fix the value of $R=20$~kpc in order to 
facilitate the comparisons with other analyses,
especially that in Refs.
\cite{Trotta:2010mx,Putze:2010zn}.
It  is anyway useful to examine whether  
the $Z_{h}$ dependence of the related observables 
can be affected by different choices of $R_{h}$
and the spatial distributions of the primary source term.
In \fig{fig:zhdependenc}, 
we also show the results for $R_{h}=15$ and 30~kpc, respectively.
In all the three cases, 
it is found that the value of $\epsilon_{1}$ is nonvanishing and 
depends  on $Z_{h}$.
For a larger $R_{h}=30$ kpc, 
the changes in $\epsilon_{1}$ for proton flux are slightly smaller,
while for a smaller $R_{h}=15$ kpc, 
the changes in $\epsilon_{1}$ are larger and
can reach $\sim 11\%$ in the $Z_{h}$ interval 2.2--4.2~kpc.
The changes in the value of $\epsilon_{1}$ in the  B/C flux ratio follow 
the similar trend.
The spatial distribution of the primary source can be  determined by 
independent observables such as the Galactic diffuse $\gamma$-rays.
In the GALPROP code, 
the source distribution is adopted to reproduce  
the Fermi-LAT $\gamma$-ray data of the 2nd Galactic quadrant
\cite{Tibaldo:2009spa
}.
For a comparison,
we consider a simplified case where 
the primary source is uniformly distributed along
the $R$-direction. 
The corresponding results on the variation of $\epsilon_{1}$ are 
shown in the lower panels of \fig{fig:zhdependenc}.
One can see that 
the $Z_{h}$ dependences are more significant than
the case where 
the source term is described by \eq{eq:source-distribution}.
These results  suggest that 
the breakdown of the $D_{0}/Z_{h}$ degeneracy  by the proton flux
may be a generic feature of the  two-dimensional diffusion models.

Complementary constraints on $Z_{h}$ can be obtained from 
the synchrotron emission of the Galaxy 
\cite{
DiBernardo:2012zu,
Orlando:2013ysa,
Bringmann:2011py,
Fornengo:2014mna
}.
The conclusion inevitably  depends on the assumption on 
the strength and distribution of the Galactic magnetic field $B$
which are largely unknown.
Assuming an uniform  $B=6.5~\mu G$, it was found that 
1 kpc $\lesssim Z_{h} \lesssim$ 15 kpc in the two-zone diffusion model
\cite{
Bringmann:2011py
}.
A GALPROP based calculation with a spatial-dependent $B$ field and 
including an anisotropic component favoured a halo size around 10~kpc
\cite{
Orlando:2013ysa
}.
An analysis using the GRAGON code with spatial-dependent $D_{xx}$ and $B$
favoured $Z_{h}\gtrsim 6$ kpc
\cite{
DiBernardo:2012zu
}.
Using the low energy  electron/positron flux,
is was found that the PAMELA data disfavoured $Z_{h}\lesssim 3$~kpc
\cite{
Lavalle:2014kca
}.
But significant uncertainties can arise from dealing with the effects of the solar modulation.

\begin{figure}
\begin{center}
\includegraphics[width=0.45\textwidth]{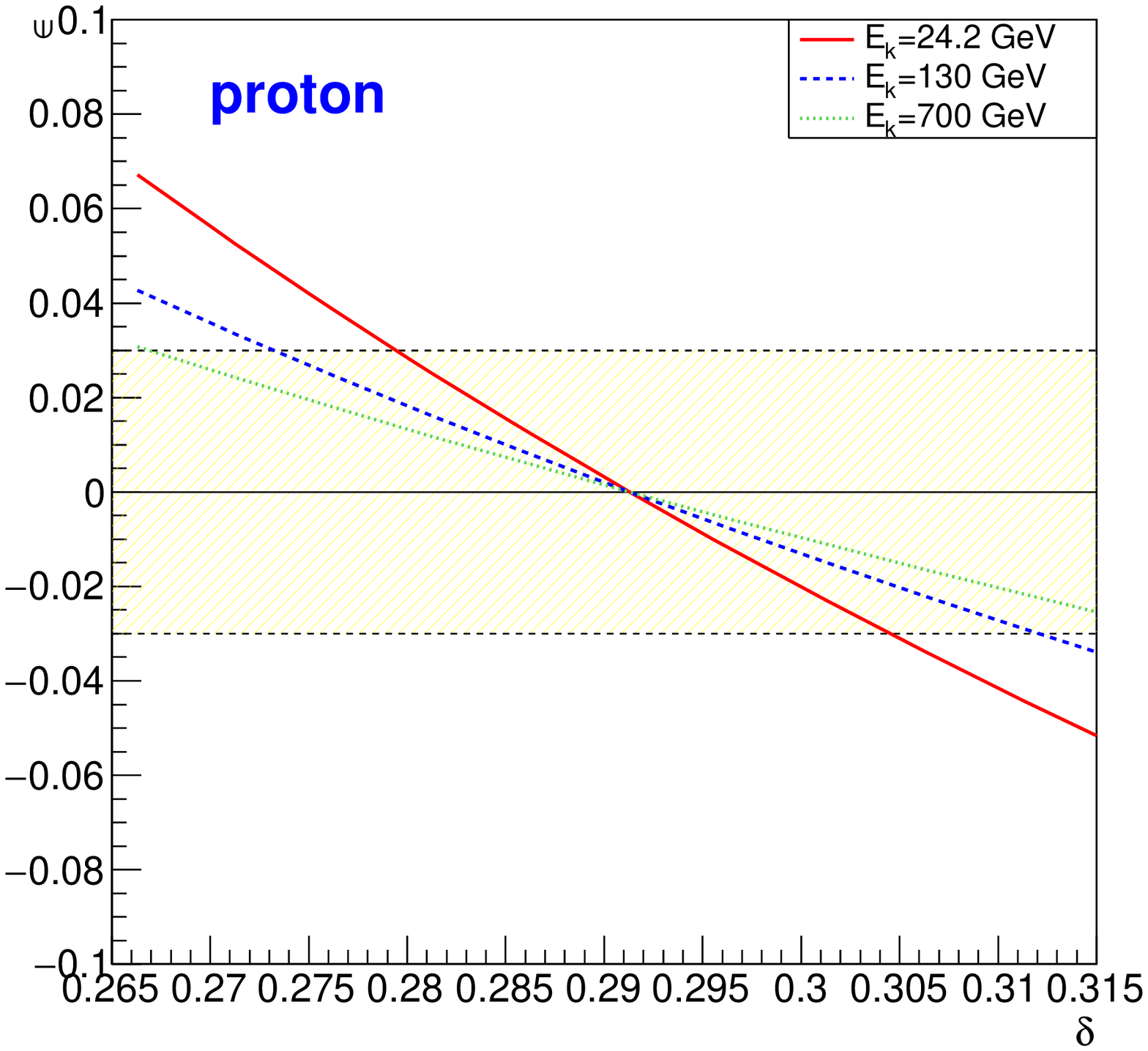}
\includegraphics[width=0.45\textwidth]{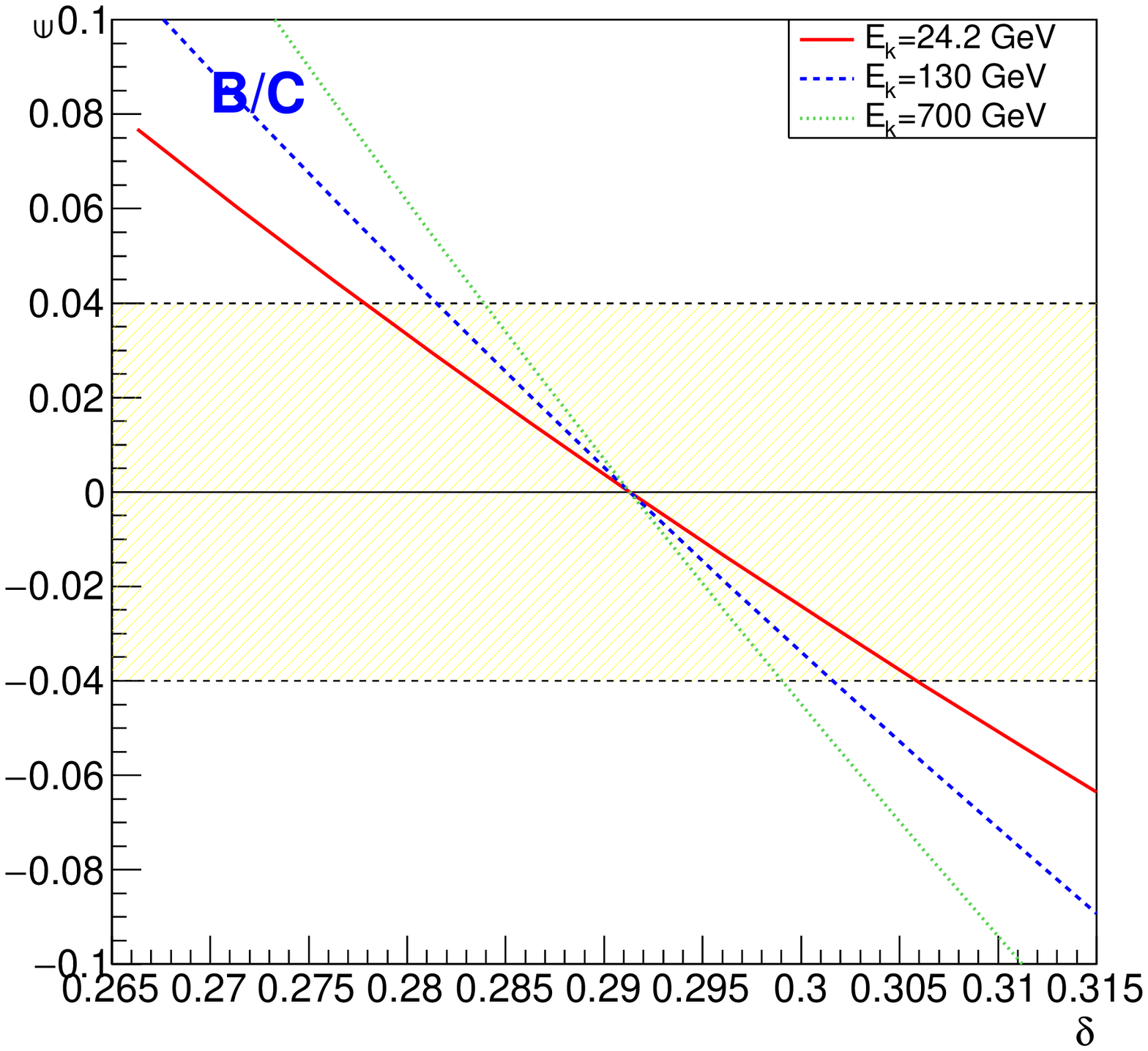}
\caption{
Left)
Relative deviation $\epsilon_{2}$ of proton flux 
as a function of the power law index $\delta$ in the diffusion coefficient,
under the condition $\gamma_{p2}+\delta=2.74$,
for three values of  kinetic energy 
$E_{\text{kin}}=24.2$, 130, and 700~GeV/n, respectively. 
Other parameters are fixed at their best-fit values given in \tab{tab:param}.
The calculations are done using the GALPROP code.
The horizontal band represents
the uncertaintie of the AMS-02 data at around 20 GeV~\cite{Haino:icrc2013}.
Right ) The same as Left but for the B/C flux ratio.
}
\label{fig:del-dependence}
\end{center}
\end{figure}

The determined power-law index in the diffusion term is 
$\delta=0.29\pm0.01$
which is smaller than $\delta\approx 0.7$ from 
the analysis based on the two-zone diffusion model
\cite{
Maurin:2001sj
},
but 
is consistent with $0.31\pm0.02$ from  
the previous GALPROP based global fit~\cite{Trotta:2010mx}
and 
is very  close to 1/3 from the  Kolmogorov-type spectrum.
Since the prior range for $\delta$ is set to be 0.1--0.6 
which is much wider than
the favoured range of $\delta$ at $95\%$~CL, 
the determined value of $\delta$ is insensitive to 
the choice of prior distribution.
The power-law indices of the nuclei source term are found to be
$\gamma_{p1}=1.78\pm0.01$ and
$\gamma_{p2}=2.45\pm0.01$, respectively.
As emphasized in section~\ref{sec:frame}, 
the low energy spectrum of the proton flux can be used to lift the degeneracy
between $\gamma_{p2}$ and $\delta$. 
Similar to the quantity $\epsilon_{1}$, 
one can define a relative change in the proton flux as a function of $\gamma_{p2}$
and $\delta$ as follows
\begin{align}
\epsilon_{2}(\gamma_{p2}, \delta)
\equiv 
\frac{\psi(\gamma_{p2}, \delta)-\psi(\hat{\gamma}_{p2}, \hat{\delta})}{\psi(\hat{\gamma}_{p2}, \hat{\delta})},
\end{align}
where $\hat{\gamma}_{p2}$ and $\hat\delta$ are the best-fit values given in \tab{tab:param}.
We show in the left panel of \fig{fig:del-dependence} 
the value of $\epsilon_{2}(\gamma_{p2}, \delta)$ as a function of $\delta$,
under the constraint $\gamma_{p2}+\delta=\hat\gamma_{p2}+\hat\delta=2.74$. 
If there exists an exact degeneracy in $\gamma_{p2}$ and $\delta$, 
it is expected that $\epsilon_{2}$ will be vanishing. 
However, 
as can be seen from \fig{fig:del-dependence},
at $E_{\text{kin}}=24.2$~GeV,
the value of $\epsilon_{2}$ is not vanishing, and
the change in $\epsilon_{2}$ is more than $\sim 10\%$ when 
$\delta$ increases from 0.265 to 0.315.
At higher energies $E_{\text{kin}}=103$ and 700~GeV, 
the changes in $\epsilon_{2}$ become smaller,
which is consistent with the fact that 
the  proton energy spectrum
is  closer to a  single power law at high energies.
A stronger $\epsilon_{2}$ dependence is found in the B/C flux ratio 
as shown in the right panel of  \fig{fig:del-dependence},
which indicates that the value of $\delta$ can be constrained by the B/C ratio.

\begin{figure}
\includegraphics[width=0.19\textwidth]{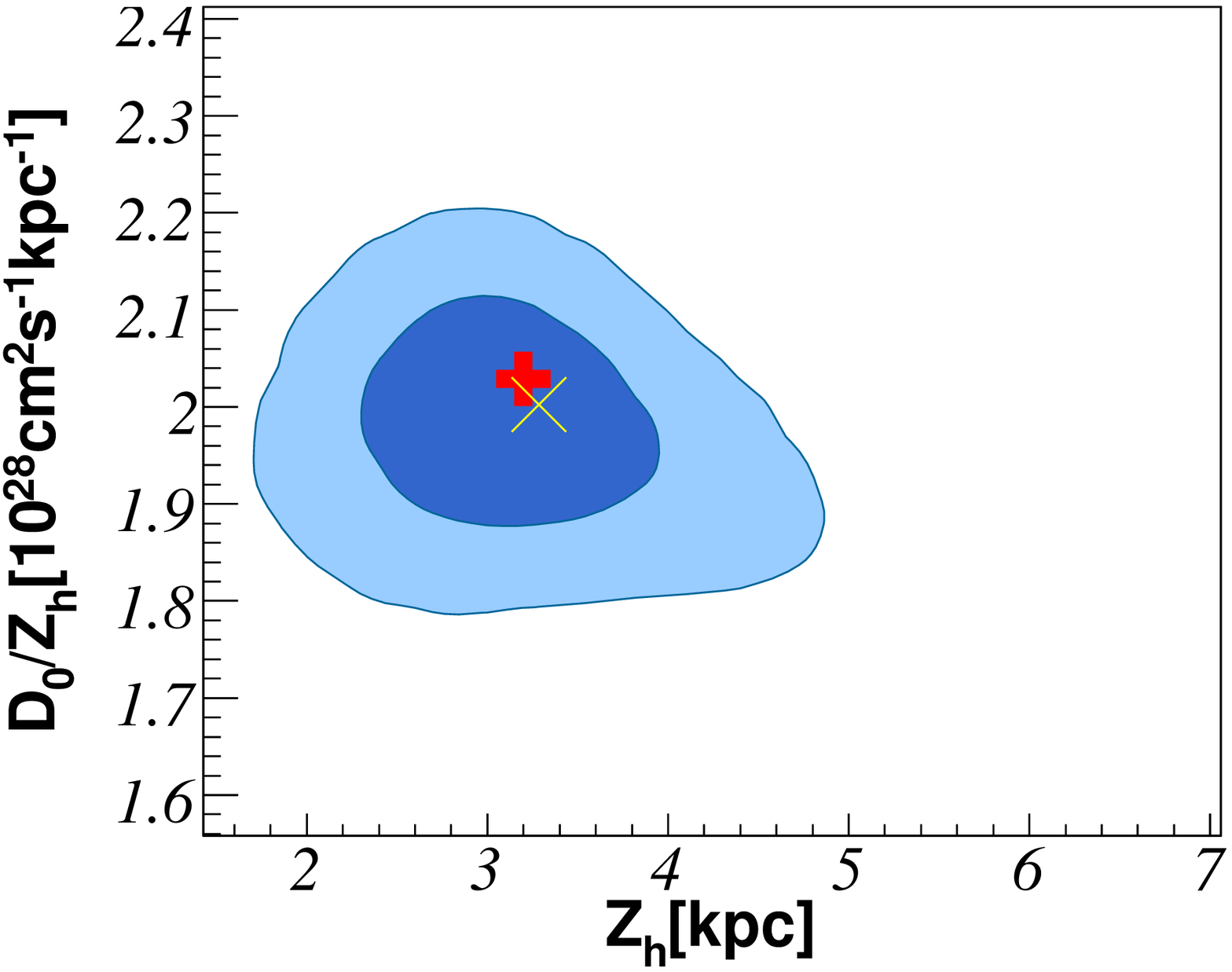}
\\
\includegraphics[width=0.19\textwidth]{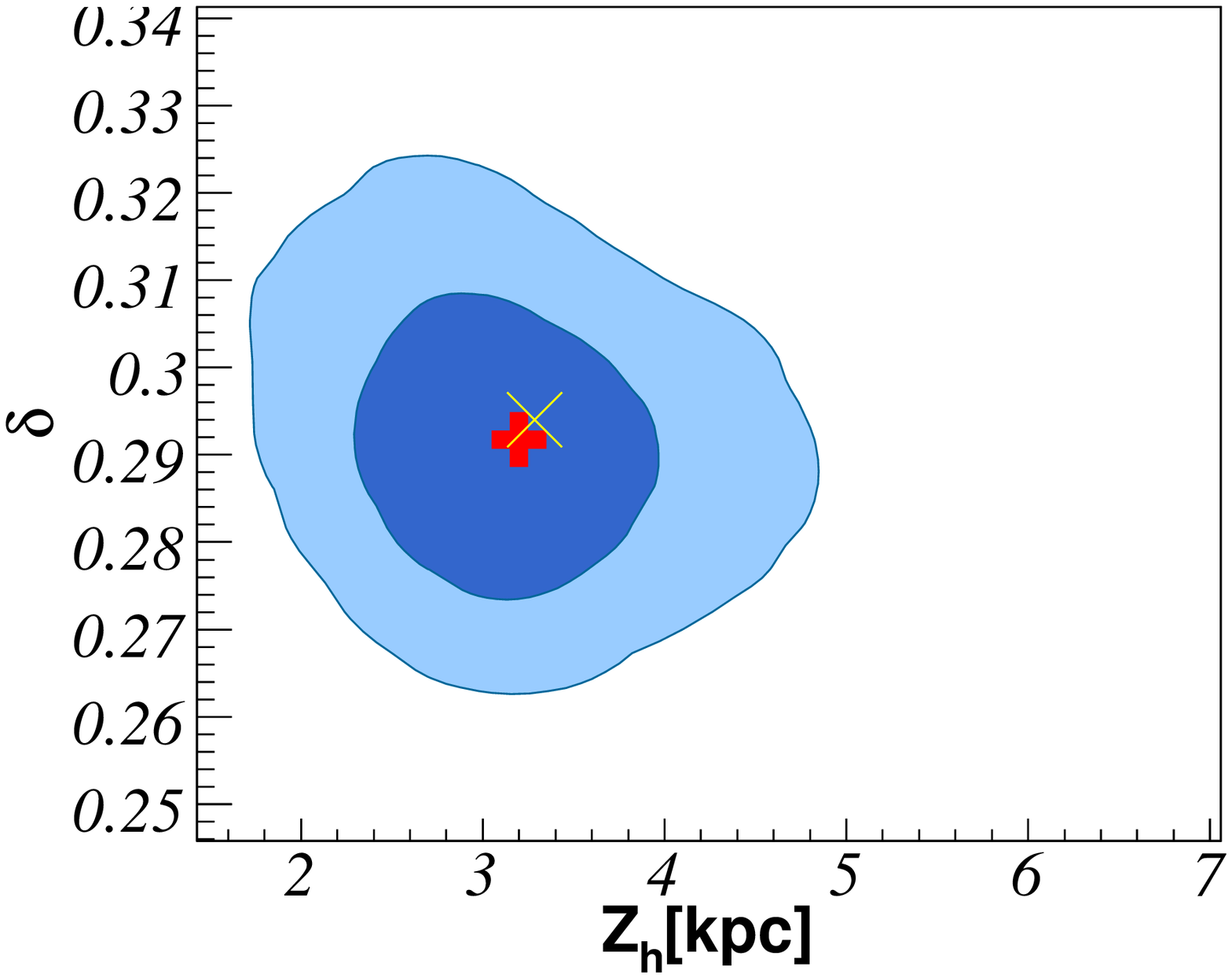}
\includegraphics[width=0.19\textwidth]{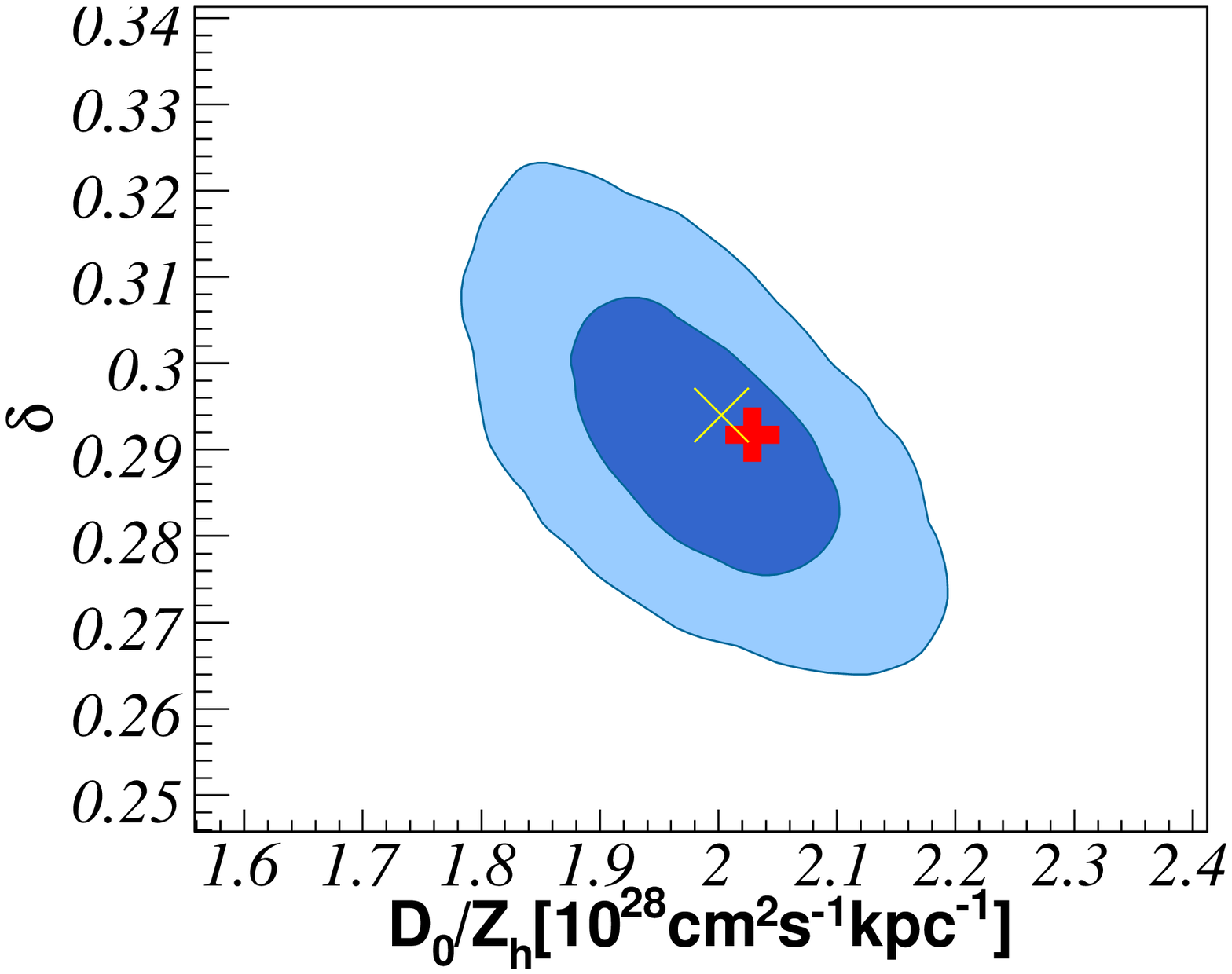}
\\
\includegraphics[width=0.19\textwidth]{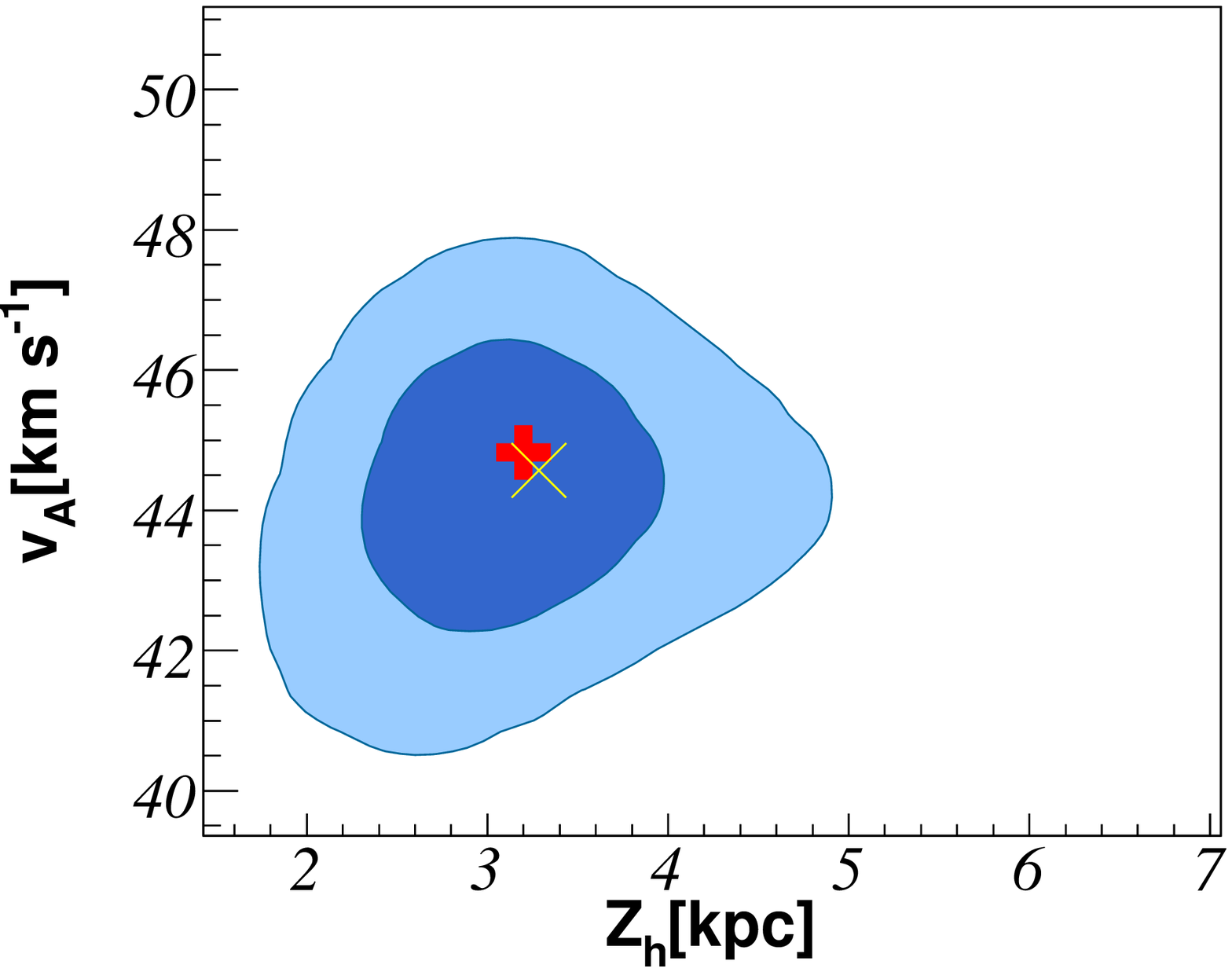}
\includegraphics[width=0.19\textwidth]{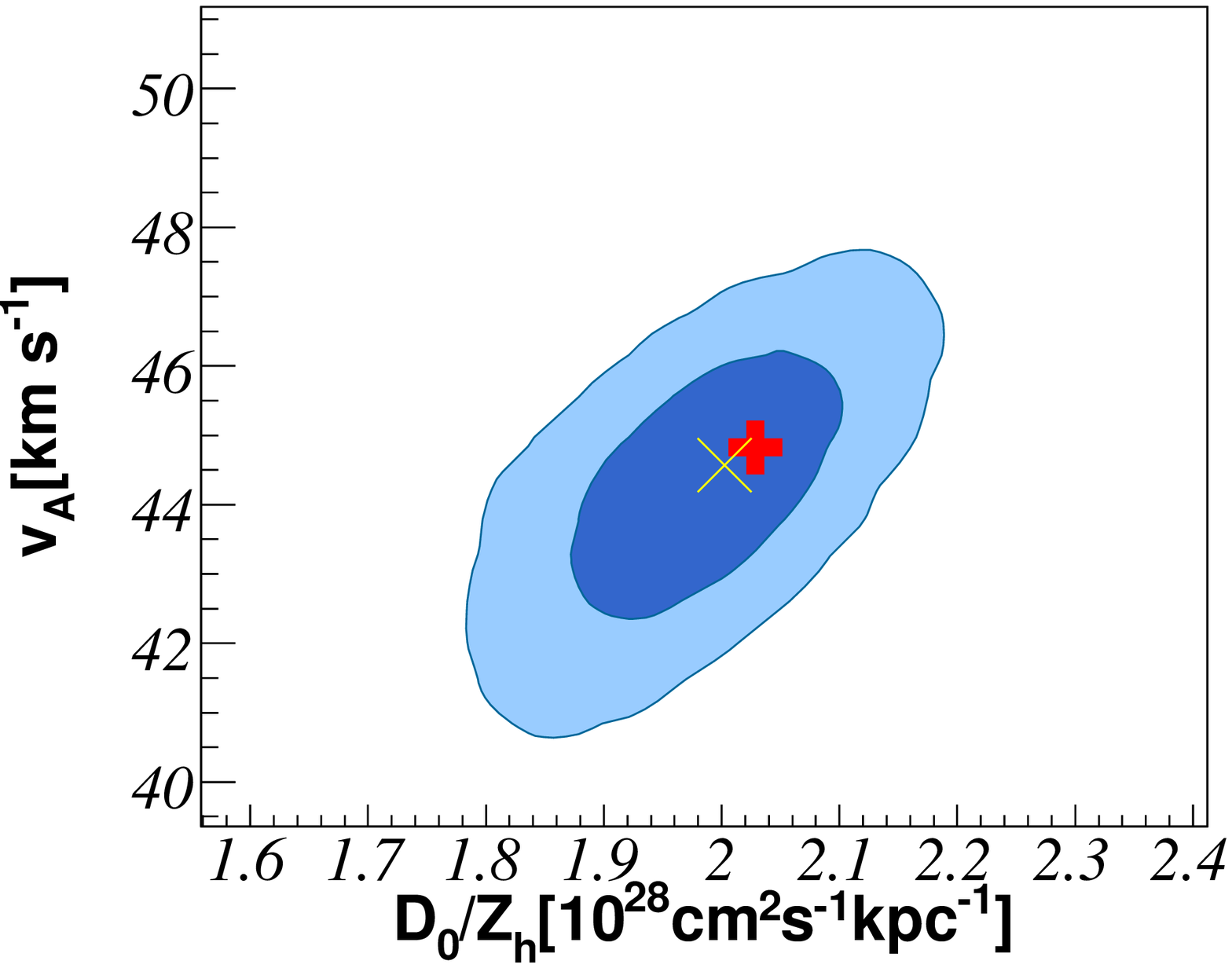}
\includegraphics[width=0.19\textwidth]{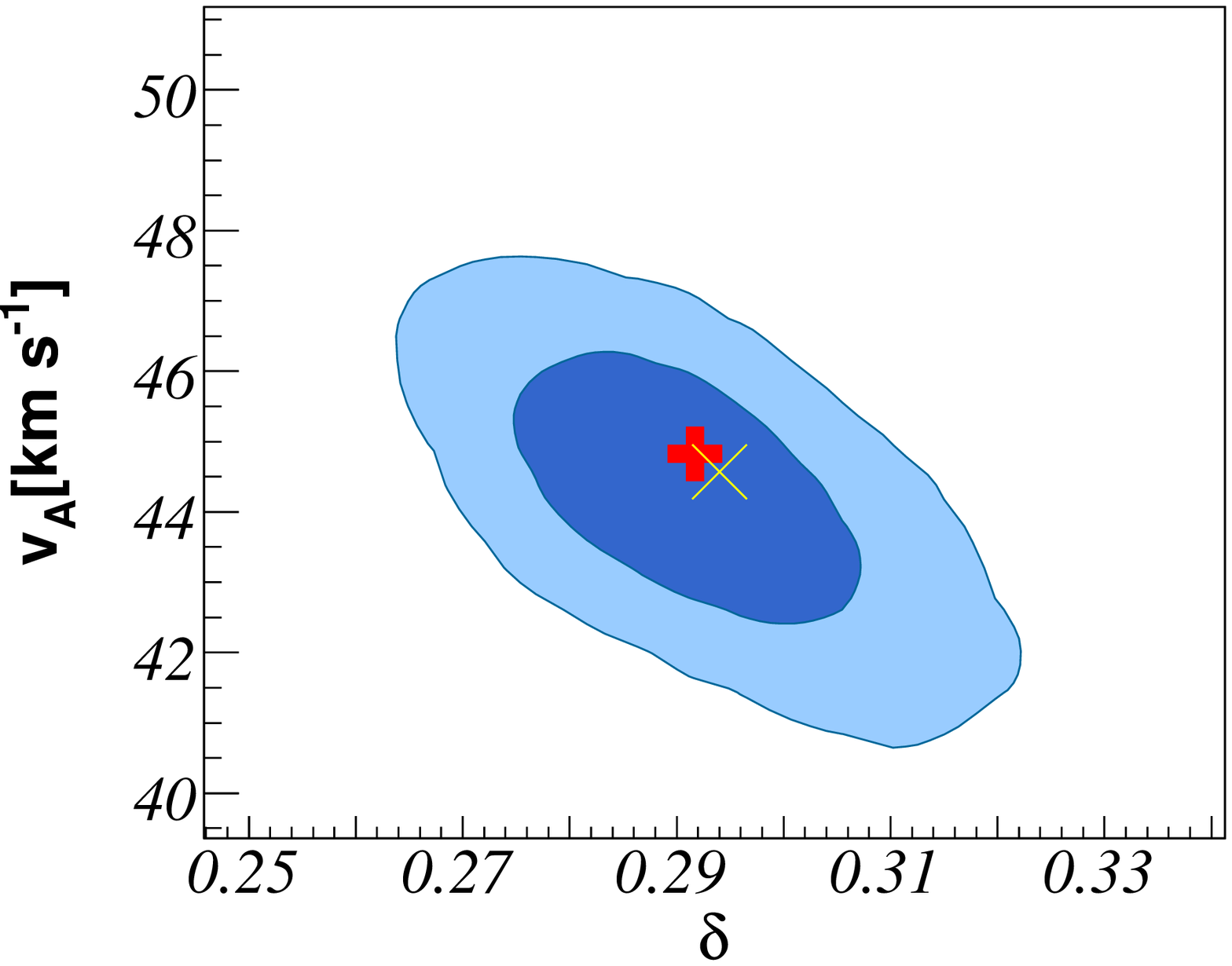}
\\
\includegraphics[width=0.19\textwidth]{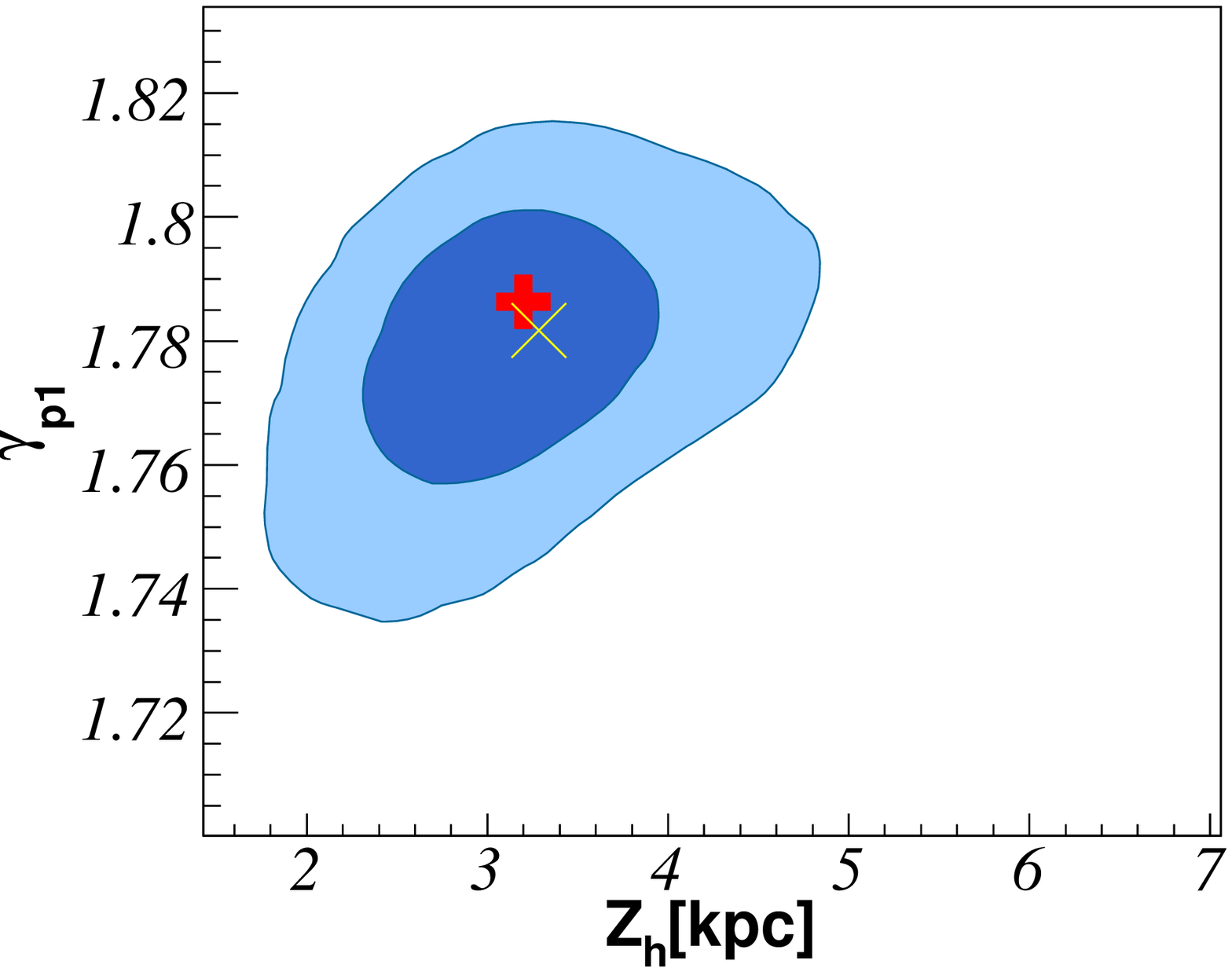}
\includegraphics[width=0.19\textwidth]{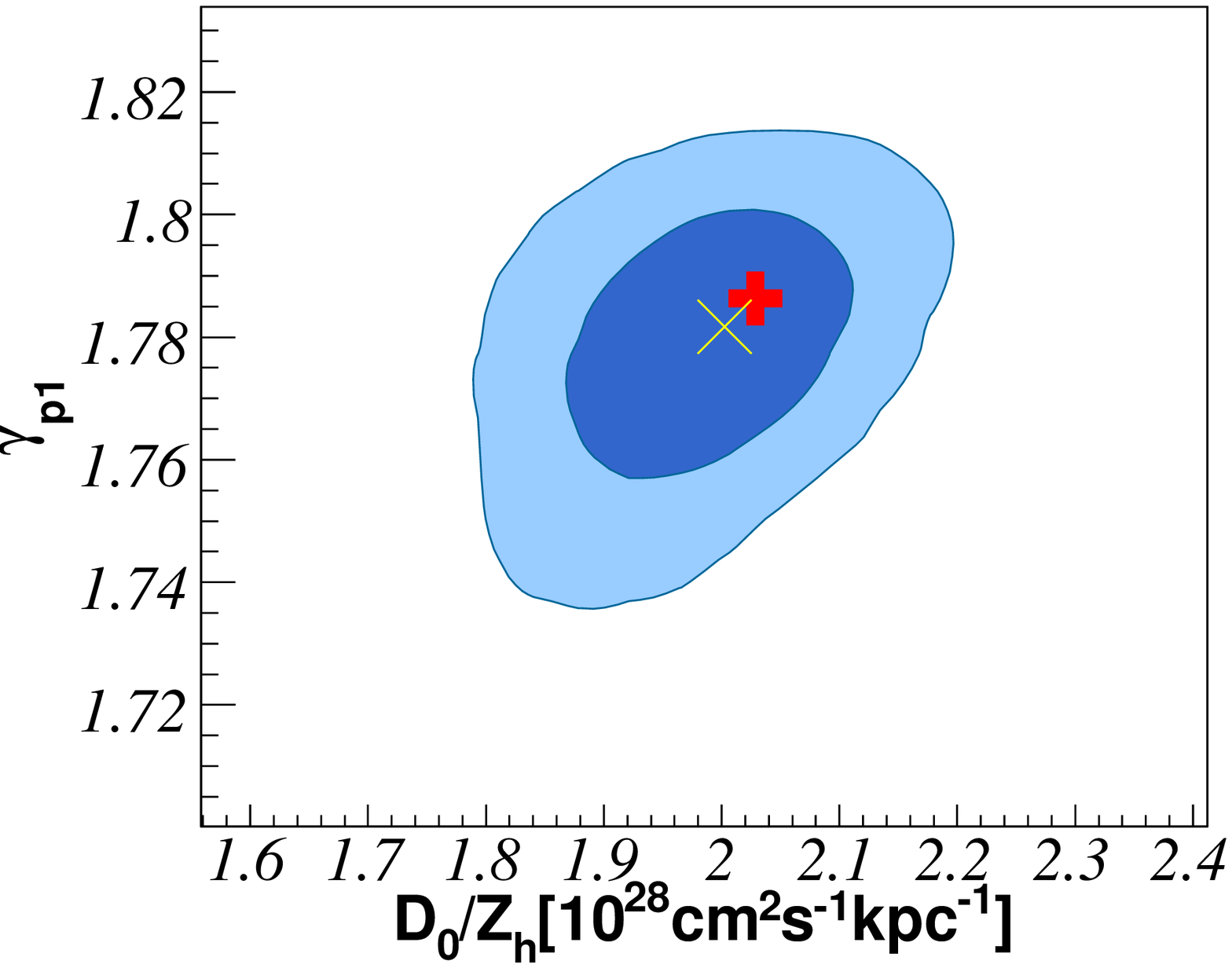}
\includegraphics[width=0.19\textwidth]{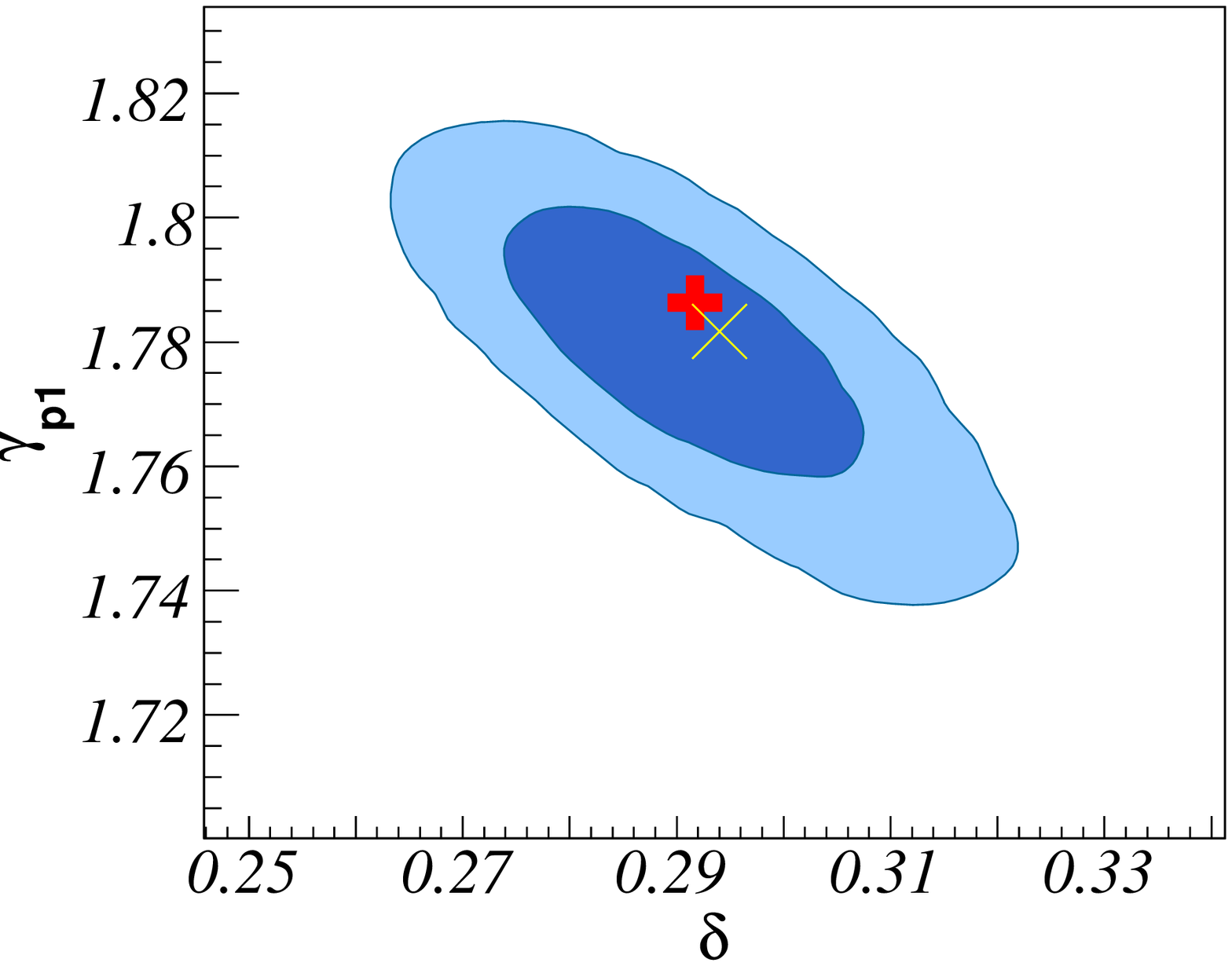}
\includegraphics[width=0.19\textwidth]{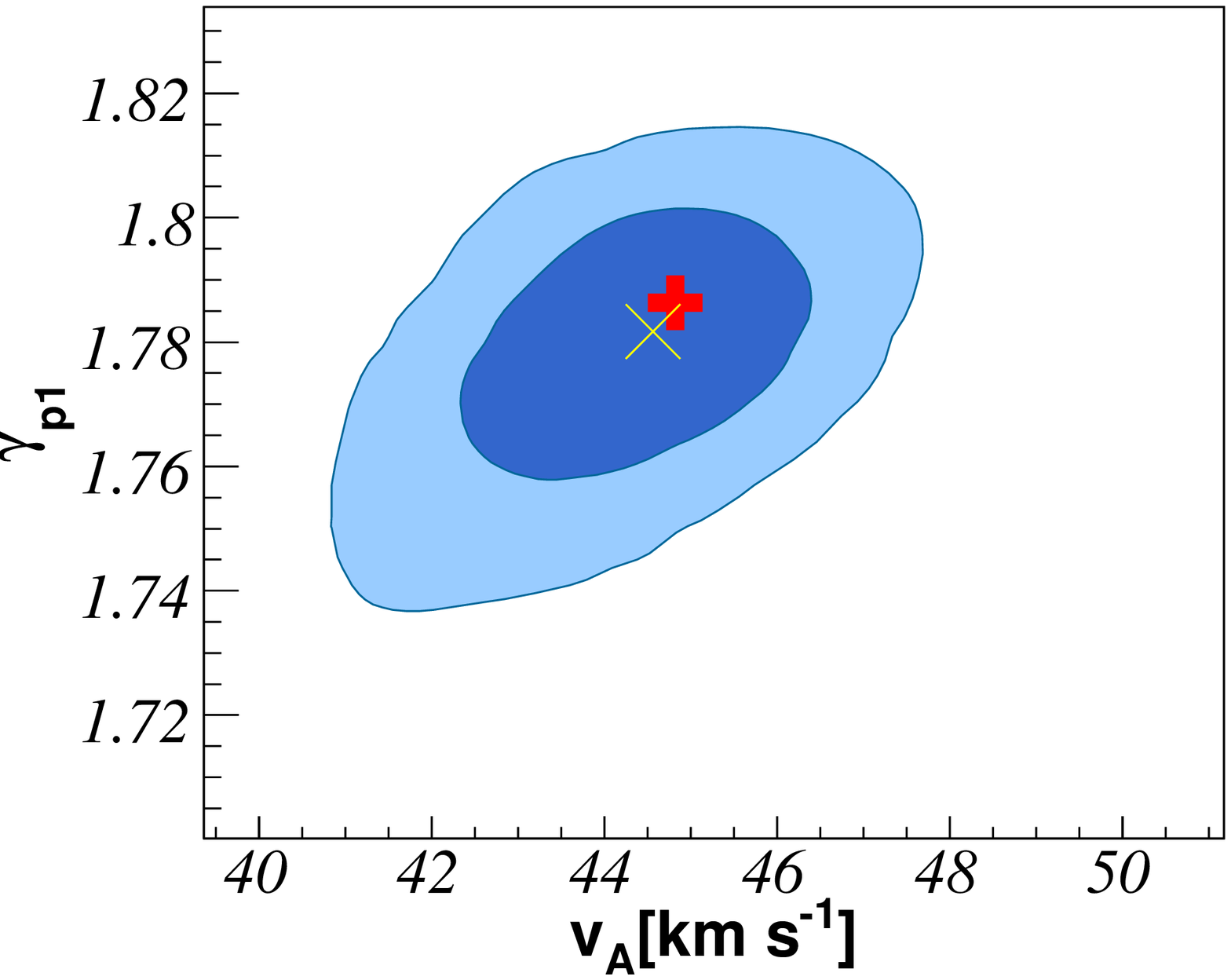}
\\
\includegraphics[width=0.19\textwidth]{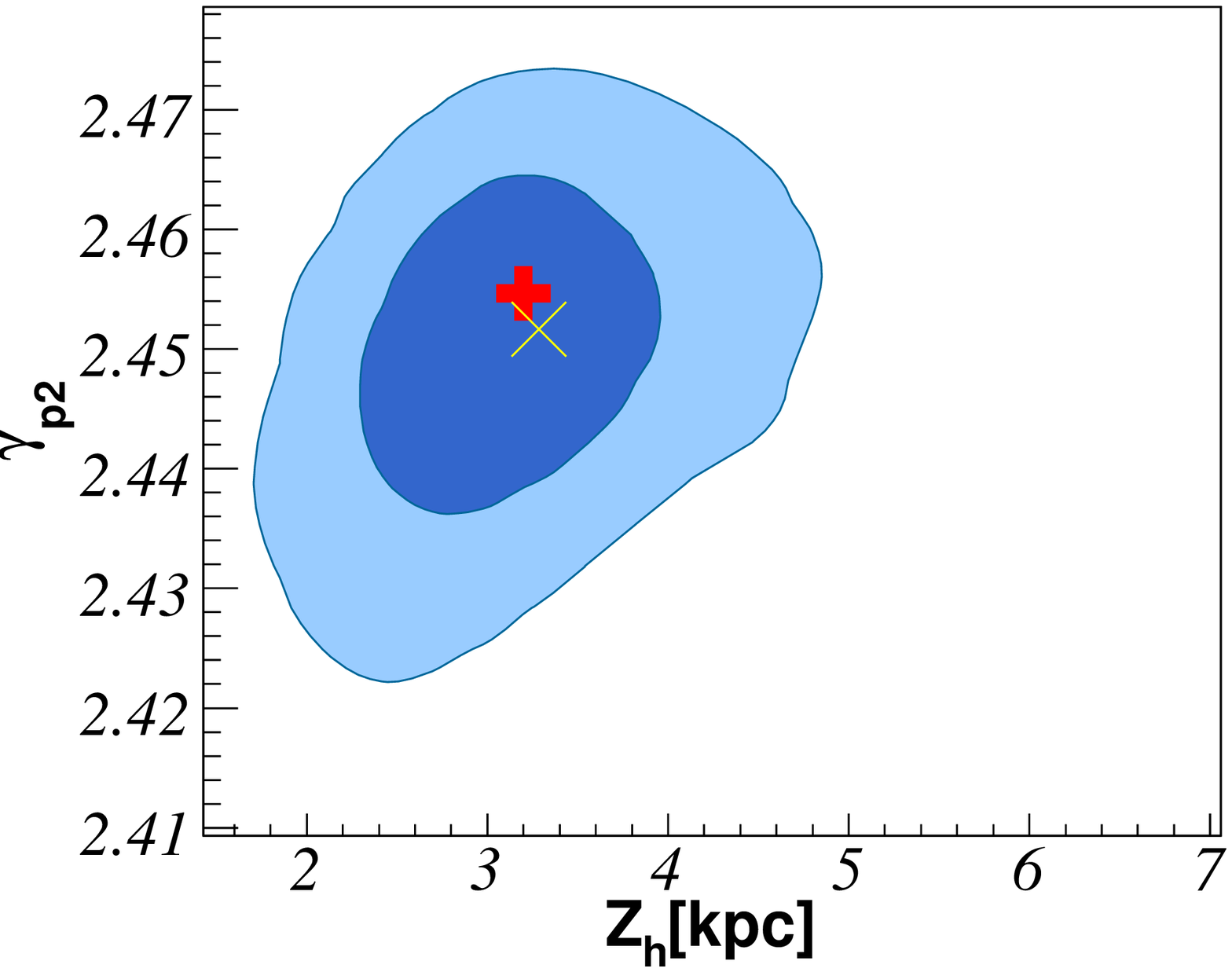}
\includegraphics[width=0.19\textwidth]{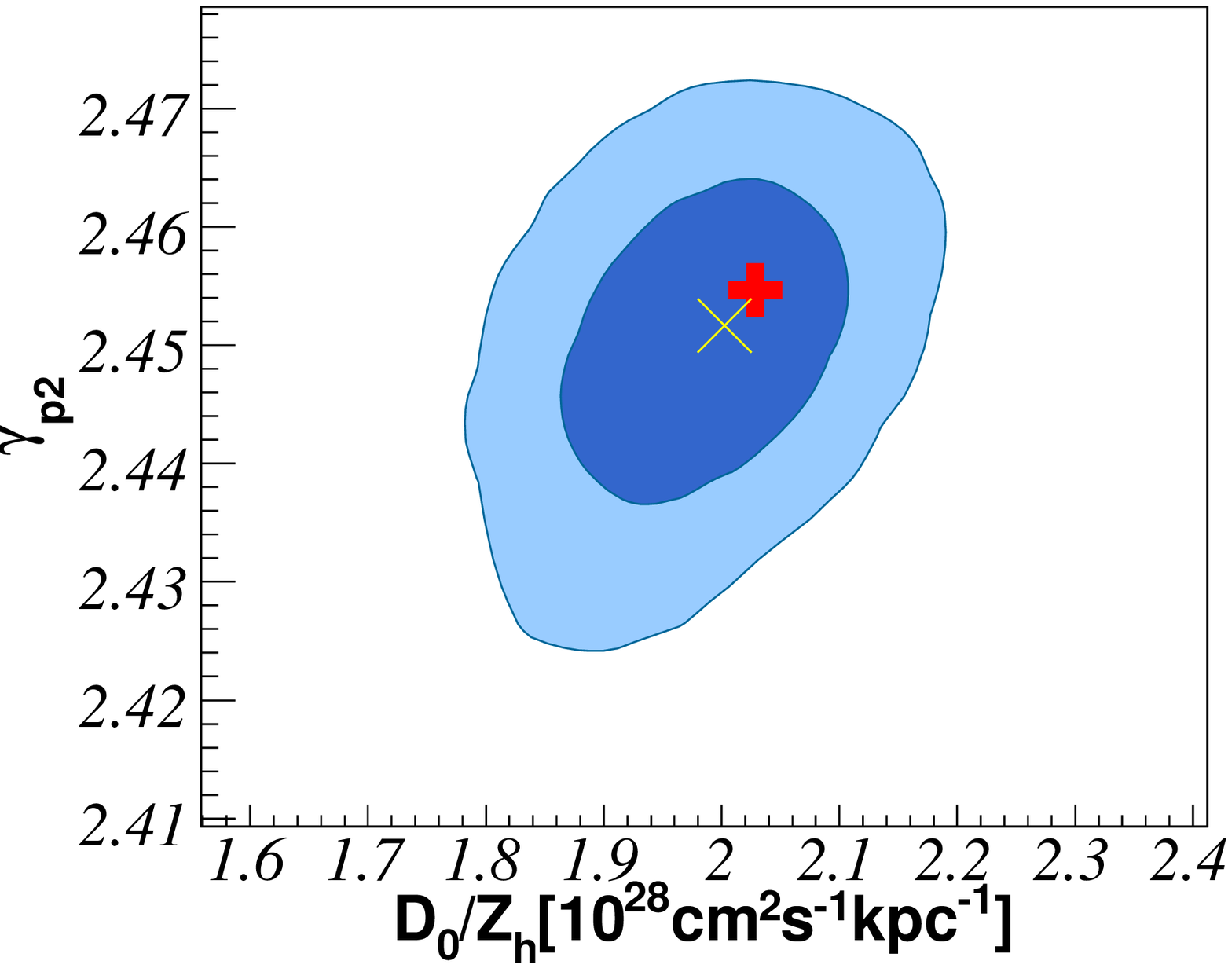}
\includegraphics[width=0.19\textwidth]{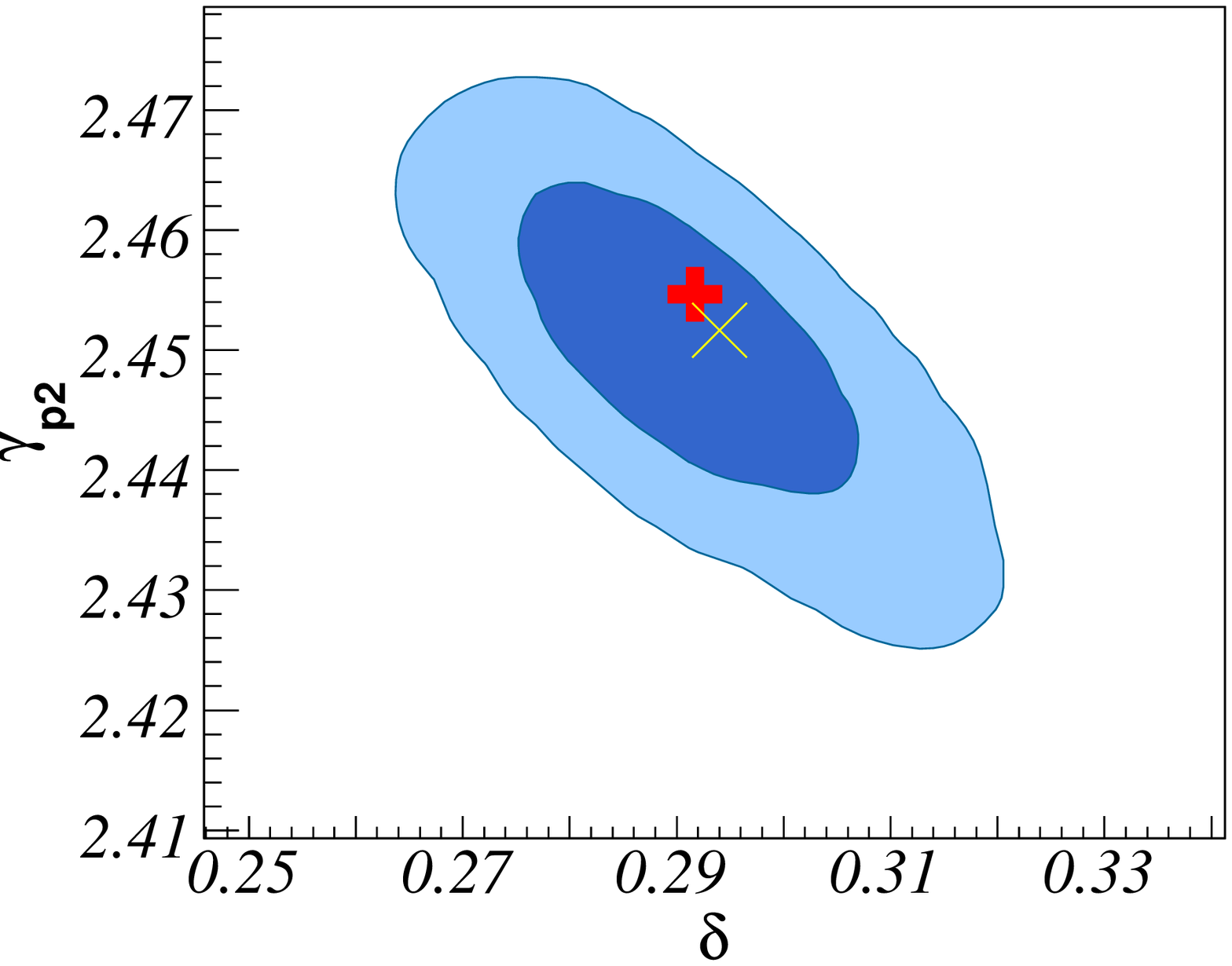}
\includegraphics[width=0.19\textwidth]{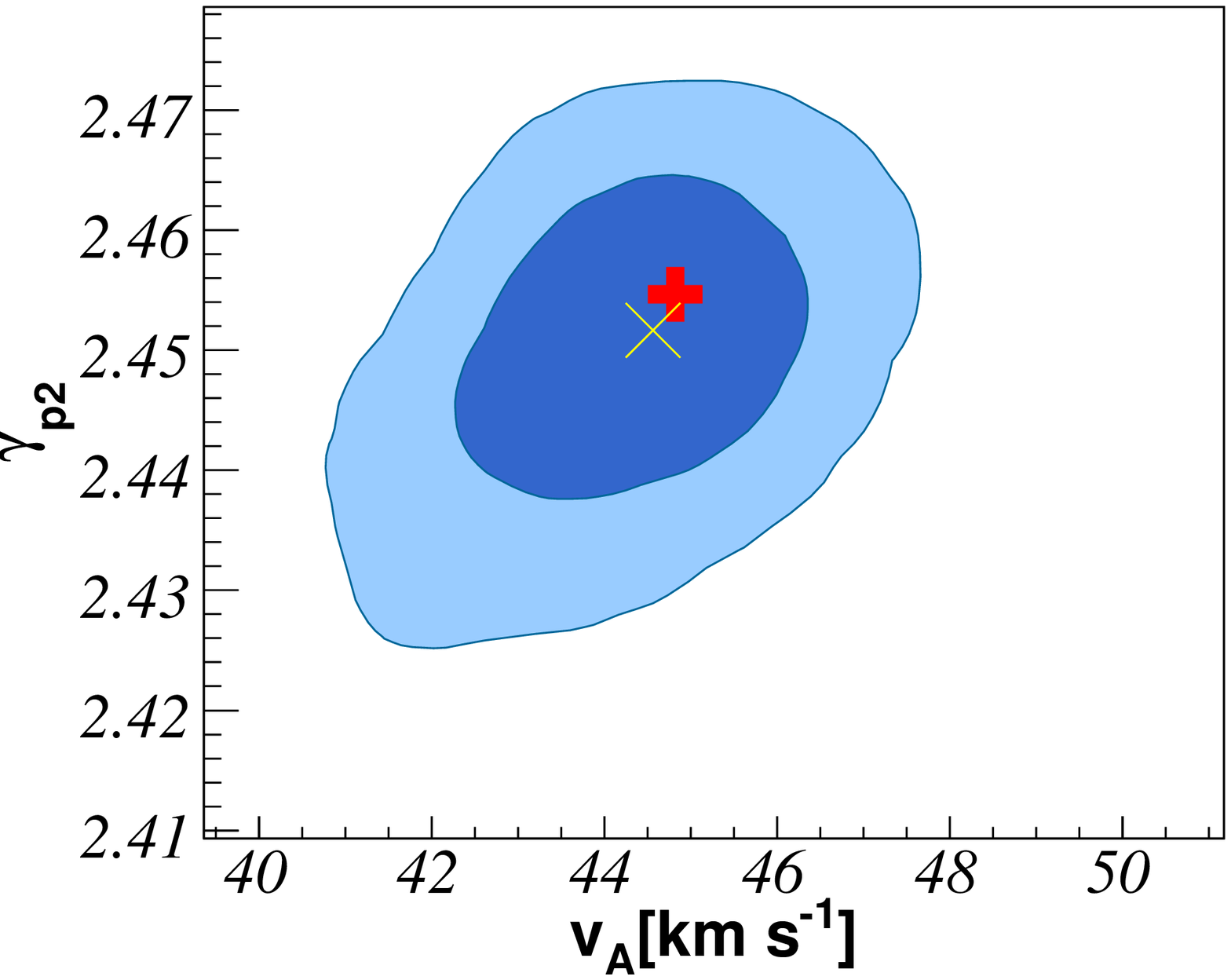}
\includegraphics[width=0.19\textwidth]{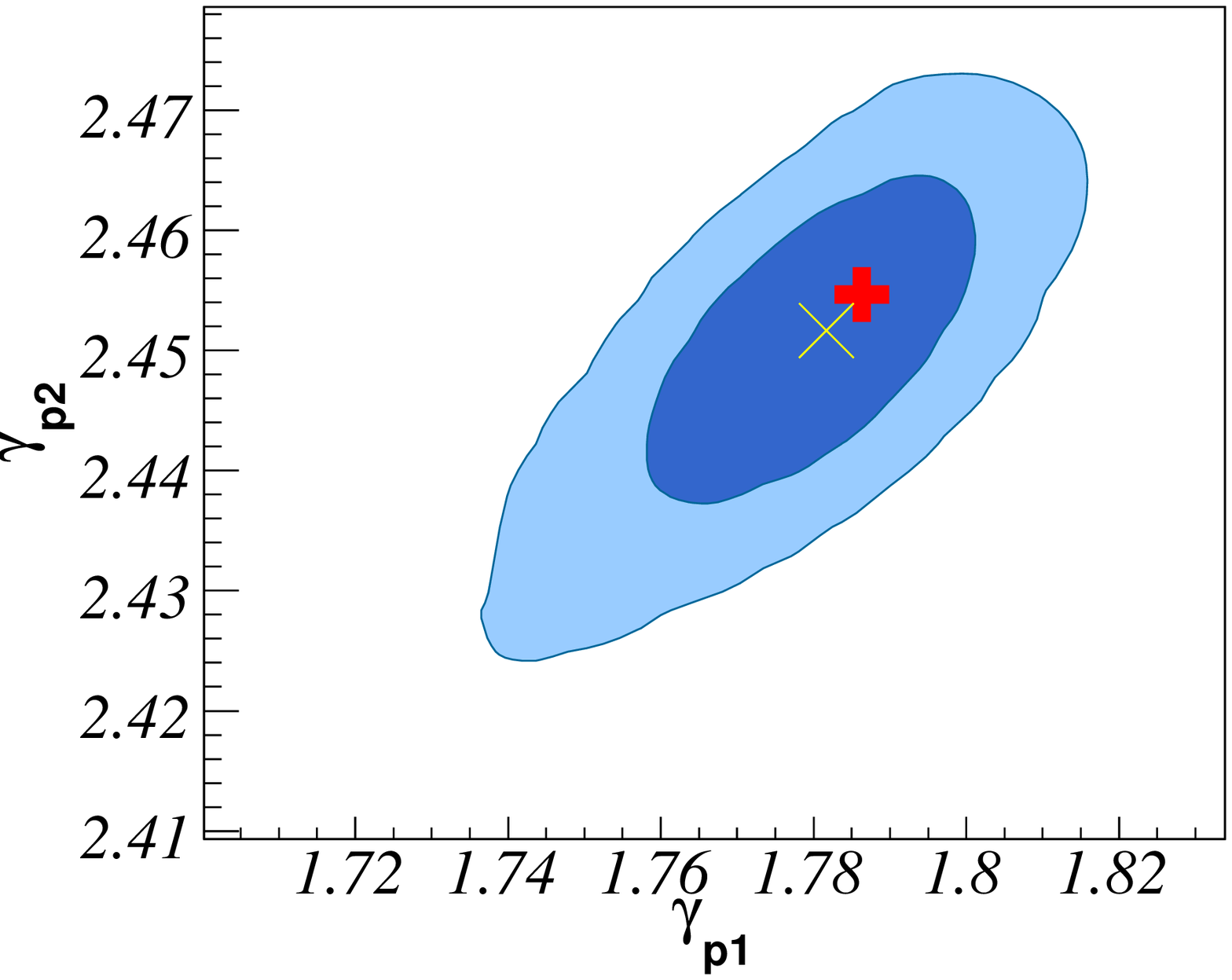}
\caption{
Two-dimensional marginalized  posterior PDFs for
the combinations of some selected parameters involving 
$Z_{h}$, $D_{0}/Z_{h}$, $\delta$, $V_{a}$ and $\gamma_{p1}$.
The regions enclosing $68\%(95\%)$ CL are shown in dark blue (blue).
The red plus (yellow cross ) in each plot indicates 
the best-fit value (statistic mean value). 
}
 \label{fig:param_2d}
\end{figure}

\begin{figure}
\begin{center}
\includegraphics[width=0.3\textwidth]{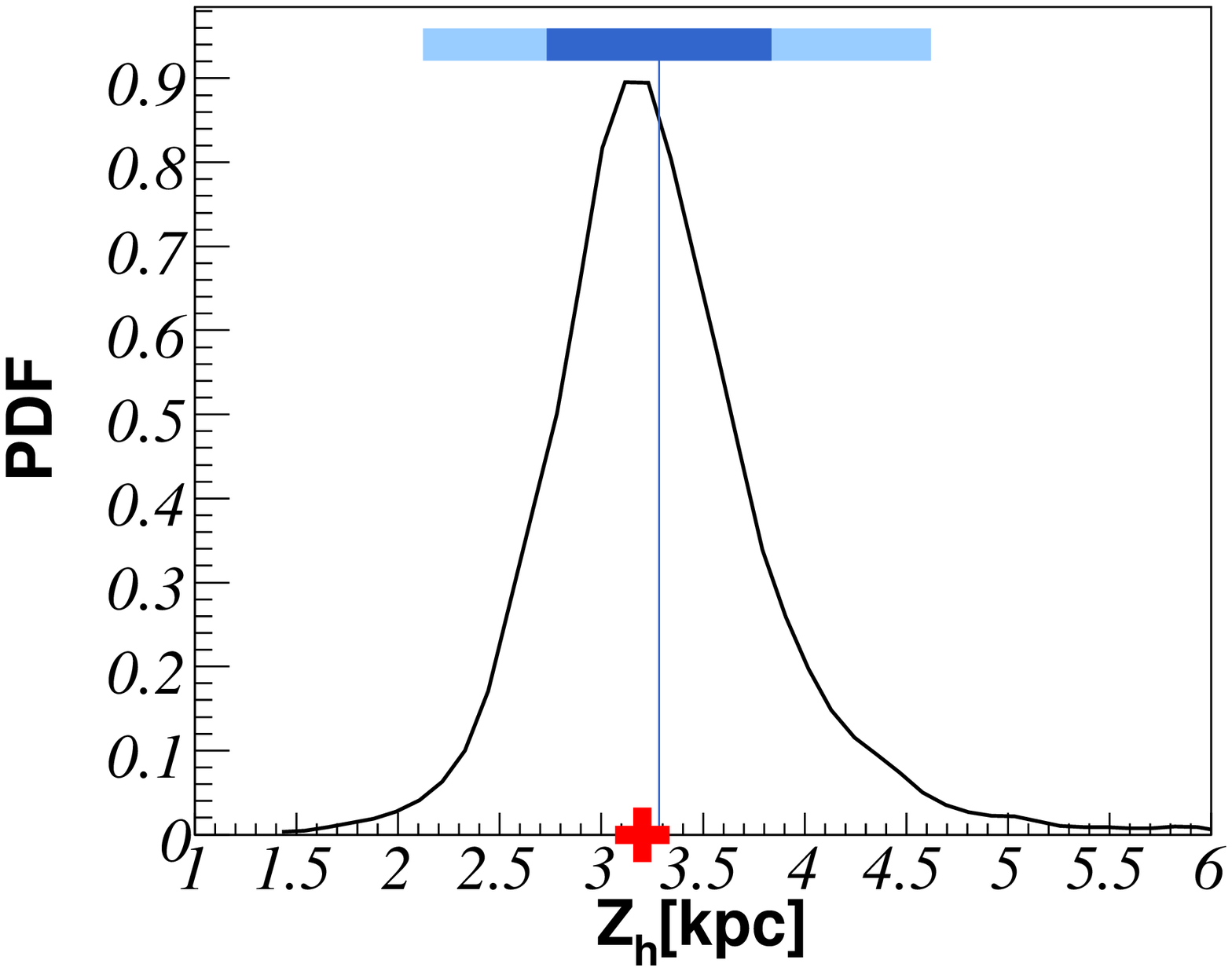}
\includegraphics[width=0.3\textwidth]{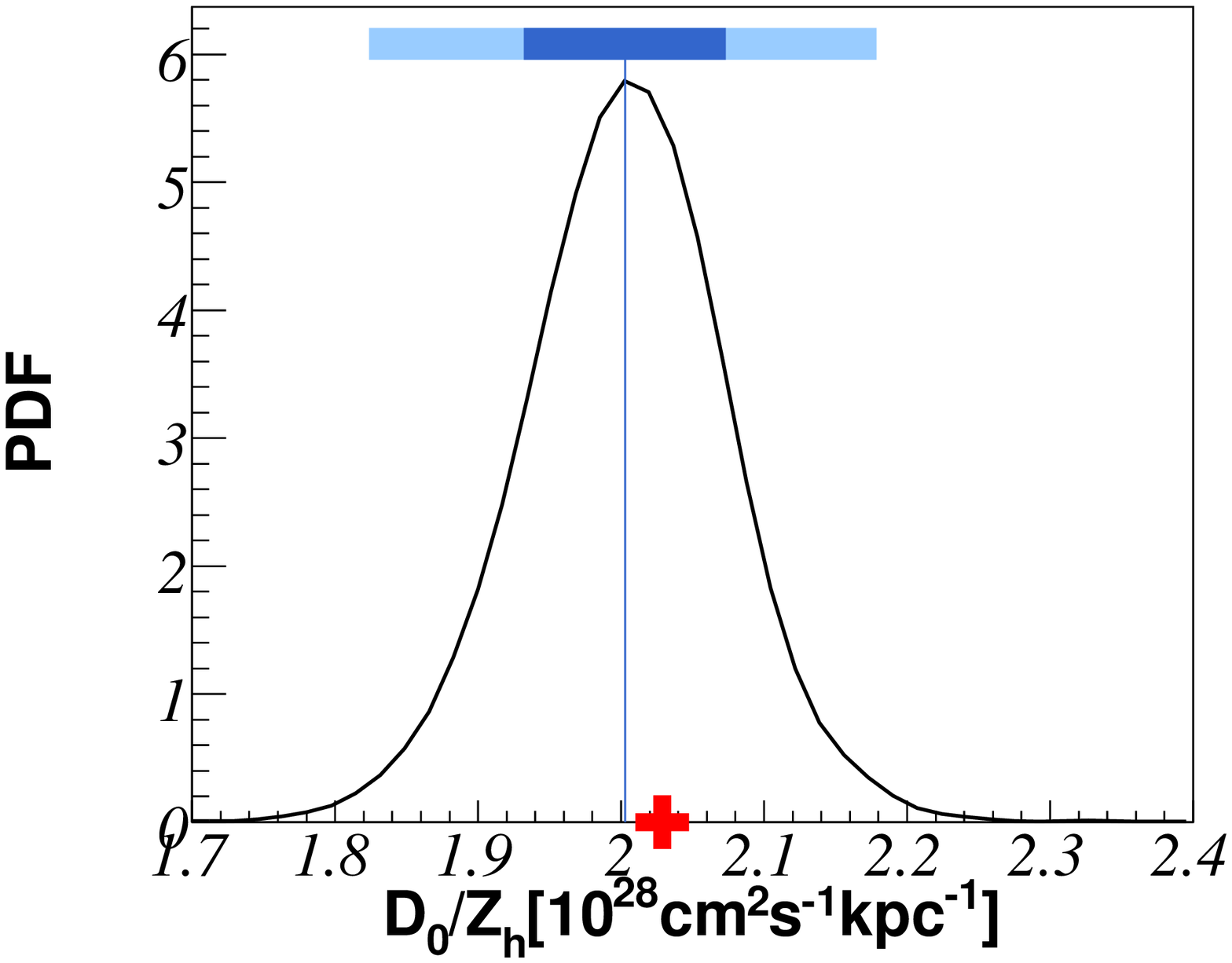}
\includegraphics[width=0.3\textwidth]{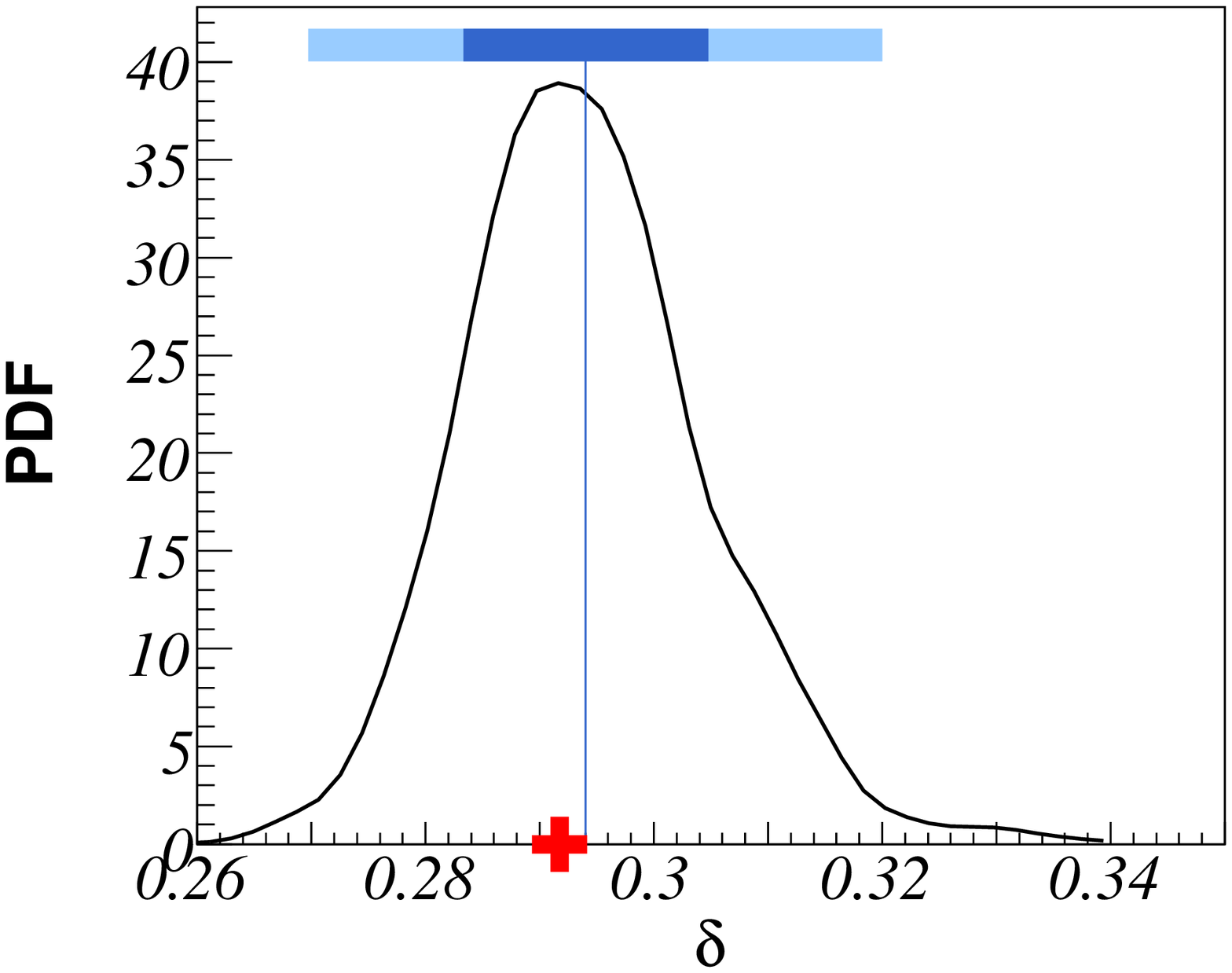}
\\
\includegraphics[width=0.3\textwidth]{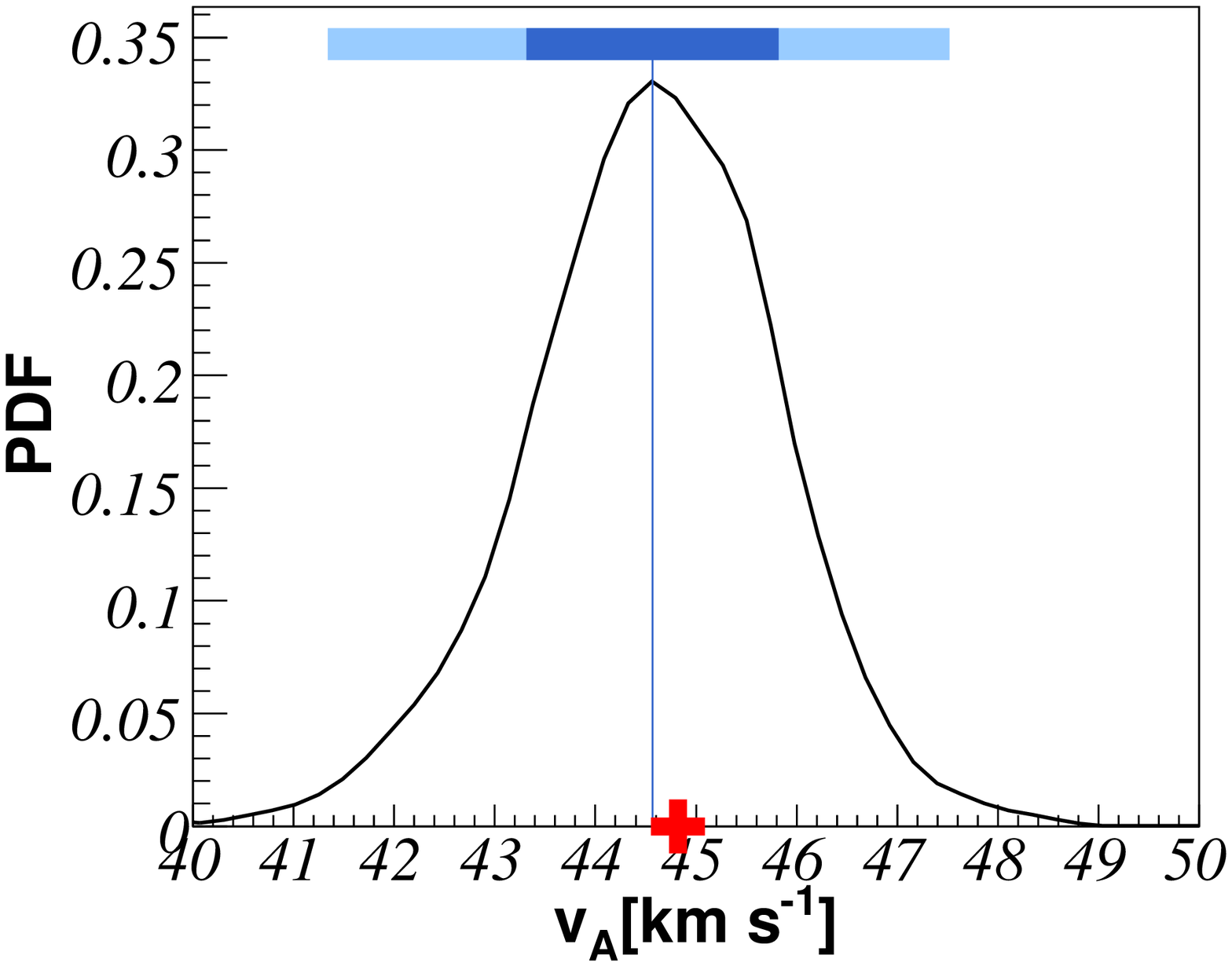}
\includegraphics[width=0.3\textwidth]{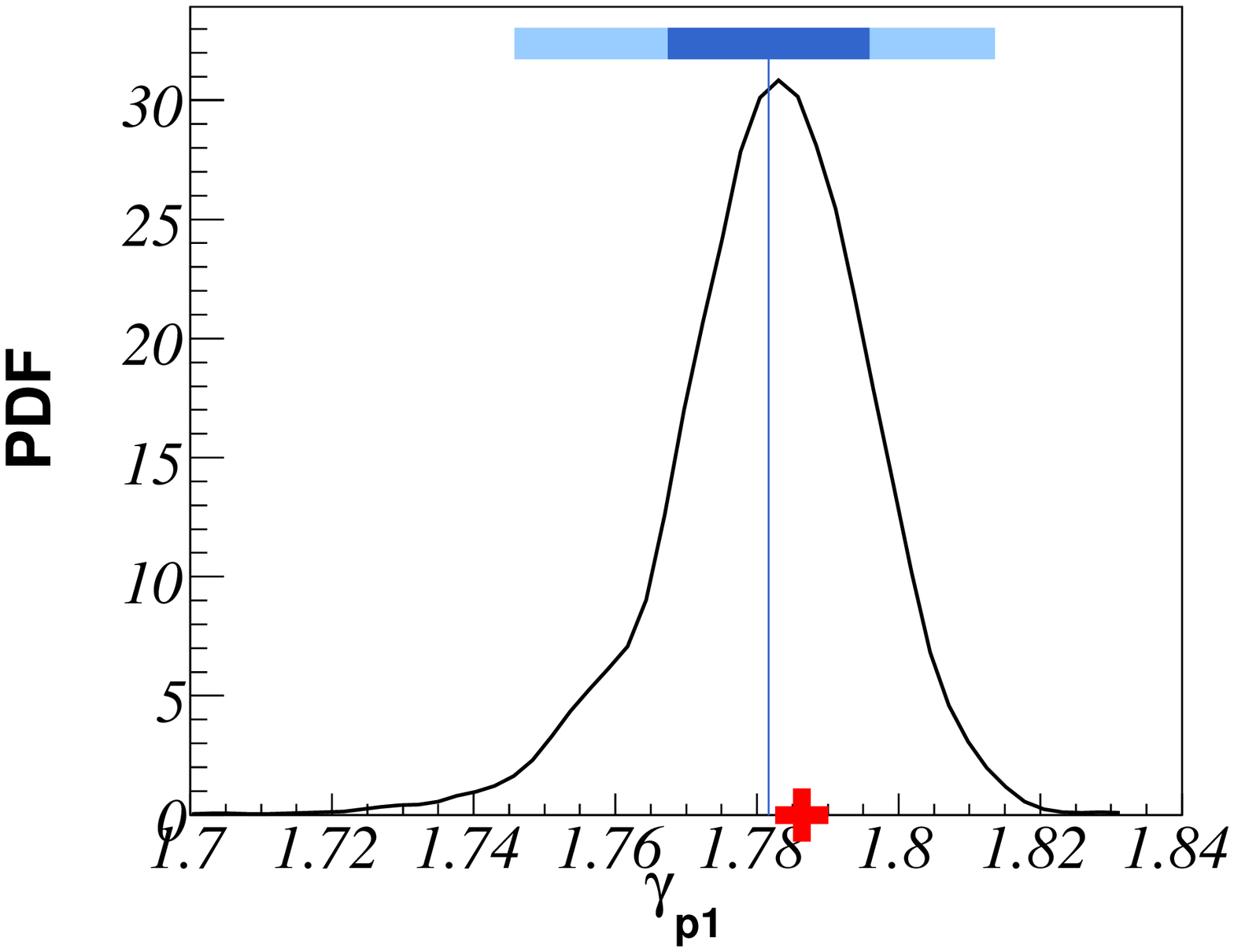}
\includegraphics[width=0.3\textwidth]{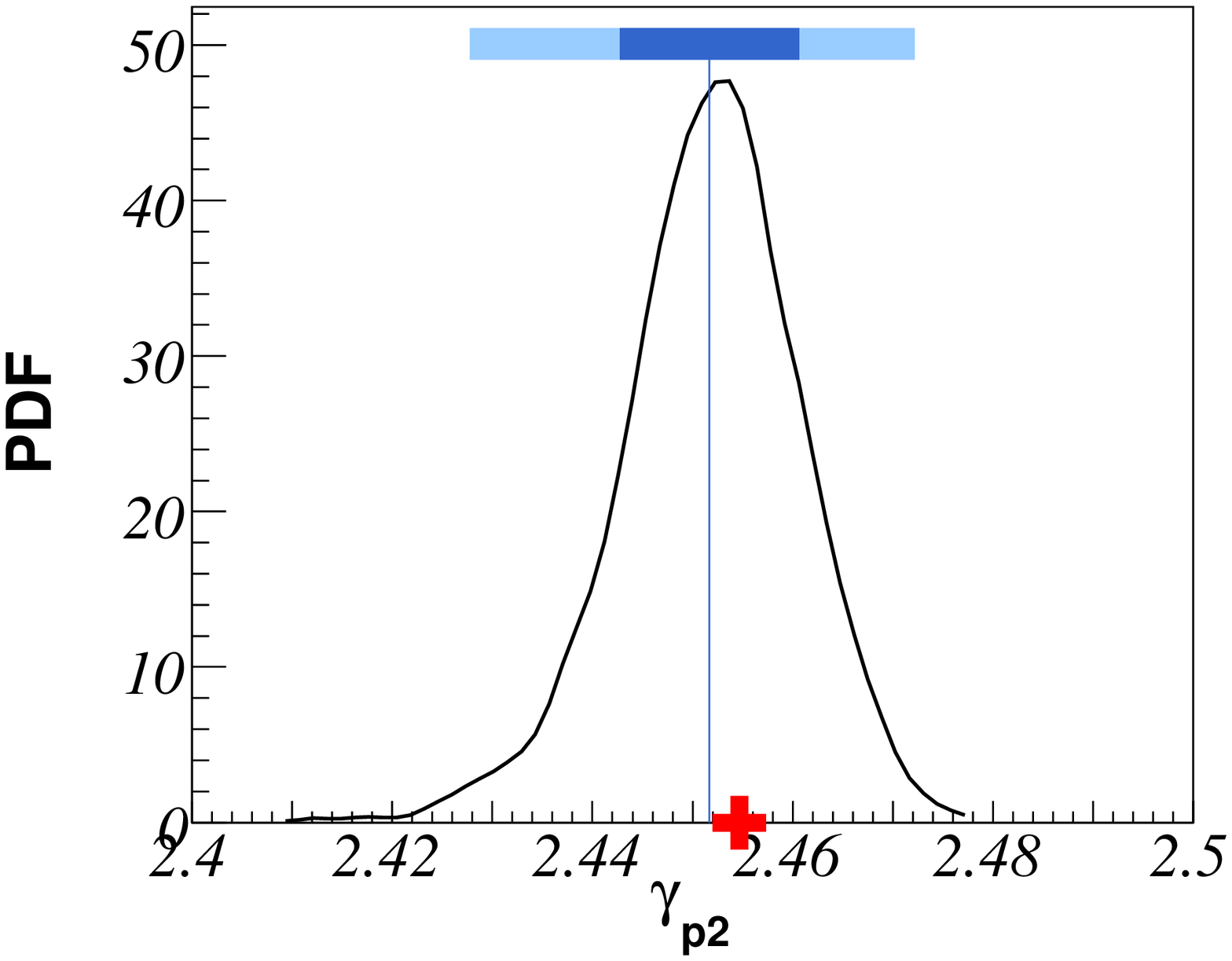}
\end{center}
\caption{
One-dimensional marginalized posterior PDFs for propagation parameters 
$Z_{h}$, $D_{0}/Z_{h}$, $\delta$, $V_{a}$, $\gamma_{p1}$, $\gamma_{p2}$.
In each panel, 
the horizontal bar indicates the $1\sigma$- and $2\sigma$-standard deviations,
with vertical line indicating the statistic mean value.
The best-fit value is shown as red plus.
}
\label{fig:param_1d}
\end{figure}

Based on the MCMC samples, 
the contours of allowed regions at $68\%$ and $95\%$ CL for 
a selection  of propagation parameters are shown in \fig{fig:param_2d}.
Some of the determined parameters are strongly correlated.
For instance, 
$D_{0}/Z_{h}$ is negatively correlated with $\delta$,
which is expected from 
the analytical solution of \eq{eq:two-zone-solution}.
The parameter $\delta$ is  negatively correlated with 
$\gamma_{p1}$ and $\gamma_{p2}$,
which is also consistent with \eq{eq:two-zone-solution}, 
as the sum $\delta+\gamma_{p1}(\gamma_{p2})$ should roughly reproduce 
the observed proton energy spectrum at low (high) energies.
The Alfv$\grave{\mbox{e}}$n speed $V_{a}$ is found to be 
positively correlated with $D_{0}/Z_{h}$, 
which can be understood from the definition of the re-acceleration term in \eq{eq:Dpp}.
Less pronounced correlations are  found between parameters 
$V_{a}$ and $\gamma_{p1,p2}$.
The one-dimensional marginal posterior PDFs for some of the  parameters
are shown in \fig{fig:param_1d}. 
In the figure,
the best-fit values, mean values with standard deviations are also shown. 

\begin{figure}
\begin{center}
\includegraphics[width=0.49\textwidth]{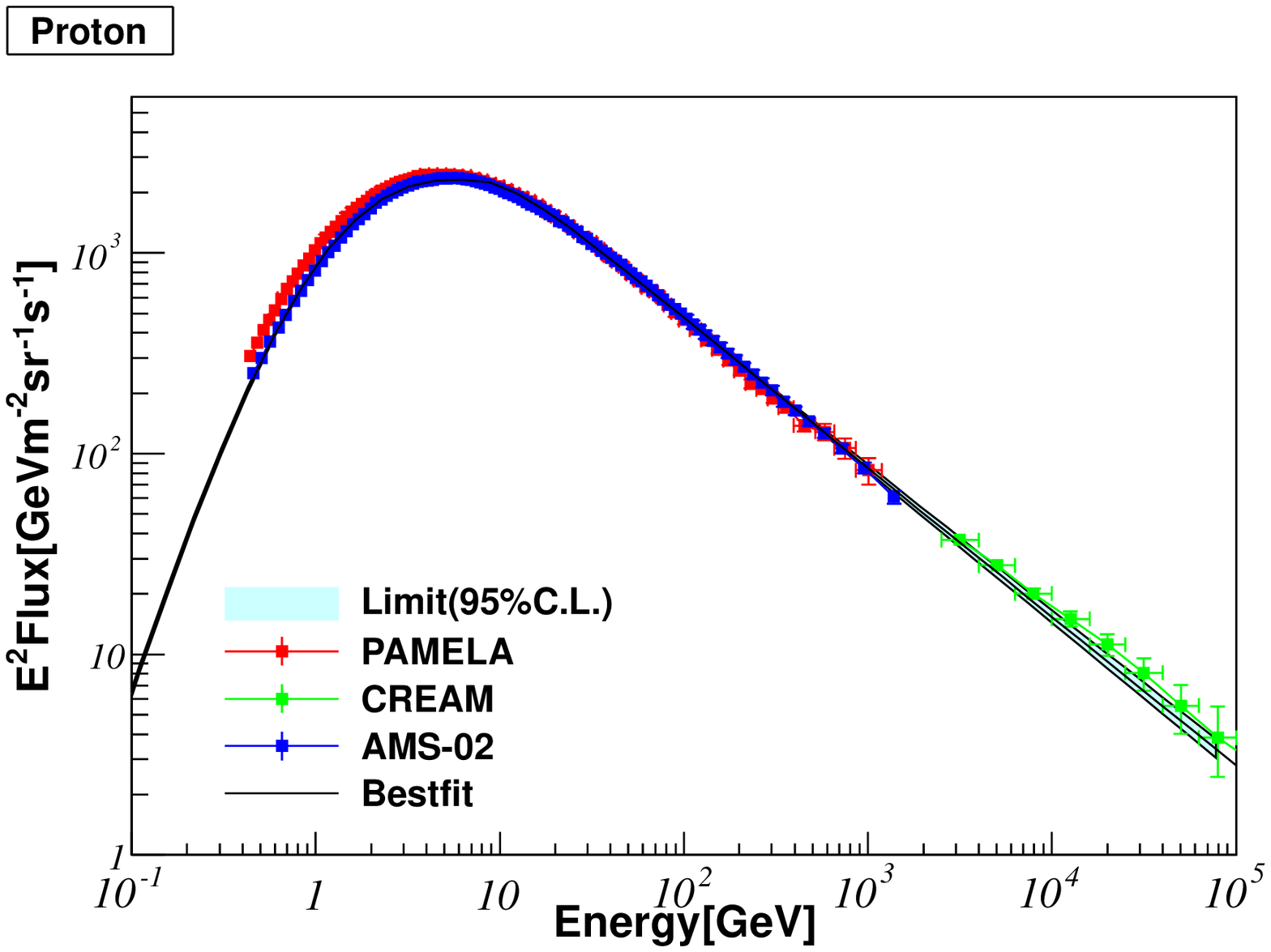}
\includegraphics[width=0.49\textwidth]{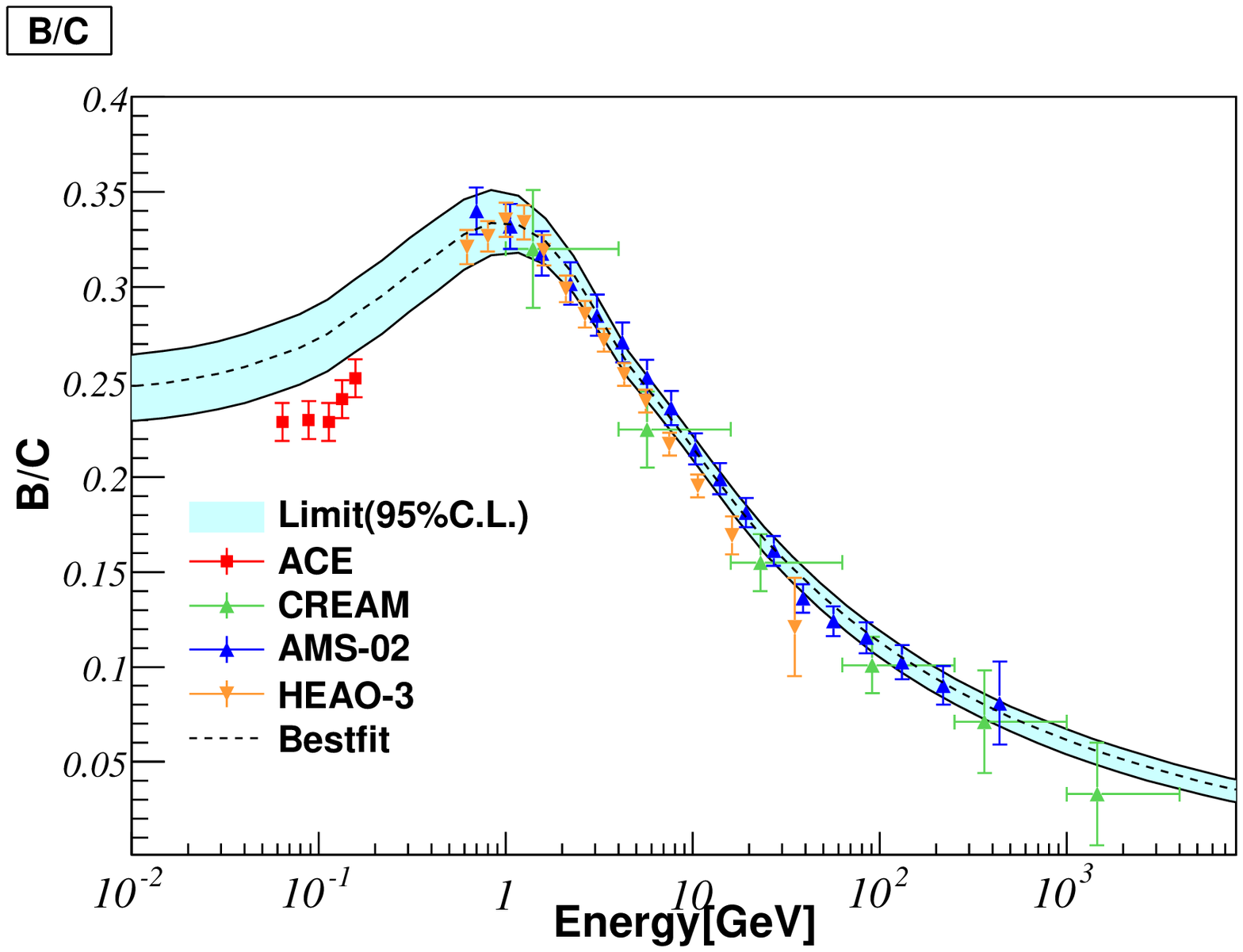}
\\
\includegraphics[width=0.49\textwidth]{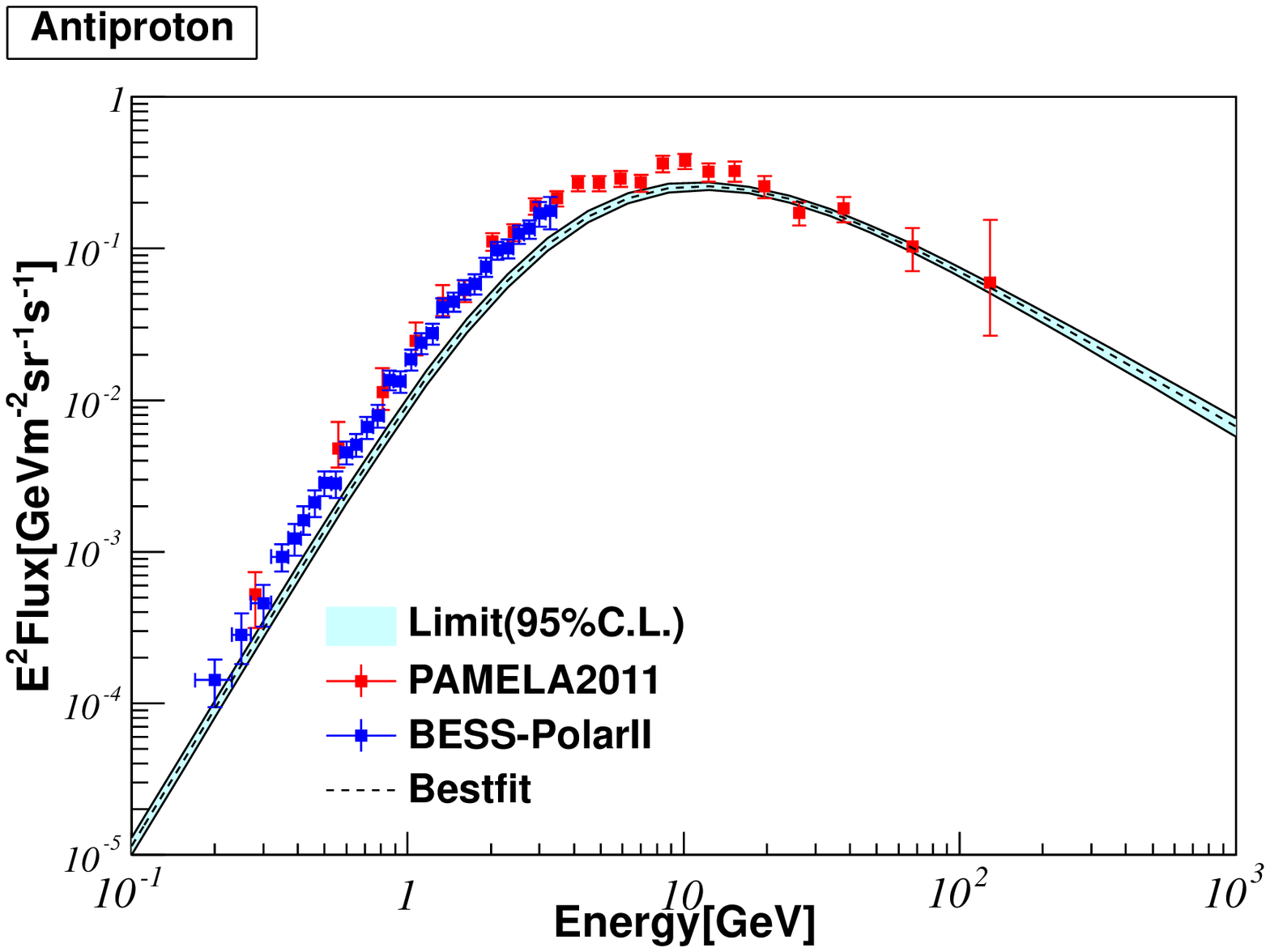}
\includegraphics[width=0.49\textwidth]{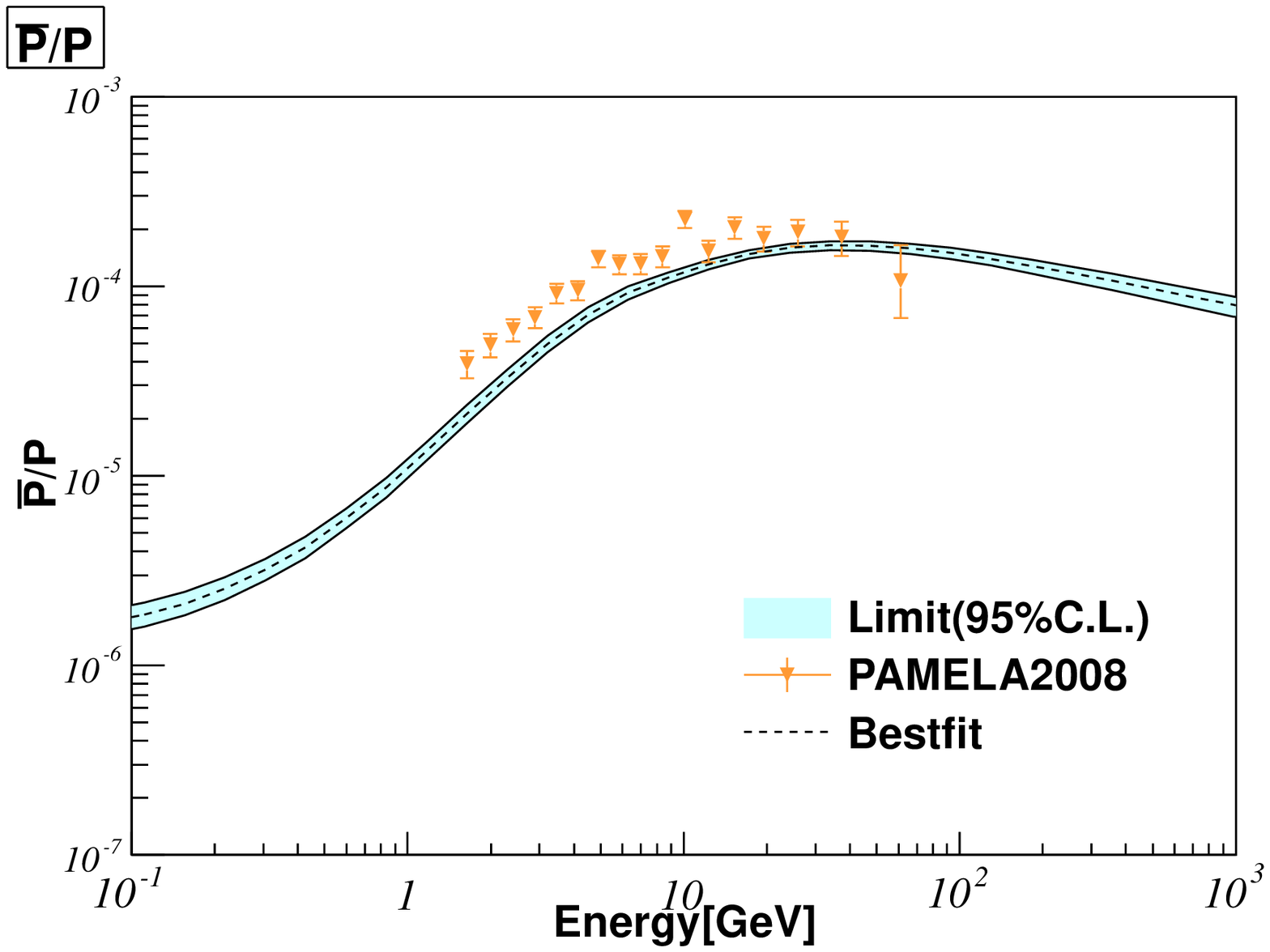}
\includegraphics[width=0.49\textwidth]{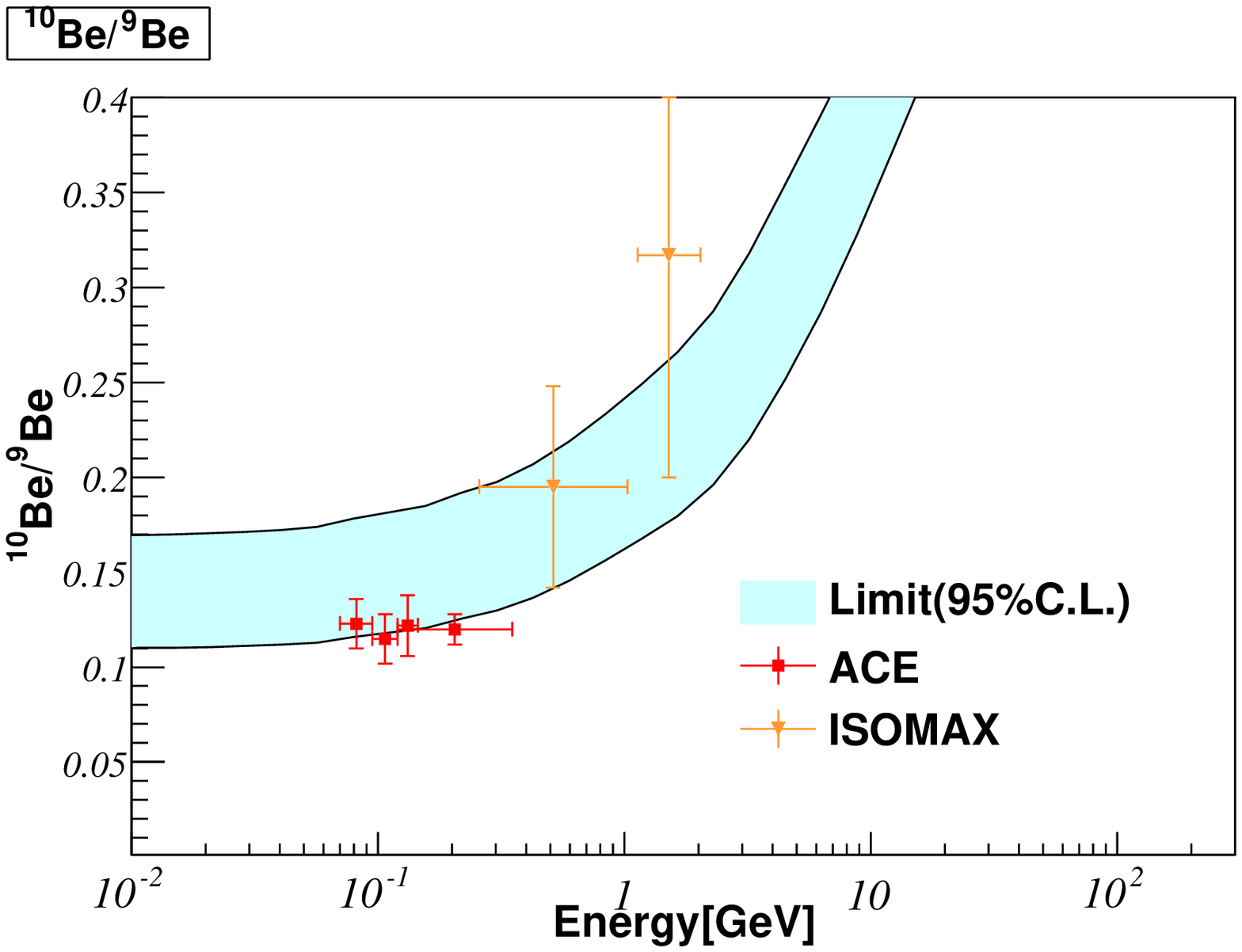}
\includegraphics[width=0.49\textwidth]{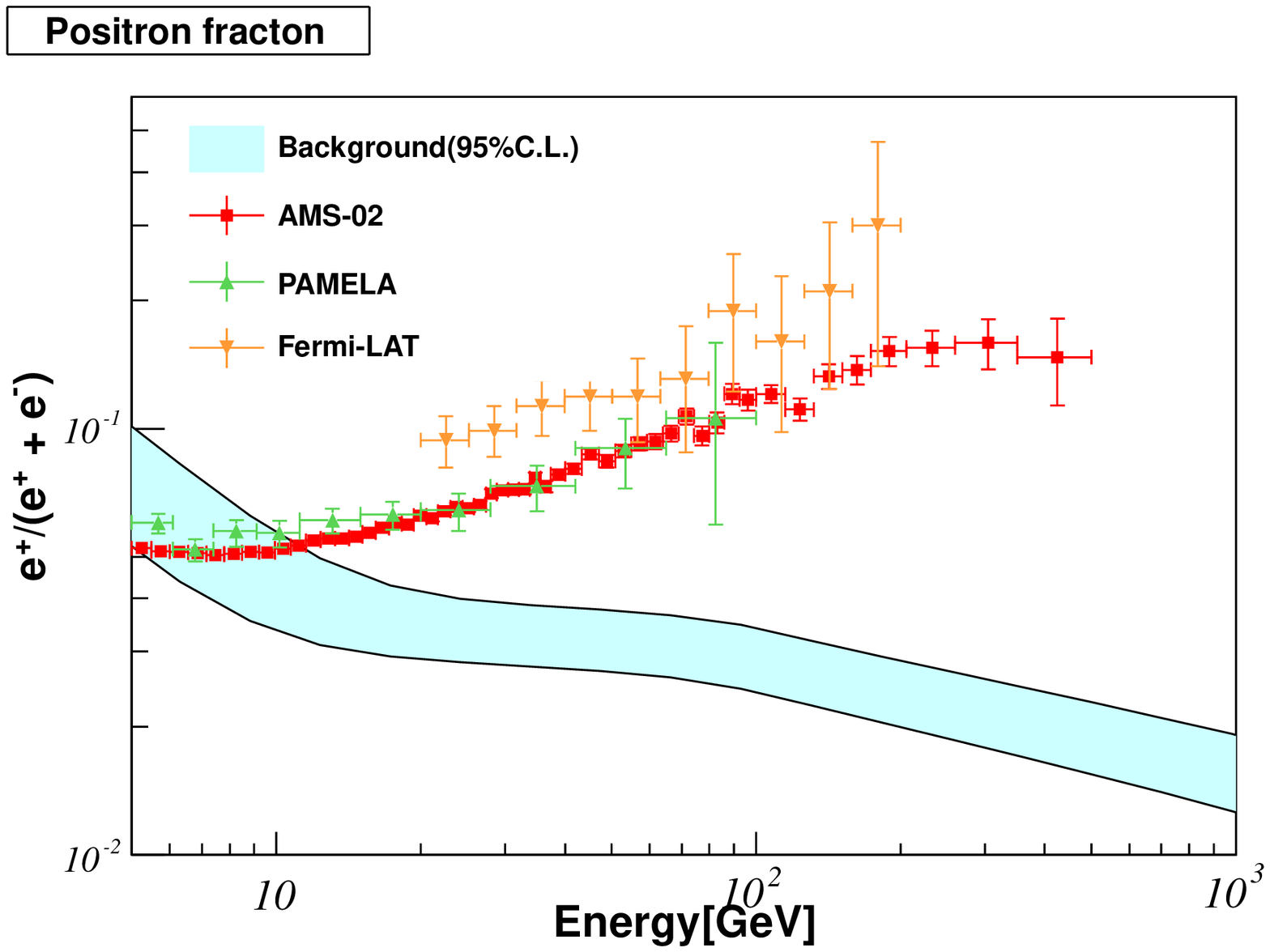}
\caption{
Cosmic ray nuclei fluxes and flux ratios from a global fit to the AMS-02 proton
and B/C data.
(Upper left)
the fitted spectra of cosmic-ray proton flux. 
The band corresponds to the values of propagation parameters  
allowed at $95\%$ CL.
The data of proton flux from 
AMS-02
\cite{Haino:icrc2013},
PAMELA
\cite{Adriani:2011cu} 
and CREAM 
\cite{Yoon:2011aa} 
are also shown.
(Upper right)
the fitted spectra of B/C ratio. 
The data of 
AMS-02
\cite{Oliva:icrc2013},
ACE
\cite{davis2000low},
CREAM
\cite{Ahn:2008my}
and HEAO-3
\cite{Engelmann:1990zz}
are also shown.
(Middle left)
the prediction for the antiproton flux at $95\%$ CL.
The data of 
PAMELA
\cite{Adriani:2010rc}
and 
BESS-Polar II
\cite{Abe:2011nx}
are shown.
(Middle right)
the prediction for the antiproton to proton flux ratio at $95\%$ CL.
The data of 
PAMELA
\cite{Adriani:2008zq} 
are shown.
(Lower left)
the prediction for $^{10}\text{Be}/^{9}\text{Be}$ flux ratio,
the data of ACE
\cite{yanasak2001measurement}
and ISOMAX
\cite{Hams:2004rz}
are shown.
(Lower right)
the prediction for positron fraction,
the data of AMS-02
\cite{Accardo:2014lma} 
PAMELA 
\cite{Adriani:2010ib} 
and 
Fermi-LAT 
\cite{FermiLAT:2011ab} 
are shown.
}
\label{fig_flux_bg}
\end{center}
\end{figure}

\fig{fig_flux_bg} shows the fitted spectra of 
the proton flux and B/C ratio, 
and the predicted antiproton fluxe, antiproton/proton ratio and 
$^{10}\text{Be}/^{}\text{Be}$ ratio 
using the parameters allowed within $95\%$ CL.
The AMS-02 data on proton flux and B/C ratio are well reproduced
by the  GALPROP  DR  models.
Although the $Z_{h}$ is determined purely by the proton flux, 
the predicted $^{10}\text{Be}/^{}\text{Be}$ ratio  is consistent 
with the data of 
ACE~\cite{yanasak2001measurement} 
and 
ISOMAX~\cite{Hams:2004rz}. 
The predicted antiproton fluxes are consistent with the PAMELA data only for the 
kinetic energies above 10 GeV. 
At lower  energies, 
the predicted  antiproton flux is about $40\%$ lower than the data of 
PAMELA and BESS-Polar II,
which is a typical  feature of the DR  models in GALPROP~\cite{Moskalenko:2001ya}.
The low energy antiproton spectrum can be correctly reproduced
if one constructs  sophisticated  GALPROP models with 
a flattening of the diffusion coefficient together with  a convection term 
and  a break in the injection spectrum
\cite{Moskalenko:2001ya}.  
Another possibility is that 
the solar modulation may have a charge sign dependence, namely,
the modulation for antiprotons is different from that of protons.

\section{Positron fraction from DM annihilation}\label{sec:positron}
Recently the measurement of the positron fraction was extended to 
the energy range up to 500 GeV by AMS-02
\cite{Accardo:2014lma}. 
For the first time, it was shown that 
the positron fraction stops to increase with energy at  $\sim 270$ GeV.
The spectral features of the positron fraction such as
the rate of increase with energy,
the energy beyond which it ceases to increase and
the rate at which it falls beyond the turning point
are of crucial importance in distinguishing the DM models.
Since the  uncertainties in the propagation parameters affect
the calculations of
$both$ the background and the DM contribution in the positron fraction,
it is necessary to consider this uncertainty in deriving the properties 
of DM particles from the positron excess.

\mnote{}
We first investigate the predicted positron fraction for the case of background only.
The result is shown in \fig{fig_flux_bg}, 
where we have chosen a reference electron primary source with 
two breaks at $\rho_{e1}=4$ GV and $\rho_{e2}=86.8$ GV, and 
three power law indices between the breaks: 
$\gamma_{e1}=1.46$, $\gamma_{e2}=2.72$ and $\gamma_{e3}=2.49$,
respectively.
The shaded bands in the figure correspond to the variation of 
the propagation parameters within $95\%$ CL.
The figure shows that
the typical uncertainties in the positron fraction can reach a factor of two
in the background-only case.
Clearly, at energies above $\sim 20$~GeV, 
the positron fraction cannot be explained by 
the background even after including the uncertainties of the propagation parameters,
which calls for exotic contributions such as halo DM annihilation.

\begin{figure}\begin{center}
\includegraphics[width=0.60\textwidth]{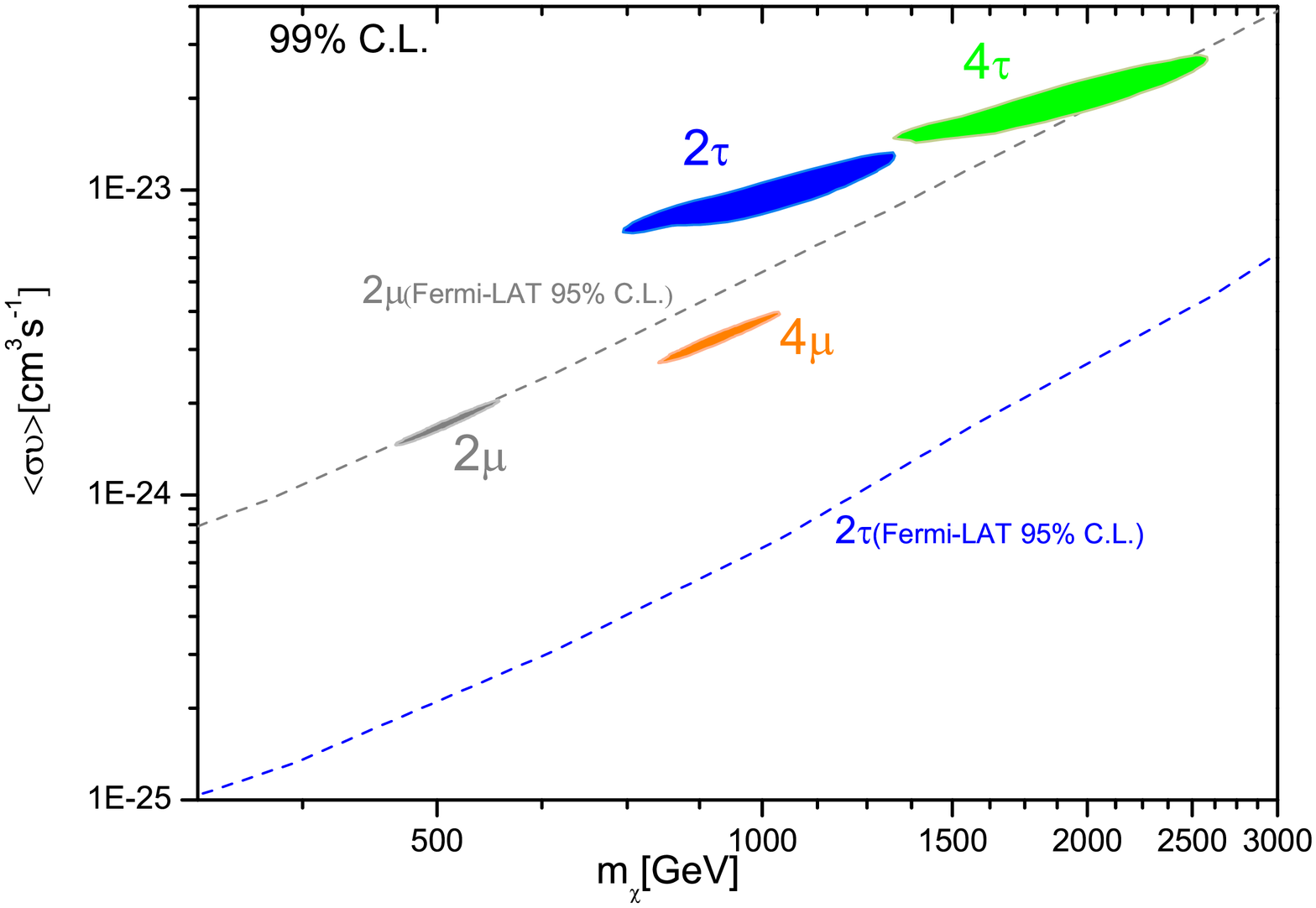}
\caption{
Allowed regions for 
DM particle mass and annihilation cross section at $99\%$ CL for 
DM annihilation into 
$2\mu$, $4\mu$, $2\tau$ and $4\tau$ final states from the global fit. 
The upper limits on 
the $2\mu$ and $2\tau$ channels from 
the Fermi-LAT 6-year gamma-ray data of 
the dwarf spheroidal satellite galaxies of the Milky Way
are also shown
\cite{Ackermann:2015zua}.
}
\label{fig:2leptons}
\end{center}\end{figure}

\begin{figure}
\includegraphics[width=0.49\textwidth]{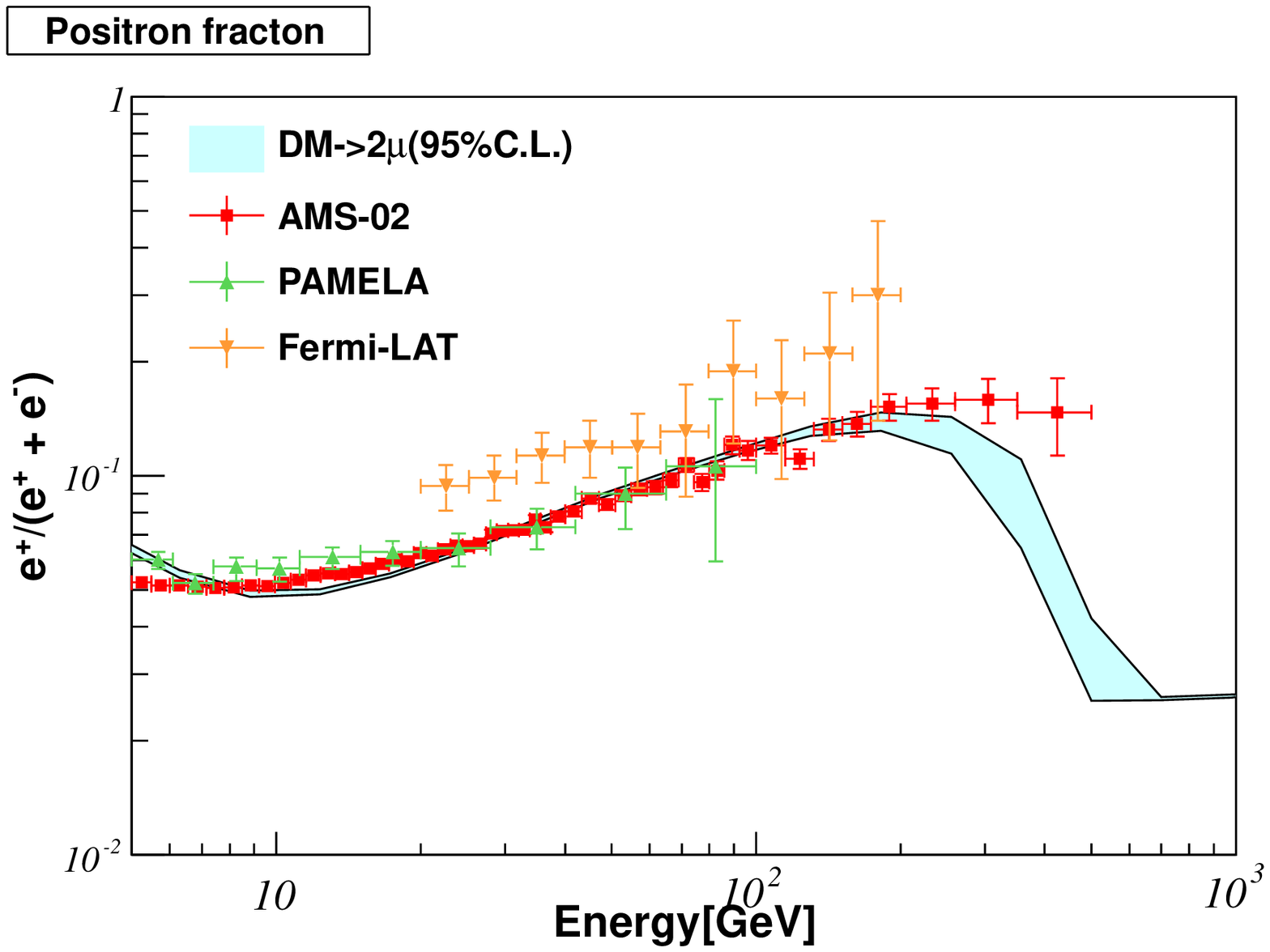}
\includegraphics[width=0.49\textwidth]{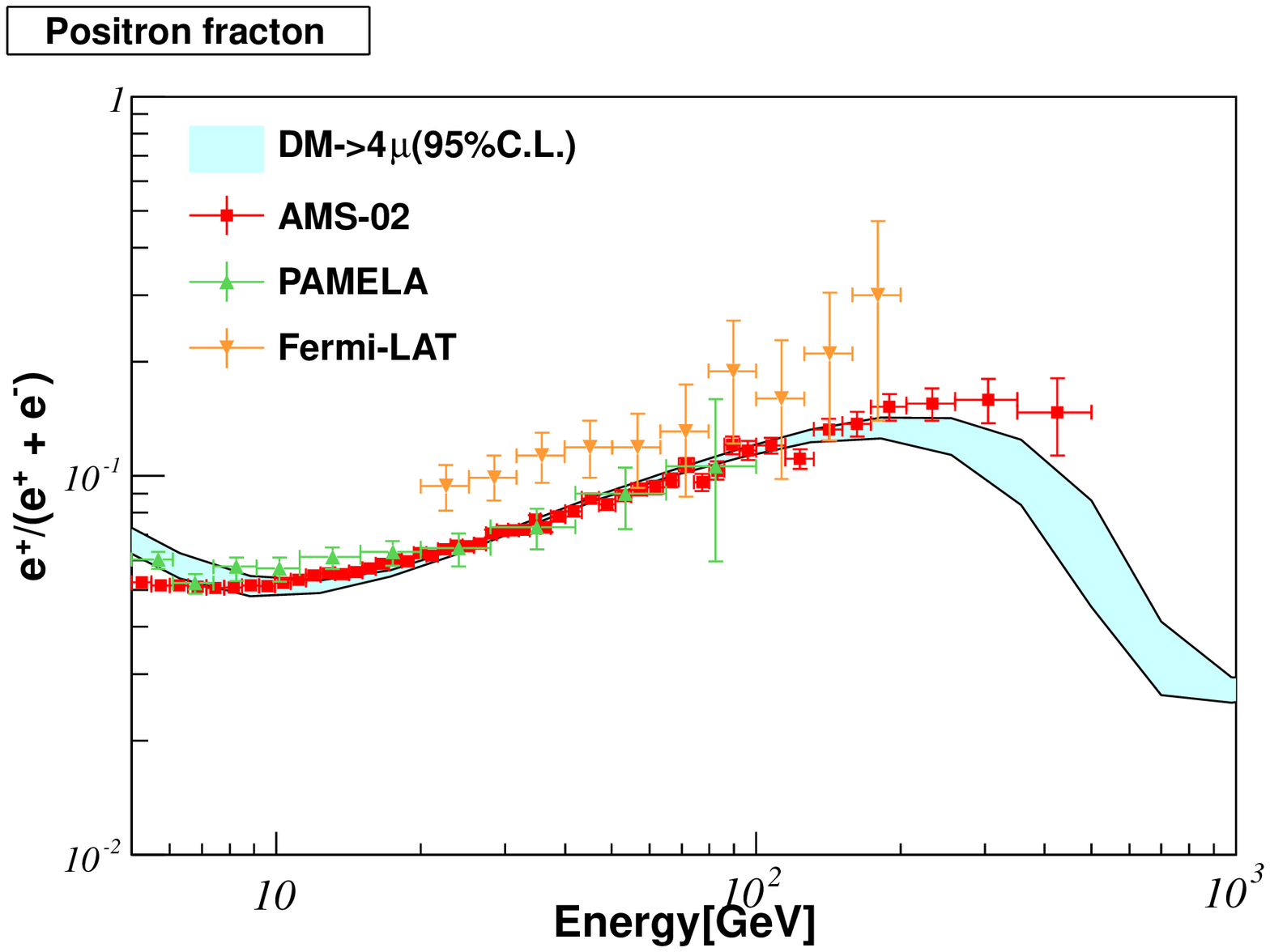}
\\
\includegraphics[width=0.49\textwidth]{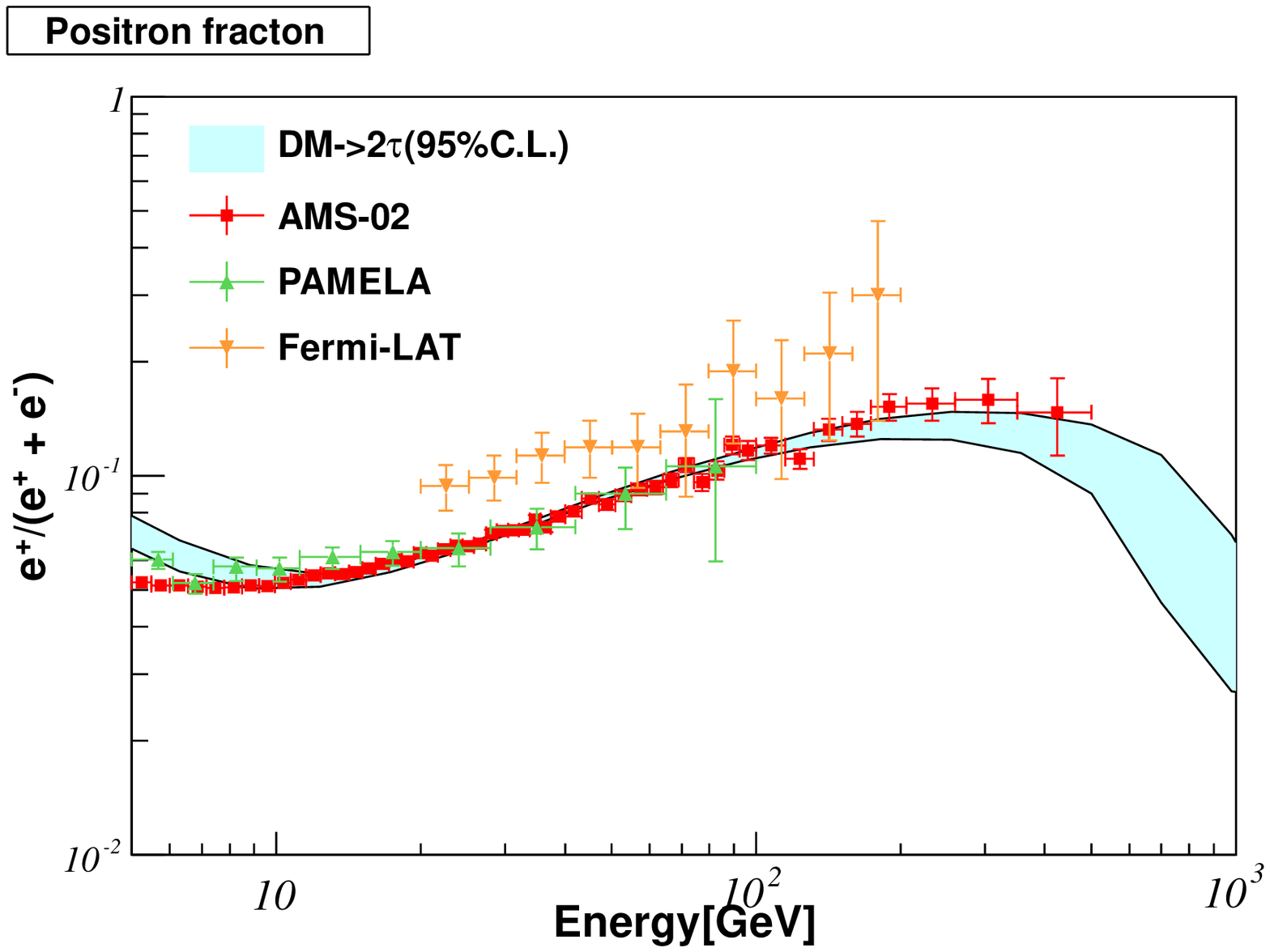}
\includegraphics[width=0.49\textwidth]{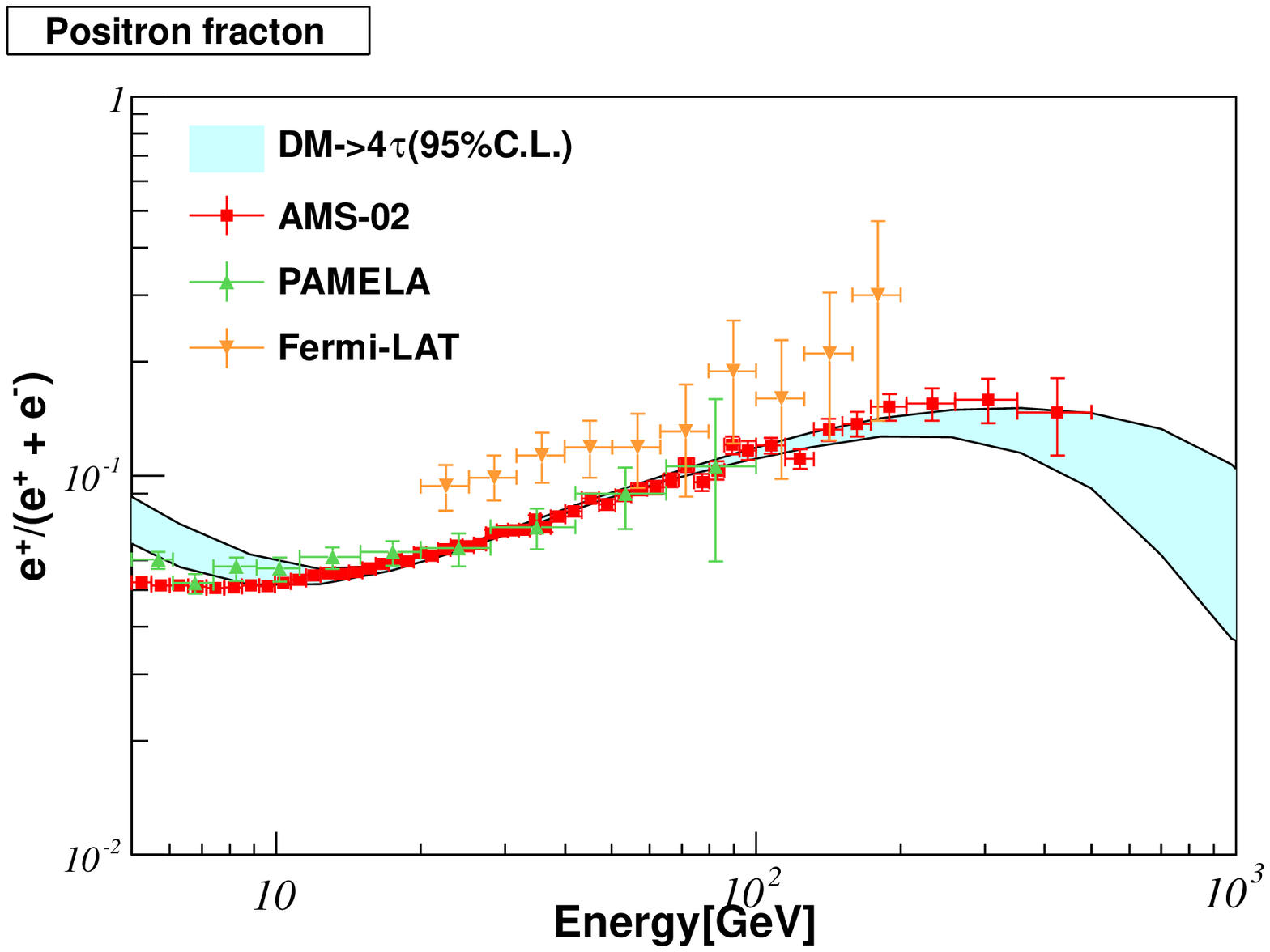}
\caption{
Predictions for cosmic-ray positron fraction from DM annihilation 
into final states $2\mu$, $4\mu$, $2\tau$ and $4\tau$.
In each plot,
the shaded band represents the uncertainties due to
that in the propagation parameters 
and the DM properties ( $m_{\chi}$ and $\langle\sigma_{v}\rangle$ )
at $95\%$ CL.
The data of AMS-02
\cite{Accardo:2014lma}, 
PAMELA 
\cite{Adriani:2010ib} 
and Fermi-LAT 
\cite{FermiLAT:2011ab} 
are also shown.
}
\label{fig:PBC_Fraction4Final}
\end{figure}

We then include the DM contribution and 
add the AMS-02 data of positron fraction into 
a similar global Bayesian fit detailed in \Sec{sec:Bayes} to determine 
the  DM particle mass $m_{\chi}$ and annihilation cross section $\langle \sigma v\rangle$
for various DM annihilation channels.
The major propagation parameters  such as
$Z_{h}$, $D_{0}/Z_{h}$, $V_{a}$, $\delta$, $\gamma_{p1}$ and $\gamma_{p2}$
are also allowed to vary  freely as nuisance parameters in the fit.
In order to avoid the uncertainties  related to the modelling of  Solar modulation, 
only the positron fraction data with kinetic energy above 20 GeV are included in the fit.
For the four typical DM annihilation channels
$\chi\bar \chi \to 2\mu$, $4\mu$, $2\tau$ and $4\tau$ 
with the Einasto DM profile, 
we find the following results
\begin{align}
2\mu\text{ :  }
&m_{\chi}= 507\pm 30\text{ GeV}, 
& \langle \sigma v \rangle = (1.72\pm0.14)\times 10^{-24}~\text{cm}^{3}\text{s}^{-1} , 
\nonumber \\
4\mu\text{ : }
&m_{\chi}=903\pm50\text{ GeV}, 
&\langle \sigma v \rangle =(3.28\pm 0.24)\times 10^{-24}~\text{cm}^{3}\text{s}^{-1}  ,
\nonumber \\
2\tau\text{ : }
&m_{\chi}=1076\pm100\text{ GeV}, 
& \langle \sigma v \rangle =(1.03\pm0.10)\times 10^{-23}~\text{cm}^{3}\text{s}^{-1}  ,
\nonumber \\
4\tau\text{ : }
&m_{\chi}=1964\pm224\text{ GeV}, 
& \langle \sigma v \rangle =(2.06\pm0.23)\times 10^{-22}~\text{cm}^{3} \text{s}^{-1} .
\end{align}
The allowed regions in the ($m_{\chi},\langle \sigma v \rangle$) plane at $99\%$~CL are shown in \fig{fig:2leptons}.
The corresponding values of $\chi^{2}/\text{d.o.f}$ 
which indicate the goodness-of-fit are 
$2.92\ (2\mu)$, $2.16\  (4\mu)$, $1.44\ (2\tau)$ and $1.27\ (4\tau)$, respectively.
In \fig{fig:PBC_Fraction4Final}, 
we show the predicted positron fraction  for 
the  four typical DM annihilation channels with 
the Einasto DM profile. 
The band in each plot indicates the uncertainties due to 
both the DM parameters and the propagation parameters at 95\%~CL. 
One can see from the the figure that 
for the channels with $\mu$ final states 
the predicted spectra are too hard to fit the AMS-02 data at high energies.
Thus the $\tau$ final states are favoured over $\mu$ final states by the AMS-02 data.
However, 
the cross sections for $\tau$ final states are very large and 
in strong tension with the gamma-ray bound from the dwraf spheroid
satellite galaxies of the Milky Way
\cite{Ackermann:2015zua} 
as can be seen from the figure.
These result are consistent with 
our previous  work using the earlier  AMS-02 data
and  a set of fixed backgrounds
\cite{Jin:2013nta}.

\section{Antiproton flux from DM annihilation}\label{sec:minmax}
Compared with cosmic-ray electrons, 
which loss energy quickly due to 
the inverse Compton scattering and synchrotron radiation, 
the cosmic-ray protons lose  much less  energy in 
the propagation process.
Thus they can travel across a longer distance in the galaxy before 
arriving at the detectors, 
which makes the proton/antiproton fluxes more sensitive to 
the propagation parameters.

In the previous section, 
we have shown that with the current AMS-02 data 
the important propagation parameters
such as $D_{0}/Z_{h}$ and $Z_{h}$ can be determined with 
better precisions,
which is useful in improving the predictions for 
the cosmic-ray antiproton fluxes induced from DM interactions. 
In this section, 
we estimate the  uncertainties in the prediction for  
antiproton flux from DM annihilation
and construct reference propagation models 
which give  rise to the typically minimal, median and maximal antiproton fluxes
within $95\%$ CL.
Such reference models are useful  for a quick estimation of 
the propagation uncertainties in future analyses.
We shall focus only on the case of DM annihilation.
It is straight forward to extend the analysis to the case of DM decay.

\begin{figure}
\includegraphics[width=0.49\textwidth]{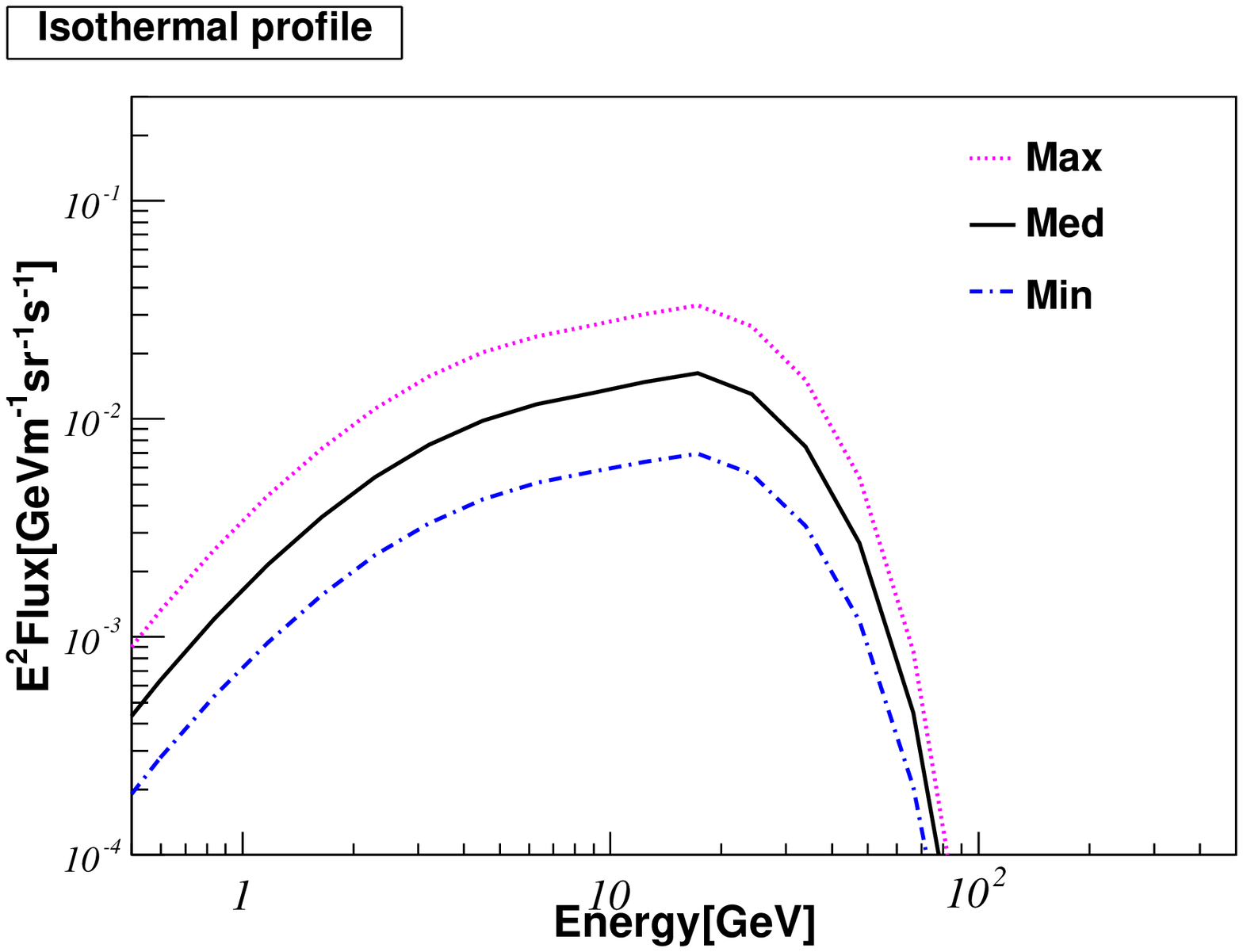}\includegraphics[width=0.49\textwidth]{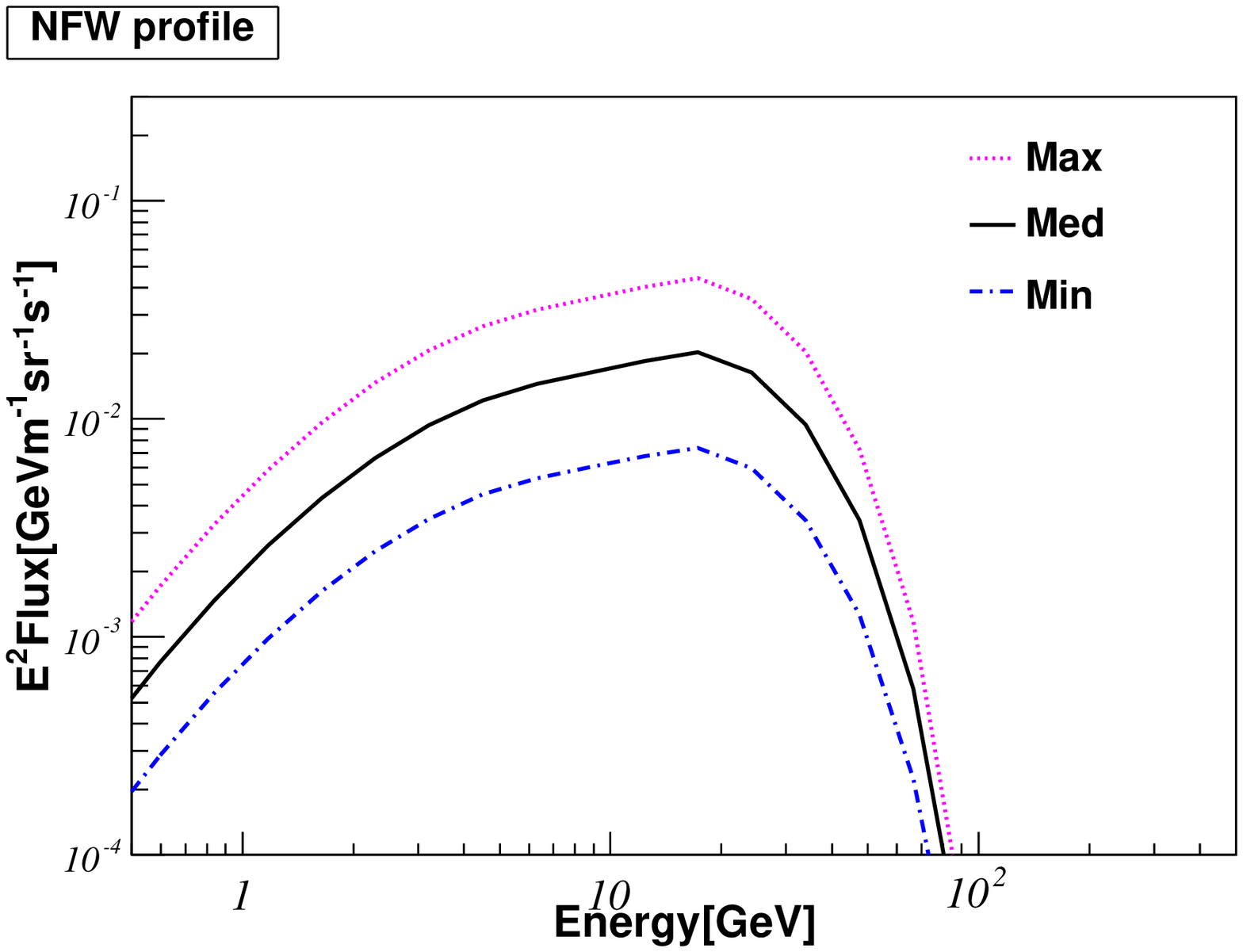}\\
\includegraphics[width=0.49\textwidth]{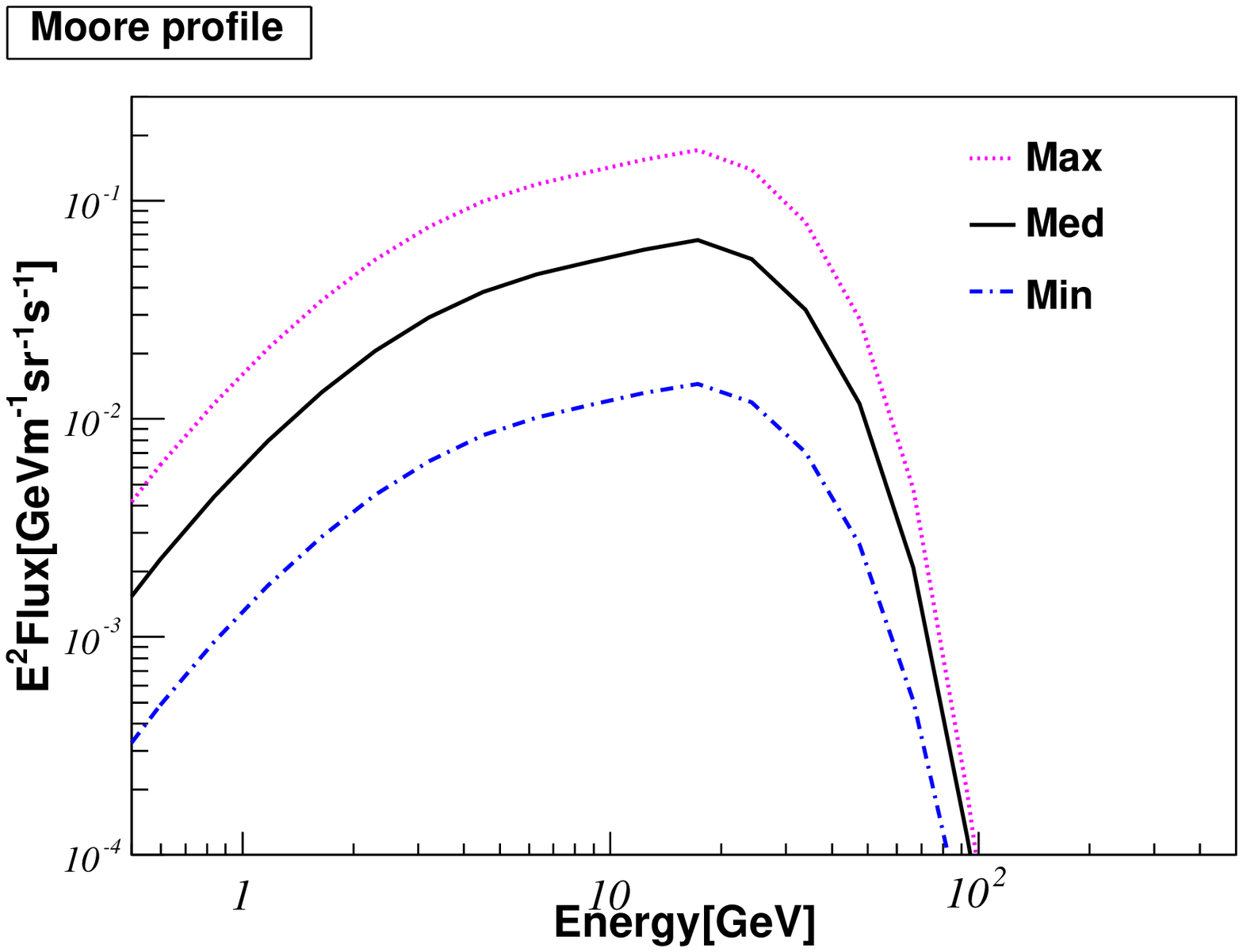}
\includegraphics[width=0.49\textwidth]{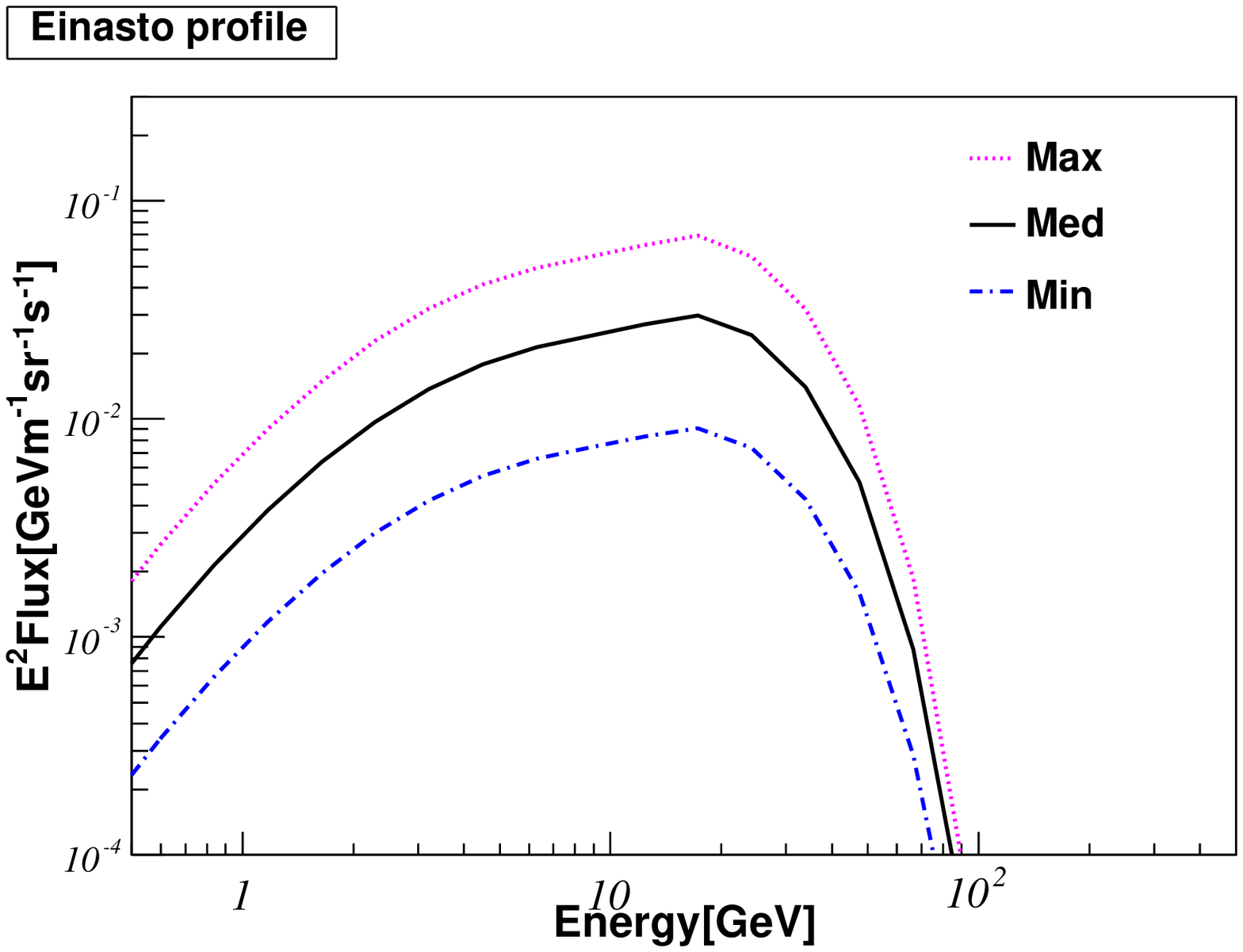}
\caption{
Prediction for the antiproton fluxes resulting from 
DM particle annihilating into $b\bar b$ final states 
in the three propagation models listed in \tab{tab:para_minMax}.
In each plot, three curves   correspond to
the typically  minimal (dot-dashed), median (solid) and maximal (dotted) antiproton fluxes
at $95\%$ CL.
The four plots corresponds to the four different DM density distribution profile
NFW (upper left)
\cite{Navarro:1996gj},
Isothermal (upper right)
\cite{Bergstrom:1997fj},
Moore (lower left)
\cite{Moore:1999nt, 
Diemand:2004wh} and 
Einasto (lower right)
\cite{Einasto:2009zd}. 
The mass of the DM particle is 130 GeV and
the annihilation cross section is fixed at 
$\langle \sigma v\rangle_{0} =3\times 10^{-26}~\text{cm}^{3}\text{s}^{-1}$.
 }
 \label{fig:EPBC_PbarMinMax}
\end{figure}

For a concrete  illustration,
we consider a reference DM model with $m_{\chi}=130$ GeV, 
and a typical WIMP annihilation cross section
$\langle \sigma v\rangle_{0} =3\times 10^{-26}~\text{cm}^{3}\text{s}^{-1}$
with final state dominated by $b \bar b$.
From the propagation models allowed by the recent AMS-02 data at $95\%$ CL,
we select reference models which give minimal, median and maximal antiproton 
fluxes. 
The values of the parameters are listed in \tab{tab:para_minMax}, 
and
the corresponding fluxes for different types of DM profiles are shown in \fig{fig:EPBC_PbarMinMax}.
\begin{table}[htb]
\begin{center}
\begin{tabular}{llll}
  \hline\hline
parameters &Min&Med&Max\\
\hline 
$Z_h (\text{kpc})$	&1.8 	&3.2		&6.0\\
$D_0/Z_h$		&1.96 	&2.03	&1.77 \\
$\delta$			&0.30 	&0.29	&0.29 \\
$V_{a} (\text{km}\cdot\text{s}^{-1})$
				&42.7 	&44.8	&43.4 \\
$\gamma_{p1}$	&1.75 	&1.79	&1.81 \\
$\gamma_{p2}$	&2.44 	&2.45	&2.46 \\
  \hline\hline
\end{tabular}
\end{center}
\caption{
Three reference propagation models 
selected from the set of propagation models allowed within $95\%$~CL
by the AMS-02 data,
corresponding to the minimal, median and maximal
antiproton fluxes from DM annihilating into $b\bar b$.
The parameter $D_{0}/Z_{h}$ is in units of $10^{28}\text{cm}^{2}\cdot\text{s}^{-1}\text{kpc}^{-1}$.
}
\label{tab:para_minMax}
\end{table}
As can be seen from the figure, 
the uncertainties due to the propagation parameters are 
within an order of magnitude.
In some previous analysis, the choice of benchmark models leads to 
an uncertainty of  $\mathcal{O}(100)$
\cite{Donato:2003xg}. 
Such a significant improvement is related to the precision AMS-02 data
on the B/C ratio.
\fig{fig:EPBC_PbarMinMax}  also shows that 
the differences due to the DM profile are typically around 
a  factor of two among the profiles of NFW, Isothermal and Einasto.
In the Moore profile, the differences are bigger and 
can reach $\mathcal{O}(20)$.
\section{Dark matter properties from current and future antiproton data}\label{sec:pbarlimits}

Taking into account the uncertainties of all the propagation parameters,
one can  derive conservative  constraints on 
the properties of DM particles from 
the current PAMELA data
and 
make projections for the sensitivity of 
the upcoming AMS-02 antiproton measurement.
Some previous  analyses based on 
simplified  assumptions of  fixed background or 
allowing part of the propagation parameters to vary can be found in 
Refs.
\cite{
Cholis:2010xb, 
Evoli:2011id, 
Cirelli:2013hv, 
Fornengo:2013xda}. 
In the Bayesian approach, 
it is straightforward to 
consider the uncertainties and correlations of 
the propagation parameters consistently,
as the  posterior PDFs of the propagation parameters obtained in \Sec{sec:fit}
can be used as the prior PDFs in 
the subsequent Bayesian analysis.
The inclusion of the new data will also update 
the ``degree of believe'' of these parameters,
as well as 
constrain the new parameters related to the properties of DM particles.
In the case of DM annihilation, 
the new parameter set related to DM annihilation is 
$\boldsymbol\theta'=\{ \langle \sigma v \rangle, m_{\chi}\}$.
The new data set of cosmic-ray antiproton is 
$D'=\{ D^{\text{PAM}}_{p}, D^{\text{PAM}}_{\bar p/p} \}$, 
where $D^{\text{PAM}}_{p}(D^{\text{PAM}}_{\bar p/p})$ stands for 
the data of antiproton flux (antiproton to proton flux ratio) from PAMELA.
The posterior PDF for the parameter set $\boldsymbol\theta'$ can be written as
\begin{align}\label{eq:Bayesian-update}
P(\boldsymbol\theta',\boldsymbol\theta| D' )
=
\frac{\mathcal{L}(D'|\boldsymbol\theta', \boldsymbol\theta ) 
\pi(\boldsymbol\theta')\tilde\pi'(\boldsymbol\theta) }
{\int \mathcal{L}(D'| \boldsymbol\theta', \boldsymbol\theta) \pi(\boldsymbol\theta') \tilde\pi(\boldsymbol\theta) d\boldsymbol\theta' d\boldsymbol\theta} ,
\end{align}
where 
$\tilde\pi(\boldsymbol\theta)$ is 
the prior PDF of the propagation parameter set $\boldsymbol\theta$ defined in \eq{eq:paramSet}, 
which has been updated from uniform distributions after 
considering the constraints from the AMS-02 data set $D$ in \eq{eq:dataSetAMS}, 
i.e., $\tilde\pi(\boldsymbol\theta)=P(\boldsymbol\theta|D)$, 
where $P(\boldsymbol\theta|D)$ is calculated using the Bayes's theorem in \eq{eq:Bayes}.

\subsection{Constraints on DM properties from PAMELA antiproton data}

We consider several reference DM annihilation channels 
$\bar\chi\chi \to X$ 
where $X=b \bar b$, $t\bar t$, $W^{+}W^{-}$, $Z^{0}Z^{0}$ and $hh$.
The energy spectra of these channels are all similar at high energies.
The main difference is in the average number of total antiprotons $N_{X}$
per DM annihilation of each channel.
For a  DM particle mass $m_{\chi}=500$ GeV, 
the values of $N_{X}$ for typical final states are 
$N_{q\bar q} = 2.97~(q=u,d)$,
$N_{b\bar b}= 2.66$,
$N_{t\bar t}= 3.20$,
$N_{WW}=1.42$,
$N_{ZZ}= 1.48$,
and
$N_{h h}= 2.18$,
respectively.
Note that some of them are related.
For instance, 
$N_{hh}\approx 2 N_{b\bar b}\cdot \text{Br}^{2}(h\to b\bar b)$.

We include the data of antiproton flux 
and antiproton-to-proton flux ratio 
from the current PAMELA experiment
\cite{Adriani:2008zq,Adriani:2010rc}.
To avoid the complicities involved in modelling  the effect of solar modulation,
we only include the data points with antiproton kinetic energy $E>10$ GeV.
In total 8 (7) data points from antiproton flux (antiproton-to-proton flux ratio) 
are included in the analysis.
\mnote{(12)}
The DM profile is chosen to be  Einasto as a benchmark profile.
Note that changes in the results with other DM profiles can be estimated from 
\fig{fig:EPBC_PbarMinMax}.
For instance, 
the limits obtained for  Isothermal  and NFW are expected by to be slightly weaker
and that for the Moore profile should more stringent.

\begin{figure}
\includegraphics[width=0.49\textwidth]{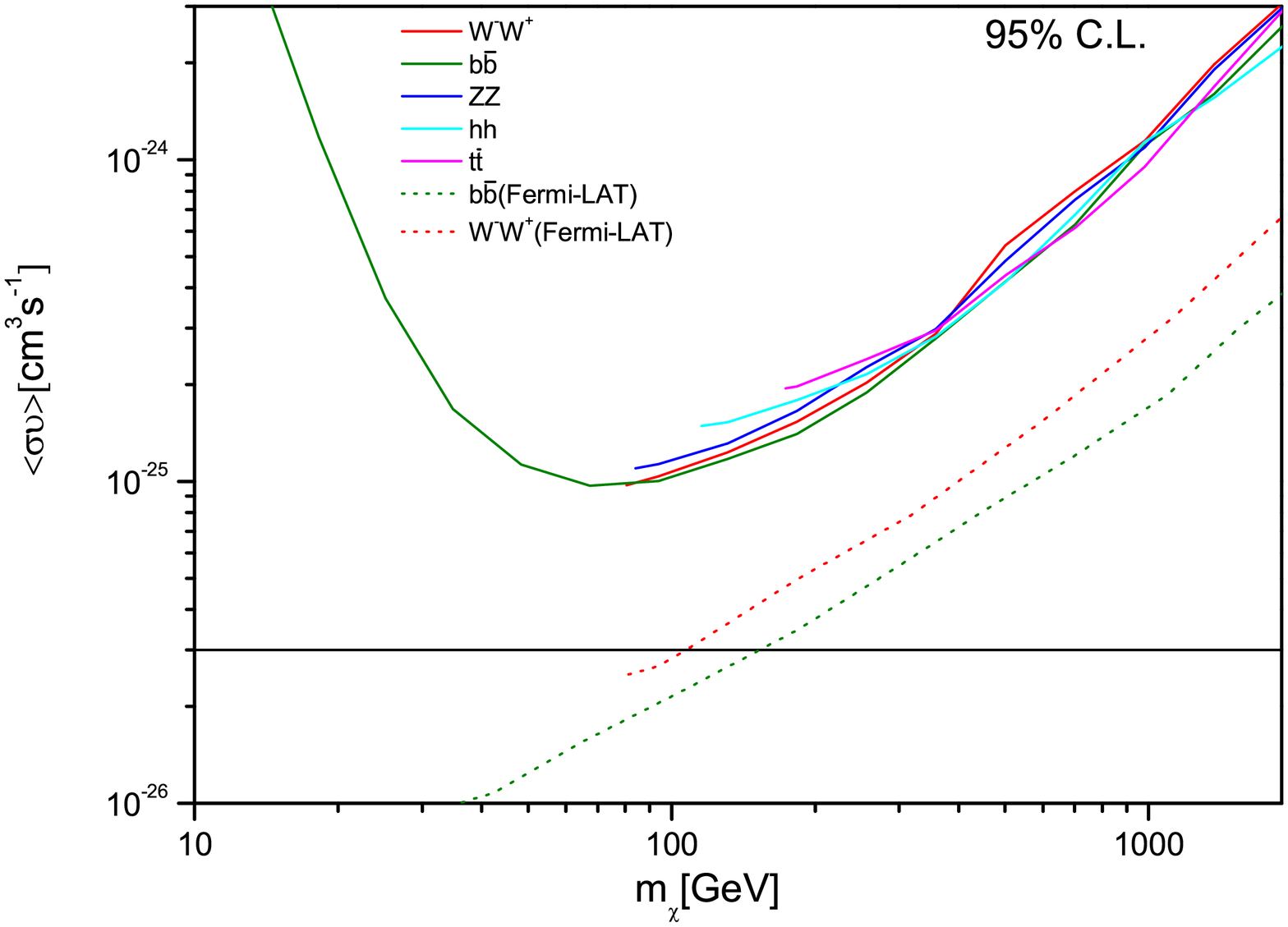}
\includegraphics[width=0.49\textwidth]{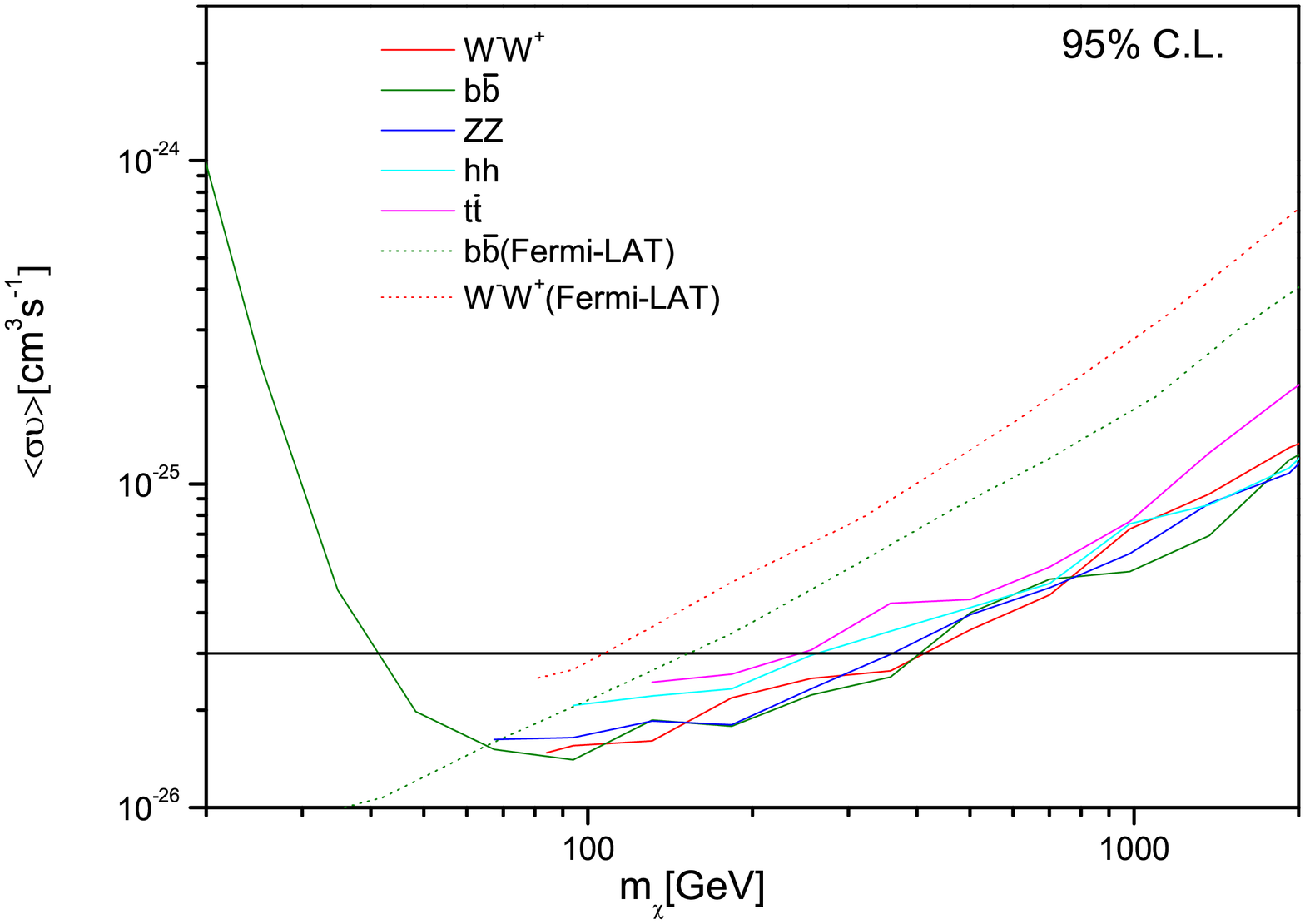}
\caption{
Left) upper limits on the cross sections for DM particle annihilating into 
$b\bar b$, $W^{+}W^{-}$, $Z^{0}Z^{0}$, $hh$ and $t\bar t$ final states at $95\%$ CL
with the uncertainties in the propagation models taken into account.
The DM halo profile is assumed to be Einasto.
The horizontal line indicates the typical thermal DM annihilation cross section
$\langle \sigma v \rangle_{0}=3\times 10^{-26}~\text{cm}^{3}\text{s}^{-1}$.
The upper limits on
the $b\bar b$ and $W^{+}W^{-}$ channels from 
the Fermi-LAT 6-year gamma-ray data of 
dwarf spheroidal satellite galaxies of the Milky Way
are also shown
\cite{Ackermann:2015zua}.
Right) the same as left, but for the mock data corresponding to
the AMS-02 three-year data taking, 
assuming background only.
}
 \label{fig:EPBC_pbar_uplimit}
\end{figure}

We use the method described in \eq{eq:Bayesian-update} to
obtain the upper limits on $\langle \sigma v \rangle$ for a give value of $m_{\chi}$, 
which takes into account of the uncertainties in the propagation parameters.
\fig{fig:EPBC_pbar_uplimit} shows the results of upper limits on 
the annihilation cross sections at $95\%$ CL.
The one-side $95\%$ CL upper  limit  is defined as 
the value of the quantity below where  
$95\%$ of the MCMC samples are found,
which corresponds to the value 0.95 of the cumulative distribution function.
When the uncertainties in the propagation parameters are included, 
the upper limits obtained are always above 
the typical thermal cross section $\langle \sigma v \rangle_{0}$
for the mass range $m_{\chi}\approx 10~\text{GeV}-1~\text{TeV}$
For $b\bar b$ final state, 
the most stringent limit is $\langle \sigma v \rangle  \lesssim 10^{-25}~\text{cm}^{3}\text{s}^{-1}$ 
at $m_{\chi}\approx 70$ GeV.
For TeV scale DM particle, 
the upper limits are around $10^{-24}~\text{cm}^{3}\text{s}^{-1}$
for all the channels.
For a comparison, 
in \fig{fig:EPBC_pbar_uplimit} we also show the upper limits on
the $b\bar b$ and $W^{+}W^{-}$ channels
obtained from 
the Fermi-LAT gamma-ray data of dwarf satellite galaxies of the Milky Way
\cite{Ackermann:2013yva}.
One can see from the figure  that 
when the uncertainties in the propagation parameters are considered,
the upper limits from the PAMELA $b\bar{b}$ data are 
slightly more stringent that from the gamma-ray data.

\subsection{Projected AMS-02 sensitivity}
The forthcoming AMS-02 data on the antiproton flux is eagerly awaited.
The AMS-02 detector has a high rejection power to distinguish antiprotons from protons,
which is extremely helpful in identifying small excesses in  the antiproton fluxes.
In this section, 
we investigate the prospect for  AMS-02 on 
reconstructing the property  of DM particle in the case where
an excess in the cosmic-ray antiproton flux over 
the conventional astrophysical background is identified  in 
the forthcoming AMS-02 antiproton data.

We generate mock data of  antiproton flux according to 
the specifications of the AMS-02 detector for the case of
an astrophysical background plus a contribution from 
DM annihilation into $b\bar b$ final states.
The binning of the kinetic energy spectrum of 
the antiproton flux is based on 
the rigidity resolution of the AMS-02 detector
which is obtained through fitting to the Fig.~2 of 
Ref. \cite{Zuccon:icrc2013}
\begin{align}
\frac{\Delta R}{R}=0.000477\times R+0.103 .
\end{align}
This value is for the observed event tracks hitting on both layer-1 and layer-9 of 
the AMS-02 silicon tracker. 
The rigidity resolution reaches $100\%$ for $R\approx 1.9$~TV, 
which roughly sets the upper limit  on 
the proton/antiproton rigidity that can be measured by the AMS-02 detector.
The relation between 
the resolution of the kinetic energy $T$ and  that of the rigidity reads
\begin{align}
\frac{\Delta T}{T}=
\left(\frac{T+2m_{p}}{T+m_{p}}\right)
\frac{\Delta R}{R},
\end{align}
where $m_{p}$ is the proton mass.
The expected number of antiprotons $N$ in
the $i$-th kinetic energy bin with kinetic energy $T_{i}$ for 
an exposure time $\Delta t$ is given by
\begin{align}
N=\epsilon a(T_{i})\phi(T_{i})\Delta T_{i} \Delta t ,
\end{align}
where 
$\epsilon$ is the efficiency  of the detector,
$a(T_{i})$ is the acceptance for antiproton at kinetic energy $T_{i}$,
$\phi(T_{i})$ is the expected antiproton flux,
and
$\Delta T_{i}$ is the width of the $i$-th kinetic energy bin.
From Ref.~\cite{Malinin:2004pw}, 
the acceptance  is 
$a(T)\approx 0.147~\text{m}^{2}$ for $1~\text{GeV}\leq T \leq~11\text{GeV}$ 
and
$a(T)\approx 0.03~\text{m}^{2}$ for $11~\text{GeV} \leq T \leq 150~\text{GeV}$. 
For $T \geq 150 ~\text{GeV}$, 
the acceptance drops very quickly with increasing kinetic energy. 
In numerical calculations, we interpolate the values of $a(T)$ from Fig.~8 of Ref.~\cite{Malinin:2004pw}.
The efficiency is assumed to be a constant $\epsilon=0.9$ in this work. 
Due to the geomagnetic effects, 
the value of $\epsilon$ becomes 
very low at kinetic energies below $1$~GeV
\cite{BAttiston:2013},
we thus only consider the mock data above 1 GeV.

Under the assumption that the distribution of the observed antiproton events is Poissonian,
the statistic uncertainty in $N$ observed events is $\Delta N=\sqrt{N}$. 
Thus the statistic uncertainty in the flux $\phi(T_{i})$ is 
\begin{align}
\Delta \phi(T_{i})_{\text{sta}}
=
\sqrt{
\frac{\phi(T_{i})}{\epsilon a(T_{i}) \Delta T_{i} \Delta t}
} .
\end{align}
The systematic uncertainties may have various sources,
such as
the misidentification of background protons and electrons as antiprotons. 
The AMS-02 detector has a rejection power of $p:\bar{p}\sim 10^{5}-10^{6}$ for protons
and $e^{-}:\bar{p}\sim 10^{3}-10^{4}$ for electrons. 
At multi-GeV energy region, the flux ratios of $p/\bar{p}$ and $e^{-}/\bar{p}$ are $\sim 10^{4}$
and $\sim 10^{2}$ respectively.
Thus   the systematic uncertainty  can reach $\sim 1-10\%$. 
In this work, we take the systematic uncertainty to be $\Delta \phi_{\text{sys}}=8\%$. 
The total uncertainty is 
$\Delta \phi(T_{i}) =\sqrt{\Delta \phi(T_{i})_{\text{sta}}^{2}+\Delta \phi_{\text{sys}}^{2}}$.

\begin{figure}
\includegraphics[width=0.49\textwidth]{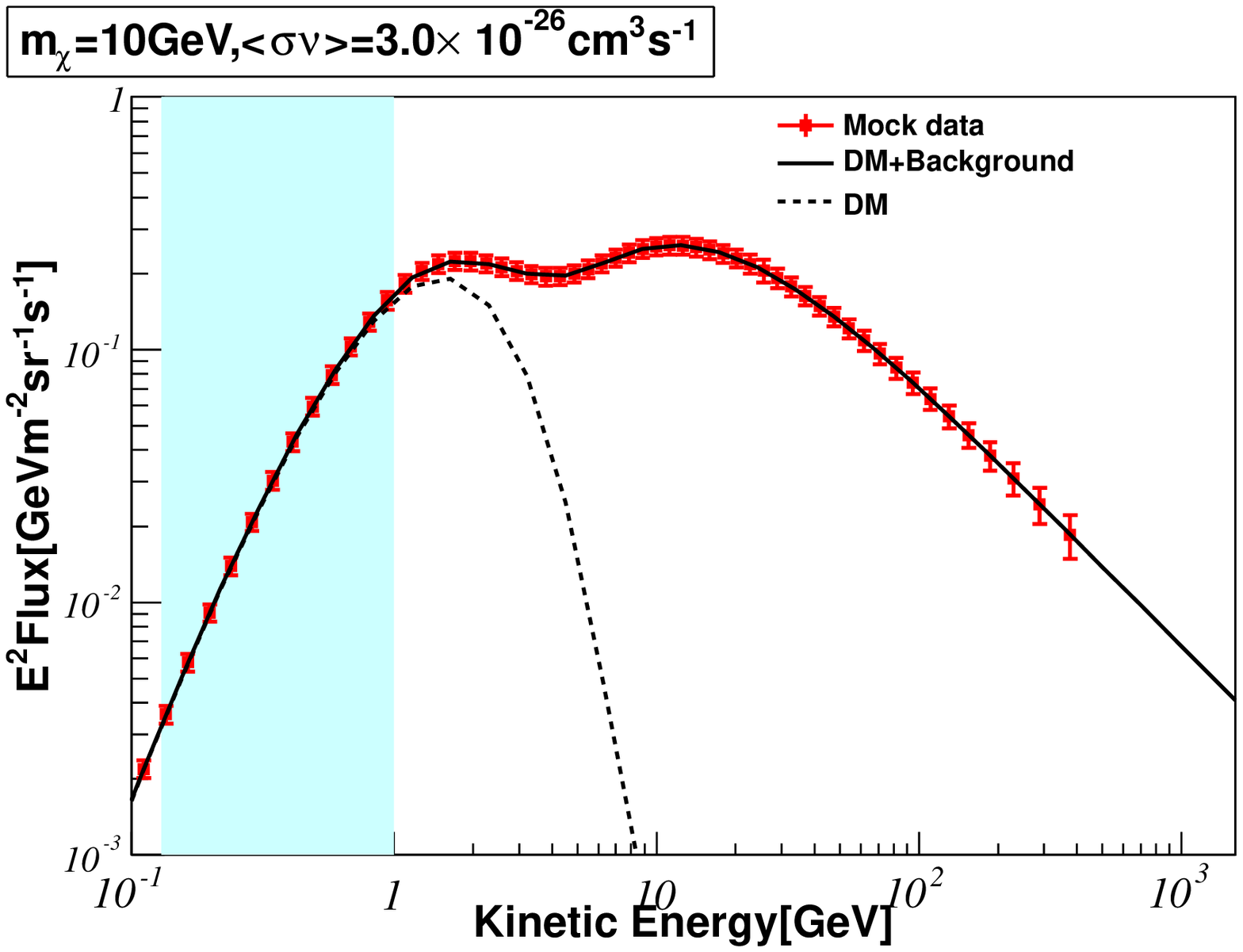}
\includegraphics[width=0.49\textwidth]{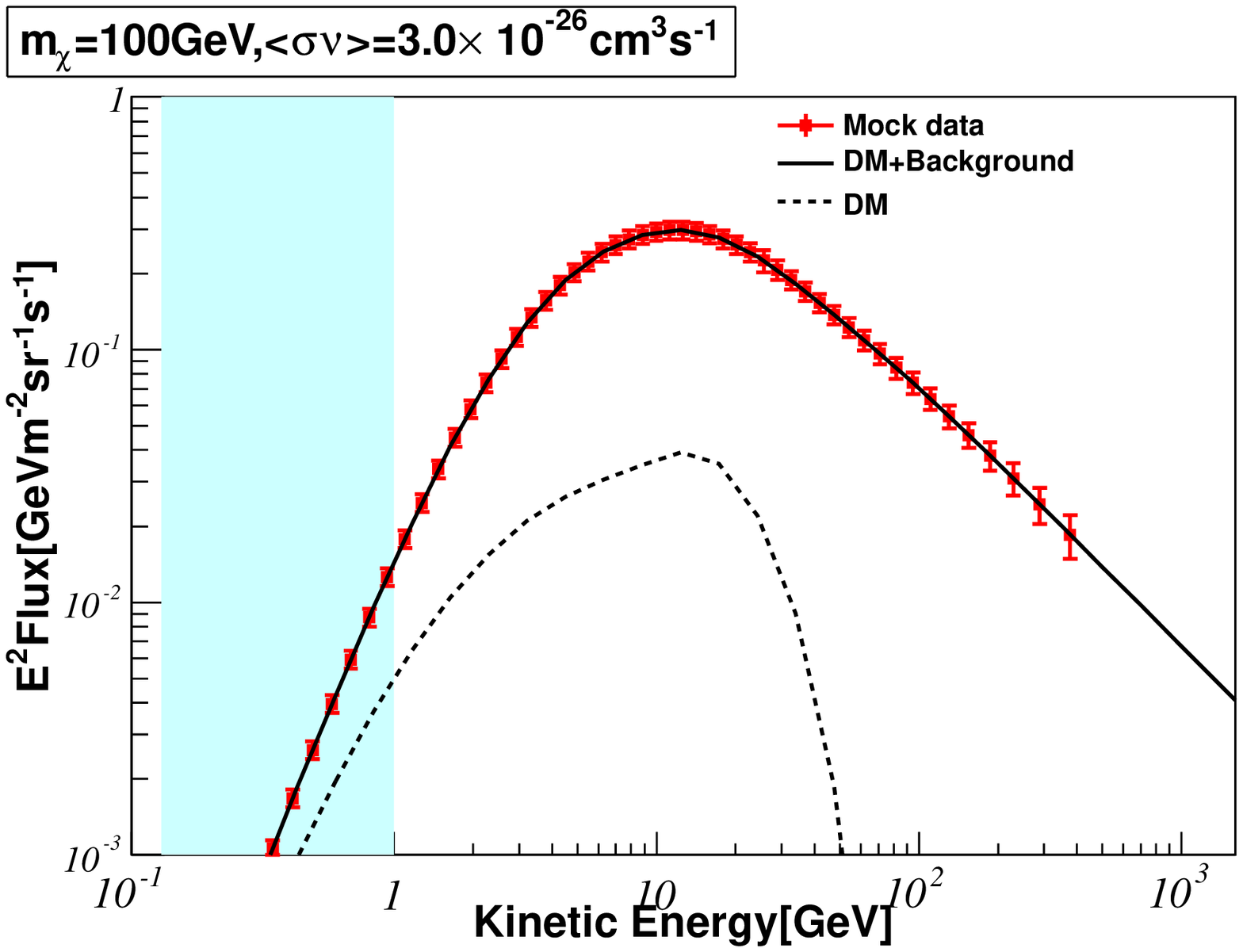}
\\
\includegraphics[width=0.49\textwidth]{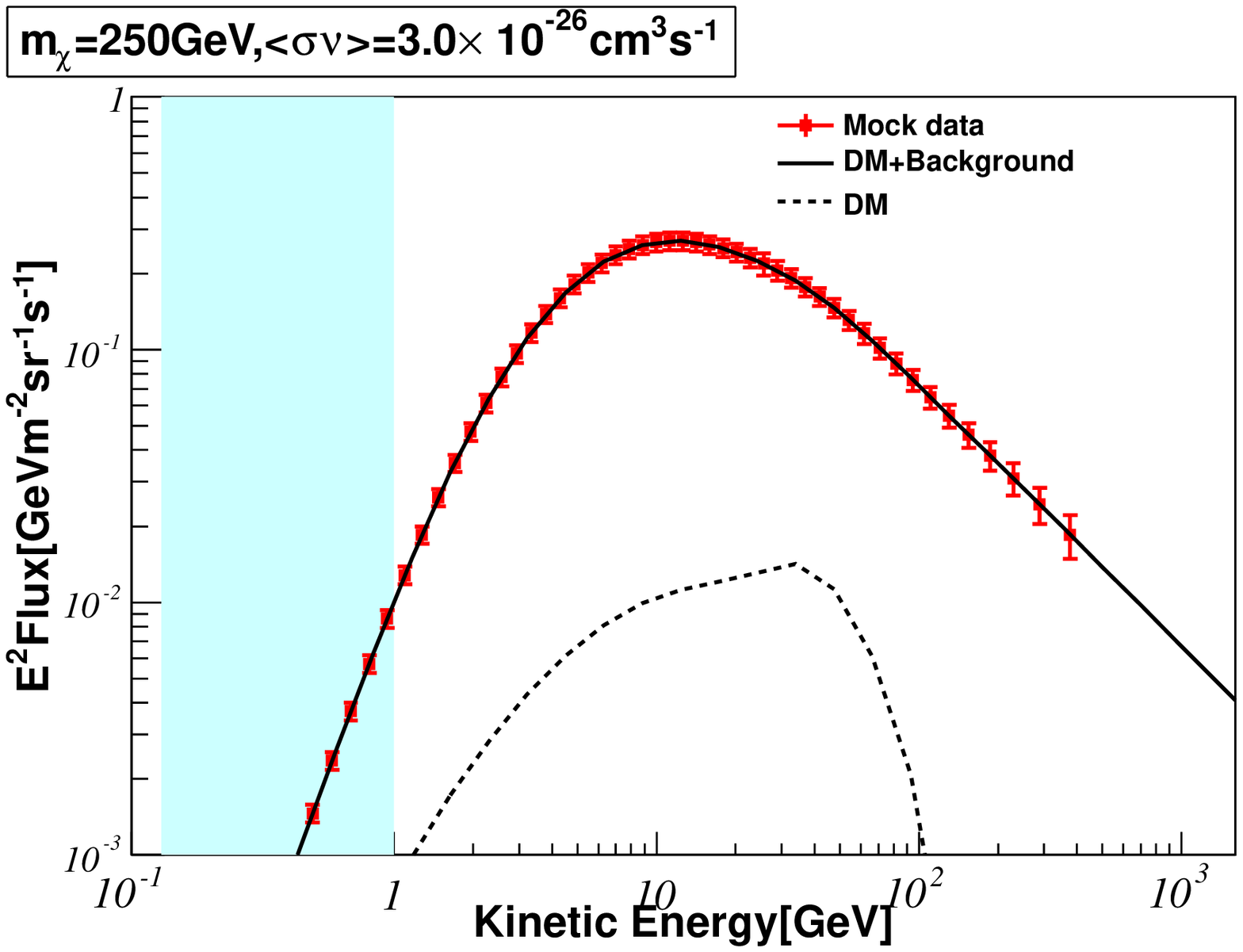}
\includegraphics[width=0.49\textwidth]{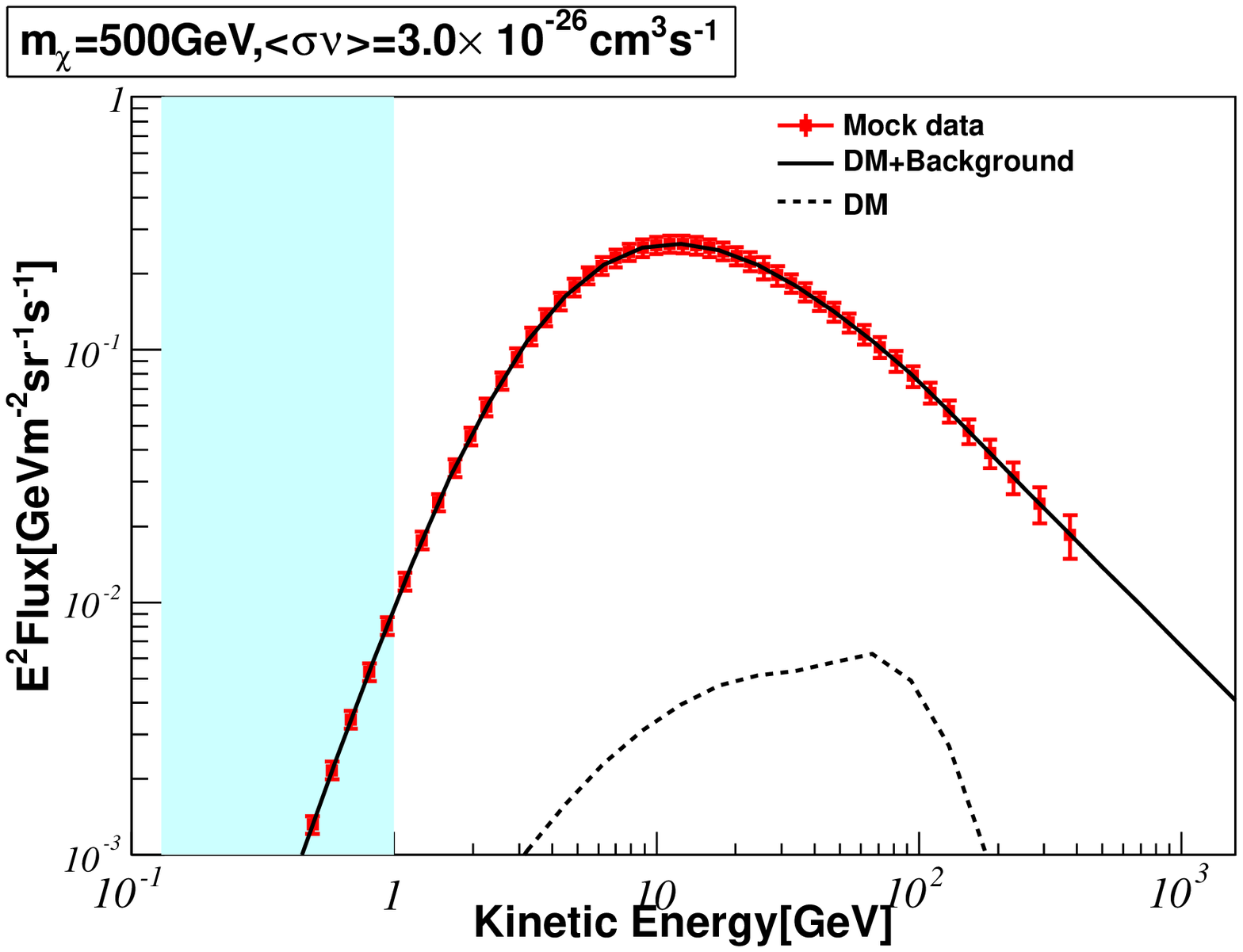}
\\
\includegraphics[width=0.49\textwidth]{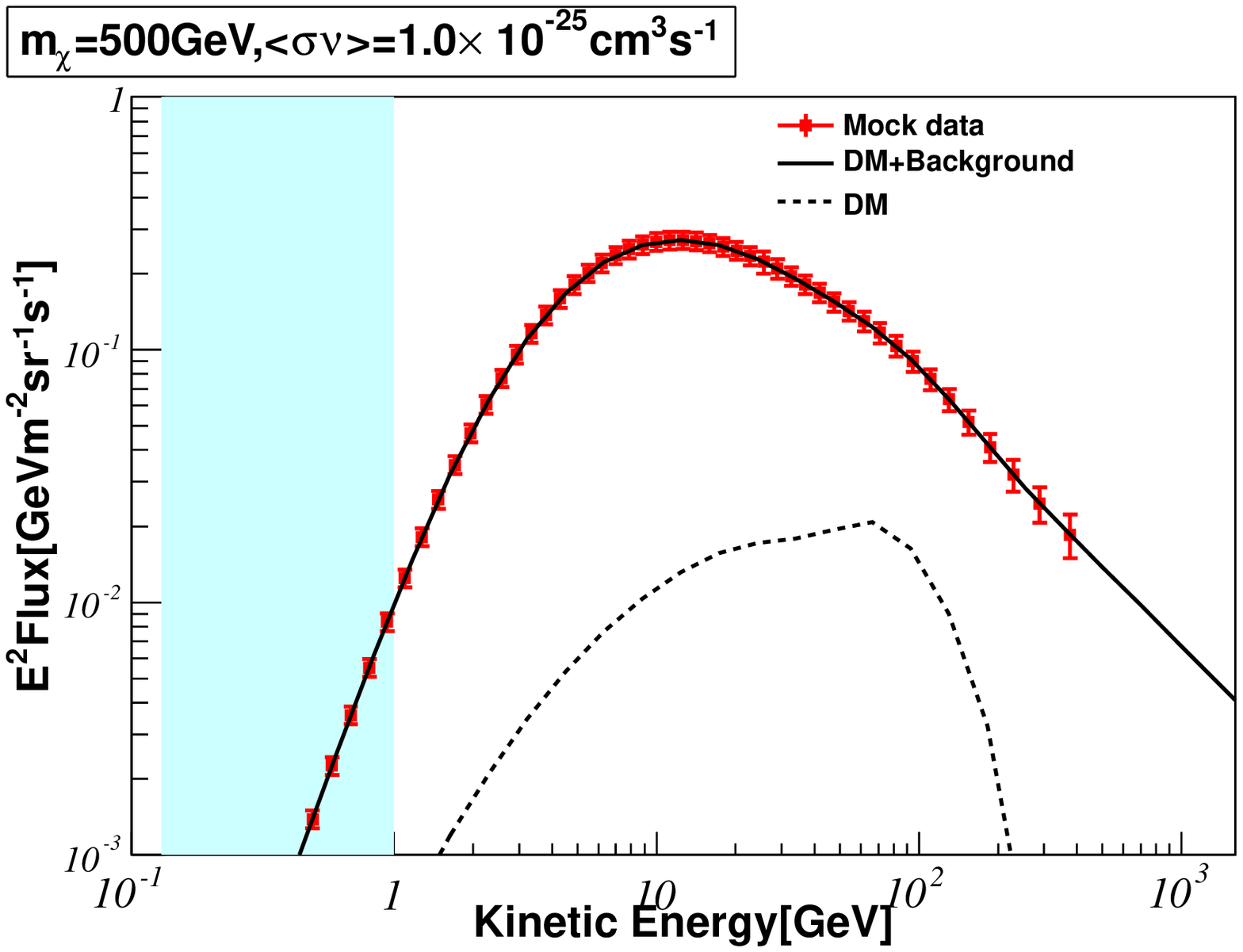}
\includegraphics[width=0.49\textwidth]{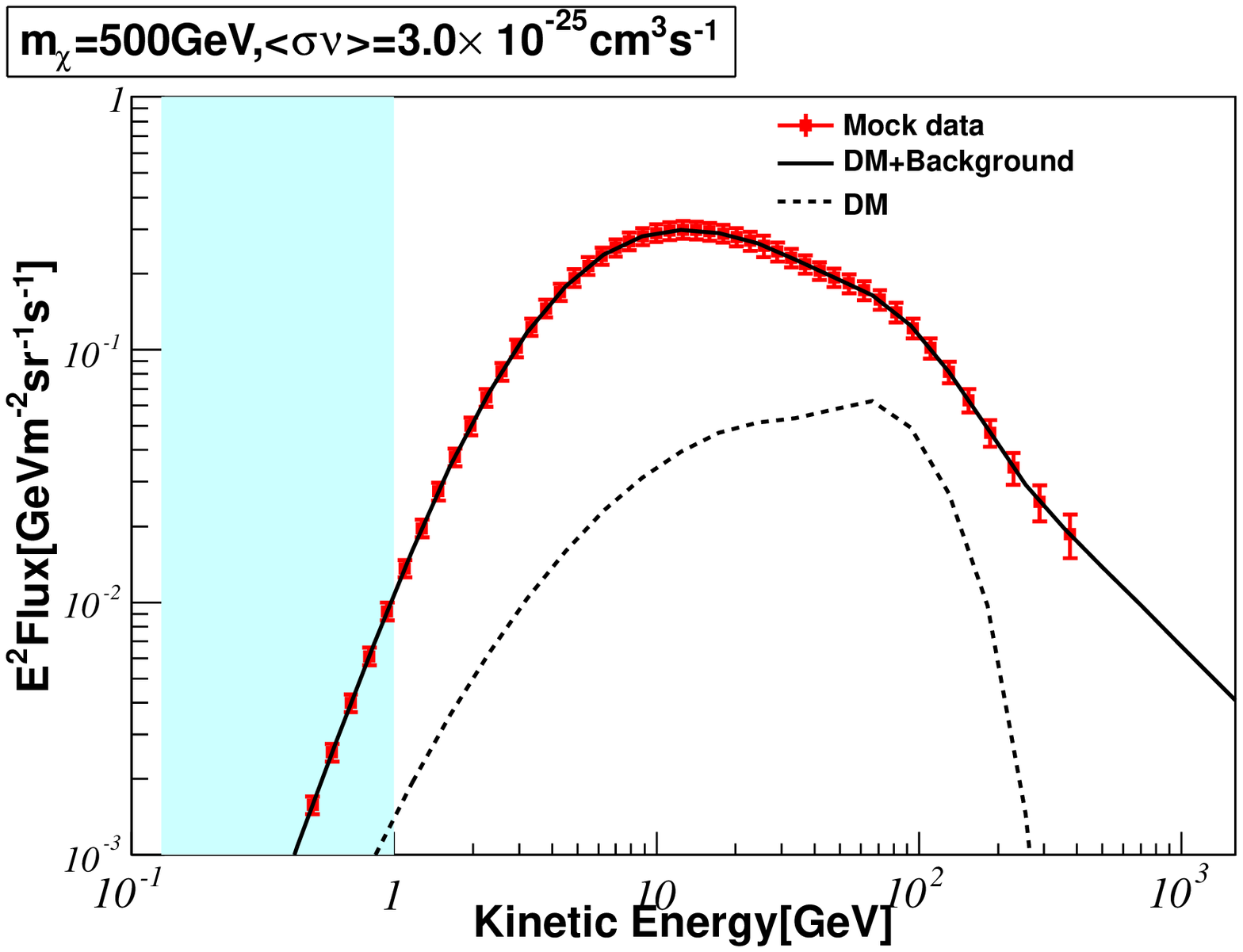}
\caption{
Mock data of 
the projected AMS-02 antiproton flux with 
3 years of data taking in the assumption of 
DM annihilating into $b\bar b$ final states 
with a typical thermal cross section  $\langle \sigma v \rangle_{0}=3\times 10^{-26}~\text{cm}^{3}\text{s}^{-1}$ for DM particle masse $m_{\chi}=10$, 100, 250, 500 GeV, respectively, 
and the cases of  large cross sections $\langle \sigma v \rangle=1 \text{ and }3\times 10^{-25}~\text{cm}^{3}\text{s}^{-1}$ for $m_{\chi}=500$ GeV.
In each plot,
the dashed line represents the contribution from DM only, 
and the solid line represents the sum of background and DM  contribution.
The background is generated from 
the best-fit propagation parameters shown in \tab{tab:param}.
The halo DM profile is assumed to be Einasto.
The mock data with kinetic energy below 1 GeV (shadowed region) is not used for 
the reconstruction of DM properties due to the geomagnetic cut off of 
the detection efficiency.
}
 \label{fig:EPBC_pbar_mockdata}
\end{figure}

\begin{figure}
\includegraphics[width=0.49\textwidth]{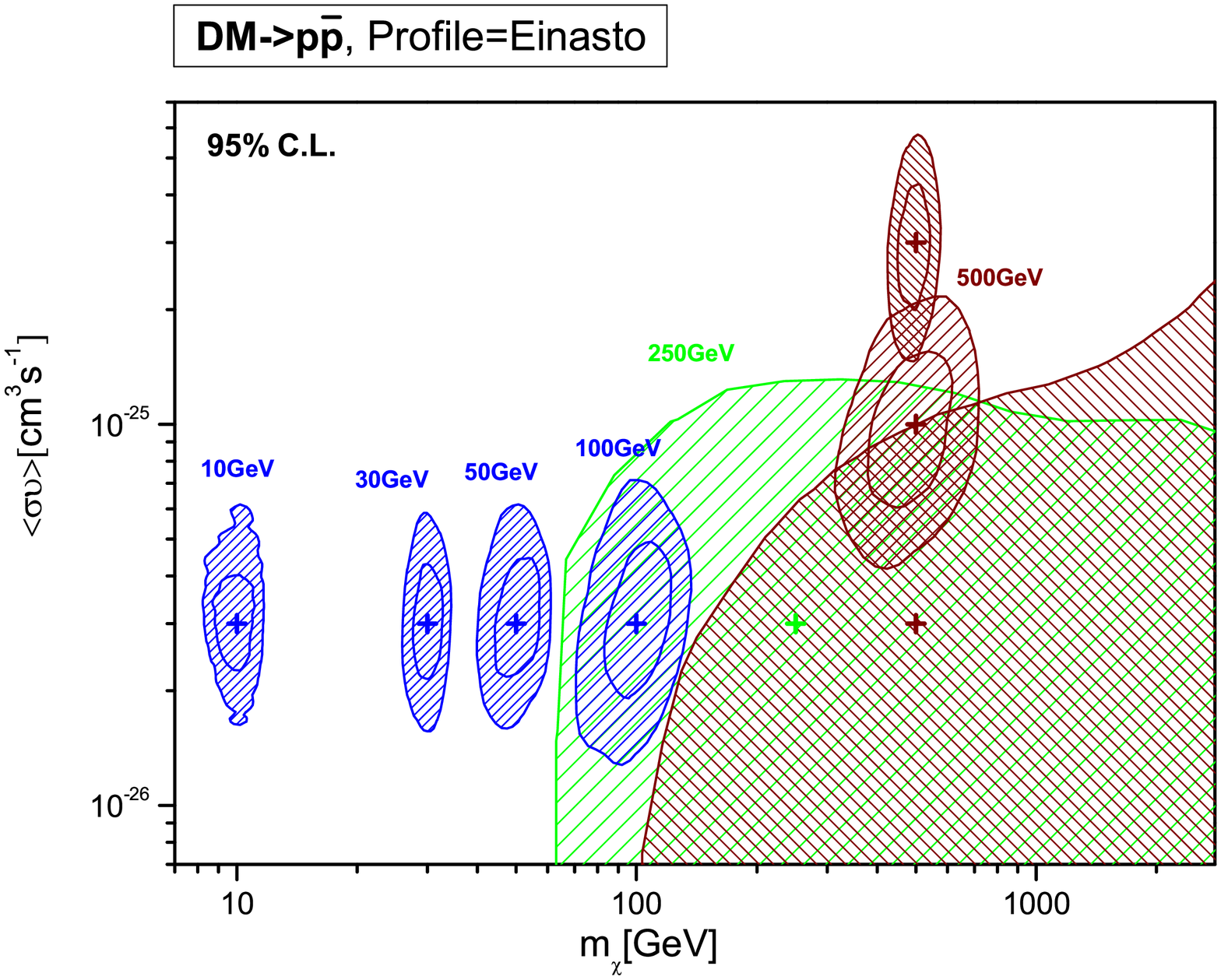}
\includegraphics[width=0.49\textwidth]{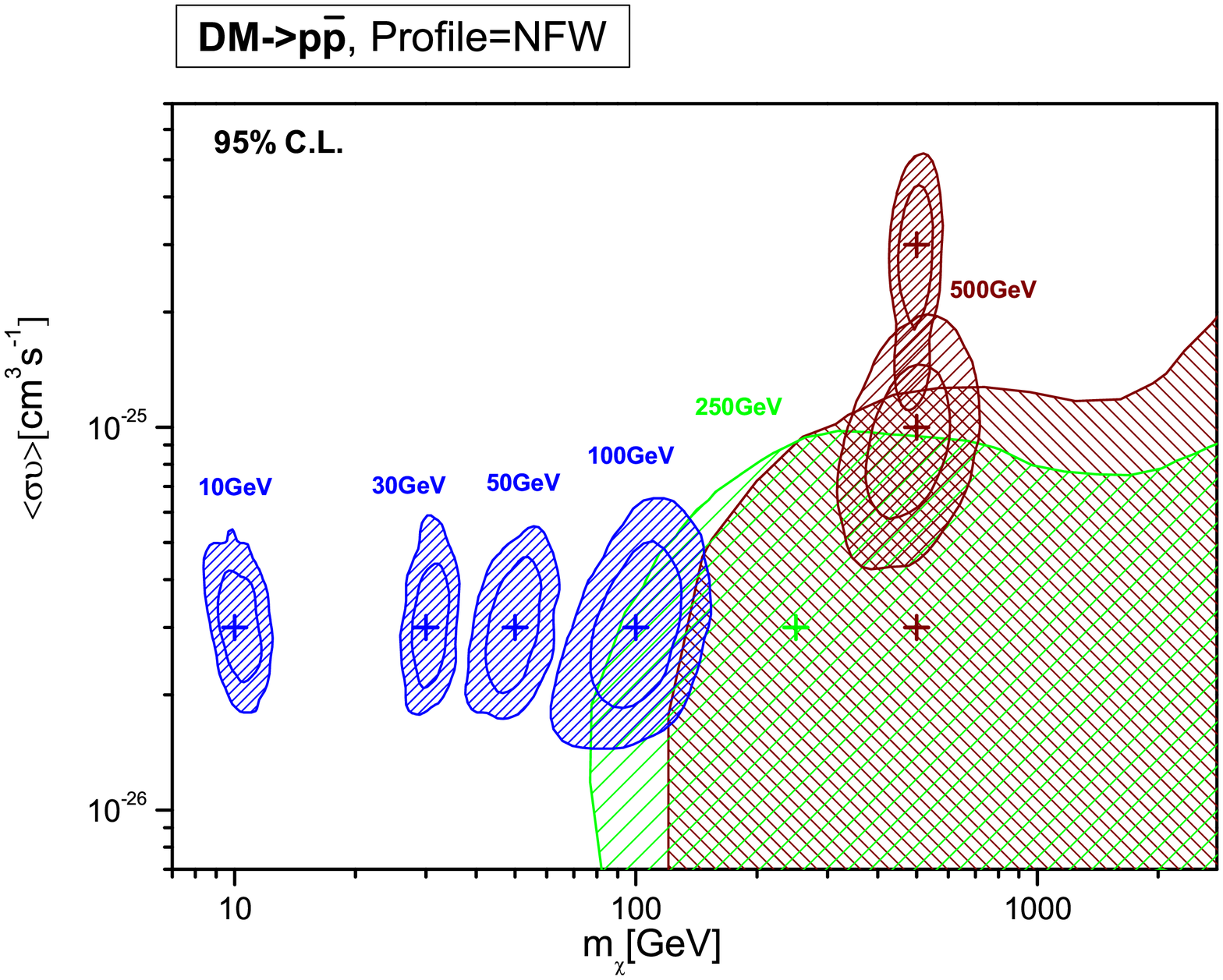}
\caption{
Left) reconstructed allowed regions of 
DM particle mass and annihilation cross section at 
$68\%$ and $95\%$ CLs  from 
the mock data of antiproton flux. 
The mock data correspond to 
the projected AMS-02 antiproton flux with 3 years of data taking 
in the assumption of DM annihilating into $b\bar b$ final states 
with a typical thermal cross section  
$\langle \sigma v \rangle_{0}=3\times 10^{-26}~\text{cm}^{3}\text{s}^{-1}$ for 
DM particle masses $m_{\chi}=10$, 30, 50, 100, 250 and 500~GeV,
and 
the cases of  large cross sections 
$\langle \sigma v \rangle=1 \text{ and }3\times 10^{-25}~\text{cm}^{3}\text{s}^{-1}$ for 
$m_{\chi}=500$~GeV.
The DM profile is assumed to be Einasto.
Right) the same as left, 
but for the NFW profile.
}
 \label{fig:EPBC_pbar_mockdataContour}
\end{figure}

In \fig{fig:EPBC_pbar_mockdata},
we show the mock data  of 
the projected AMS-02 antiproton flux with 3-year data taking.
The antiproton background is generated according to 
the best-fit propagation parameters listed in \tab{tab:param}.
We assume that the DM particles annihilate into $b\bar b$ final states 
with a typical thermal cross section  
$\langle \sigma v \rangle_{0}=3\times 10^{-26}~\text{cm}^{3}\text{s}^{-1}$ for 
different  masse $m_{\chi}=10$, 100, 250 and 500 GeV, respectively, 
and the cases of  large cross sections 
$\langle \sigma v \rangle=1 \text{ and }3\times 10^{-25}~\text{cm}^{3}\text{s}^{-1}$ 
for a large $m_{\chi}=500$ GeV.
The halo DM profile is assumed to be Einasto.
As can be seen from the figure,
only in the cases where 
a light 10 GeV DM particle with typical thermal cross section or 
a heavy 500 GeV DM particle with a large cross section,
the DM contribution can lead to a visible change in the antiproton flux.
However, 
it is still possible that 
a tiny change in the spectrum of antiproton flux can be
identified by the AMS-02 experiment.
\mnote{15.}
We first consider the case without DM contribution,
i.e., the future AMS-02 data is consistent with the background.
In this case, 
upper limits can be derived as a function of $m_{\chi}$
as it is done for the PAMELA data.
We follow the same treatment to apply a 10 GeV cut to the mockdata
as in the case of PAMELA.
The result as shown in 
the right panel of \fig{fig:EPBC_pbar_uplimit} indicates that
much stronger limits can be obtained for the AMS-02 three-year data taking,
which can be compatible with that from 
the current Fermi-LAT gamma data.
We then investigate the reconstruction capability for two specific cases
in the Einasto and NFW profiles.
In one case
the DM annihilation cross section is fixed to 
the standard thermal cross section, i.e.,
$\langle \sigma v\rangle=\langle \sigma v\rangle_{0}$ 
and 
the DM particle mass is allowed to vary in the range $\sim 10-500$ GeV. 
In \fig{fig:EPBC_pbar_mockdataContour}, 
we show the results of the reconstruction for  
$m_{\chi}=10$, 30, 50, 100, 250 and 500 GeV, respectively.
The figure shows that 
for $m_{\chi} \lesssim100$ GeV, 
the annihilation cross section  can be reconstructed with 
uncertainties about a factor of two for both 
Einasto and NFW profiles.
For a fixed annihilation cross section, 
the reconstruction becomes difficult for heavier DM particle, 
as the source term  is suppressed by  $m_{\chi}^{2}$.
As shown in \fig{fig:EPBC_pbar_mockdataContour},
when $m_{\chi} >250$ GeV, 
only an upper limit is obtained from the mock data.
In the other case, 
the DM particle mass  $m_{\chi}$ is fixed at $500$ GeV and 
$\langle \sigma v \rangle$  differs significantly from $\langle \sigma v\rangle_{0}$.
For large annihilation cross sections 
$\langle \sigma v \rangle=1\times 10^{-25}~\text{cm}^{2}$ 
and $3\times 10^{-25}~\text{cm}^{2}$, 
we find that 
the cross section can still be well reconstracted with 
uncertainty typically about a factor of two.
In both the cases, 
we find that the DM particle mass can be well reconstructed 
with uncertainties less than $\sim 30\%$.

\section{Conclusions}\label{sec:conclusion}
The AMS-02 experiment is measuring 
the spectra of cosmic-ray nuclei fluxes with 
unprecedented accuracies,
which is of crucial importance in 
understanding the origin and propagation of 
the cosmic rays and 
searching for dark matter.
We have performed a global Bayesian analysis of
the constraints on the cosmic-ray propagation models from 
the recent AMS-02 data on 
the ratio of Boron to Carbon nuclei and proton flux
with the assumption that 
the primary source is a broken power law in rigidity.
The analysis is based on the method of MCMC sampling.
The result has shown  that 
the propagation parameters can be well determined by the AMS-02 data  alone. 
For instance, 
the ratio of  the diffusion coefficient  to the diffusive halo height is found to be 
$D_{0}/Z_{h}\simeq 2.0~\text{cm}^{2}\text{s}^{-1}\text{kpc}^{-1}$
with uncertainty less than $5\%$. 
The best-fit value of the halo width is  
$Z_{h}\simeq 3.3$ kpc with uncertainty less than $50\%$.
Other parameters such as 
the Alfv$\grave{\mbox{e}}$n speed and 
the power law indices of the primary sources have also been determined. 
Such results can be used to improve the prediction of the antiproton flux
from DM interactions.
Using the allowed regions of parameter space, 
we have estimate the uncertainties in 
the positron fraction and 
antiproton fluxes predicted by DM annihilation.
We have shown that  
the uncertainty in the predicted 
positron fraction is within a factor of two and
that in the antiproton flux is within an order of magnitude,
which are much smaller than the estimations in the previous analyses  prior to AMS-02.
With all the uncertainties and correlations in 
the propagation parameters taken into account,
we have derived conservative upper limits on 
the cross sections for DM annihilating into 
various standard model final states from 
the current PAMELA antiproton data.
We have also investigated 
the  reconstruction capability  of the future AMS-02 antiproton data on the 
DM properties.
The result have shown that 
if the  DM particles are lighter than  100 GeV and 
the annihilation cross section is the typical thermal  cross section,
the annihilation cross section can be well reconstructed  with 
uncertainties around  a factor of two
for the AMS-02 three-year data taking.

\vskip 1cm
\subsection*{Acknowledgments}
We are grateful to  S. Ting for warm hospitality and 
insightful discussions during our visit to the AMS-02 POCC at CERN.
We thank P. Zuccon, A. Kounine, A. Oliva and S. Haino for 
helpful discussions on the details of the AMS-02 detector. 
YLW and YFZ also thank 
R.K. Su, Z.W. Li, Y.Q. Ma, D. Jin, X. Cai and X.M. Zhang for 
their strong support in the 973 program and useful discussions.
This work is supported in part by
the National Basic Research Program of China (973 Program) under Grants 
No. 2010CB833000;
the National Nature Science Foundation of China (NSFC) under Grants 
No. 10821504,  
No. 10905084,
No. 11335012 and
No. 11475237;
The numerical calculations were done  using
the HPC Cluster of SKLTP/ITP-CAS.

\bibliography{amsfit_inspire,misc}
\bibliographystyle{JHEP}

\end{document}